\documentclass[11pt]{article}

\usepackage[toc,page,header]{appendix}
\usepackage{soul}
\usepackage{amsmath}				
\usepackage{amssymb}				
\usepackage{amsfonts,commath,bm} 	
\usepackage[version=3]{mhchem}		
\usepackage{wasysym}				
\usepackage{setspace}				
\usepackage{overcite}				
\usepackage{graphicx}				
\usepackage{wrapfig}					
\usepackage{multirow}				
\usepackage{tocloft}
\usepackage{color}				
\usepackage{caption, setspace}
\usepackage[dvipsnames]{xcolor}	
\usepackage[braket]{qcircuit}
\usepackage{enumitem}   
\usepackage{hyperref}
\usepackage{multirow}
\usepackage{multicol}

\usepackage[top=1in, bottom=1in, left=1in, right=1in]{geometry} 	
\usepackage[compact]{titlesec}								
\titlespacing*{\section}
{0pt}{0pt}{0.05in}
\titlespacing*{\subsection}
{0pt}{5pt}{0.05in}

\usepackage[font=footnotesize,labelfont=bf,labelsep=period,width=0.95\textwidth]{caption} 	
\usepackage[font=footnotesize,labelfont=bf,labelsep=period]{subcaption} 	
\DeclareCaptionSubType*[arabic]{figure} 						
\DeclareCaptionLabelFormat{subfiglabel}{Figure #2} 						
\usepackage[capitalize]{cleveref}
\crefname{figure}{Fig.}{Figs.}
\Crefname{figure}{Figure}{Figures}
\crefname{table}{Tab.}{Tabs.}
\Crefname{table}{Table}{Tables}
\crefname{equation}{Eq.}{Eqs.}
\Crefname{equation}{Equation}{Equations}
\crefname{section}{Sec.}{Secs.}
\Crefname{section}{Section}{Sections}

\begin{document}
\begin{center}
{\LARGE Future Directions of the Cyberinfrastructure for Sustained Scientific Innovation (CSSI) Program}\\

\vspace{5pt}
{\large CSSI 2019 Workshop Report}

\vspace{10pt}
Ritu Arora, Texas Advanced Computing Center (Co-Chair)\\
Xiaosong Li, University of Washington (Co-Chair)\\
\vspace{10pt}
Bonnie Hurwitz, University of Arizona (Steering Committee Member)\\
Daniel Fay, Microsoft (Steering Committee Member)\\
Dhabaleswar K. Panda, The Ohio State University (Steering Committee Member)\\
Edward Valeev, Virginia Tech University (Steering Committee Member)\\
Shaowen Wang, University of Illinois (Steering Committee Member)\\
Shirley Moore, Oak Ridge National Lab (Steering Committee Member)\\
Sunita Chandrasekaran, University of Delaware (Steering Committee Member)\\
Ting Cao, University of Washington (Steering Committee Member)\\

\vspace{10pt}
\emph{Additional Contributors to Report Preparation:}

Holly Bik, University of California, Riverside\\
Matthew Curry, Sandia National Laboratories\\
Tanzima Islam, Texas State University\\

\vspace{10pt}
\vspace{10pt}
\vspace{10pt}
\vspace{10pt}
\vspace{10pt}
\emph{This report was submitted in April 2020 to the National Science Foundation (NSF).}

\end{center}
\vspace{10pt}
\vspace{10pt}

\tableofcontents
\newpage
\section{Executive Summary}
The CSSI 2019 workshop was held on October 28-29, 2019, in Austin, Texas. The main objectives of this workshop were to (1) understand the impact of the CSSI program on the community over the last 9 years, (2) engage workshop participants in identifying gaps and opportunities in the current CSSI landscape, (3) gather ideas on the cyberinfrastructure needs and expectations of the community with respect to the CSSI program, and (4) prepare a report summarizing the feedback gathered from the community that can inform the future solicitations of the CSSI program. 

The workshop participants included a diverse mix of researchers and practitioners from academia, industry, and national laboratories. The participants belonged to diverse domains such as quantum physics, computational biology, High Performance Computing (HPC), and library science. Almost 50\% participants were from computer science domain and roughly 50\% were from non-computer science domains. As per the self-reported statistics, roughly 27\% of the participants were from the different underrepresented groups as defined by the National Science Foundation (NSF).

The workshop brought together different stakeholders interested in provisioning sustainable cyberinfrastructure that can power discoveries impacting the various fields of science and technology and maintaining the nation’s competitiveness in the areas such as scientific software, HPC, networking, cybersecurity, and data/information science. The workshop served as a venue for gathering the community-feedback on the current state of the CSSI program and its future directions. 

Before they arrived at the workshop, the participants were encouraged to take an online survey on the challenges that they face in using the current cyberinfrastructure and the importance of the CSSI program in enabling cutting-edge research. The workshop included 16 brain-storming sessions of one hour each. Additionally, the workshop program included 16 lightning talks and an extempore session. The information collected from the survey, brainstorming sessions, lightning talks, and the extempore session are summarized in this report and can potentially be useful for the NSF in formulating the future CSSI solicitations. The workshop fostered an environment in which the participants were encouraged to identify gaps and opportunities in the current cyberinfrastructure landscape, and develop thoughts for proposing new projects.

\section{Overview of the NSF CSSI Program}
The CSSI program funds projects that:
\begin{itemize}
\item support the development and deployment of robust, reliable and sustainable data and software cyberinfrastructure, 
\item bring innovative capabilities towards sustained scientific innovation and discovery,
\item provide a cross-directorate opportunity to advance common approaches to sustain and innovate research cyberinfrastructures,
\item follow accepted data management and software development practices.
\end{itemize}
The successful projects should be science-driven, innovative, collaborative, and where possible, should build on existing capabilities. The projects should include management plans and metrics that encourage measurement of progress and sharing of results with a broad community. The projects should produce widely accessible, long-term community cyberinfrastructure.
Following is the list of the CSSI proposal evaluation criteria:

\begin{enumerate}
\item Intellectual Merit/s: Can the project advance, if not transform, the frontiers of knowledge? Is the project innovative?
\item Does the project have the potential of creating ``Broader Impacts'' (societal outcomes)?
\item Is the project science-driven?
\item Does the proposal demonstrate close collaboration with the different stakeholders?
\item Does the proposal clearly describe the role of the collaborators? 
\item Does the project builds on existing capabilities?
\item Is the project plan and the software engineering process clearly articulated?
\item Does the proposal include evaluation metrics, including the community-usage metrics for each year of the award?
\item Does the proposal mention the license under which the software/data product will be released to the public?
\item Does the proposal include the information on the dissemination of the product?
\item How will the project sustain itself after the funding is over?
\item Is the data management plan appropriate?
\end{enumerate}
Currently, a Principal Investigator (PI), Co-PI, or Senior Personnel can be named on only one proposal submitted to the CSSI program for each evaluation window.

\section{Recommendations}
The discussions held during, before, and after the CSSI 2019 workshop, and the survey responses provided by the workshop participants were used to formulate the recommendations presented in this section. The future CSSI solicitation could potentially include additional guidance on the following topics: evaluation metrics, scope of the projects, letters of collaboration, community building, best practices and policies, sustainability, and project staffing. These topics are detailed in the sub-sections below.
\subsection{Evaluation Metrics}
The CSSI program requires that the proposals discuss the evaluation metrics, including the community-usage metrics for each year of the awards. The CSSI program very realistically allows the evaluation metrics to change or evolve with the progress of the projects, and the PIs are allowed to include additional indicators of success as the projects’ progress. Some of the workshop participants commented that the need for robust evaluation metrics should be emphasized further in the solicitation by providing some general but concrete examples. They suggested that the metrics should be classified as short-term and long-term. The short-term evaluation metrics are the yearly evaluation metrics that the CSSI solicitation already advises the PIs to provide. In addition to these short-term metrics, the already established projects --- older than 5 years --- should also be required to propose long-term metrics for assessing their success in case they go in for the second round of funding. Having both short-term and long-term metrics is important because unlike research projects, the CSSI projects typically focus on product development and community-building. The PIs of the CSSI projects could be spending the majority of the duration of their projects for product development. By the time a product is ready, there may not be sufficient time left for collecting the data on its usage by the community. In such cases, perhaps, it may be useful to allow the PIs to report the metrics on the adoption of their products even after the project is over. Such metrics are long-term and could be useful for evaluating the projects if they seek a second round of funding, and can also help in evaluating the long-term impact of the projects. Such metrics could be submitted as an addendum to the final report or the outcomes report. In fact, a third report --- that is, a report in addition to the final project report and the project outcome report --- specifically related to the metrics could also be requested from the PIs to quantify the impact that the projects have created on the community. Examples of the short- and long-term metrics that could be provided are as follows:
	\begin{enumerate}[label=(\roman*)]
	\item Short-term metrics (yearly, as already mentioned in the CSSI solicitation, for all projects requesting funding): Number of users, frequency of product usage, number of downloads, and website visits (if applicable), bug reports, support forum posts, number of contributing external developers, number of case studies or new sciences enabled, community-building activities
	\item Long-term metrics (for projects that are $\ge$ 5 years old): Self-conducted user-studies, highlighted success stories, number of software citations, performance and usability of the products, and continued community-building activities
	\end{enumerate}

\subsection{Scope of the Projects}
The CSSI program is very broad is terms of the disciplines and scope of the projects that it funds. The program, very fairly, lays emphasis on the advancement of science through the innovations in the CyberInfrastructure (CI). Some of the workshop participants felt that proposing an innovative and sustainable CI that helps in accomplishing novel scientific outcomes can be challenging as an Elements project. They were unclear if the projects that propose innovative CI but include a limited number of science use-cases in the proposal (as compared to showing transformative and sustainable impact on the scientific disciplines) could be considered competitive or not. Moreover, it was not clear to them if the level (or amount) of expected innovation and sustainability from an Elements project is different from that of a Frameworks project. While the CSSI solicitation already clarifies on the scope of the Elements and Frameworks projects, it would be nice if the solicitation further clarifies whether there is any difference in the expected level of sustainability between these two categories of projects.

\subsection{Letters of Collaboration}
Currently, the CSSI program requires that the letters of collaboration from unfunded collaborators be provided in a standard format. While this brings uniformity, the prescribed format of the letter is very general and precludes the collaborators from explaining the value that they agree to add to the project. It is recommended that the unfunded collaborators briefly describe their specific contributions to the projects in their letters of collaboration, and their resume be added to the proposal so that their role and level of contributions to the project can be evaluated unambiguously and objectively. This will also imply that the collaborators’ role is not over after providing the letters, and that they really have the resources to support all the projects that they are providing the letters for.

\subsection{Community Building}
An engaged user-community is needed for supporting and evolving the products after the funding is over. Therefore, it is commendable that the community-development activities are encouraged by the CSSI program. The CSSI solicitation encourages PIs to demonstrate an existing community for the proposed products. Going forward, it is recommended that all the CSSI projects are required to engage in some level of community-building effort, and the proposals clearly include the metrics for measuring the community engagement. The CSSI program could also consider encouraging cross-agency (\emph{viz.,} NSF, DOE, and NIH) and international collaborations to help form broad communities of the funded products. It could be worth exploring the feasibility of leveraging the existing resources, networks, forums, and initiatives supported by other agencies and government laboratories, and proposing new joint workshops and funding opportunities. Some examples of the programs and groups that the PIs could be encouraged to look into for cross-agency and cross-border collaboration are:
	\begin{enumerate}[label=(\roman*)]
	\item \href{https://www.nsf.gov/funding/pgm_summ.jsp?pims_id=505584}{NSF AccelNet program} 
	\item \href{https://www.nsf.gov/funding/pgm_summ.jsp?pims_id=505038}{NSF PIRE program}
	\item \href{http://csmd.ornl.gov/group/research-software/}{ORNL Research Software Engineering Group}
	\item \href{http://www.cs.sandia.gov/ccr-01424 }{Sandia National Laboratories Center of Computing Research}
	\end{enumerate}
Some examples of the community-building activities include:
	\begin{enumerate}[label=(\roman*)]
	\item Where feasible, the projects that produce domain-specific software (\emph{e.g.}, a computation chemistry or computational physics software), can consider developing education and training content, and explore the option of incorporating the software in the University or online courses. 
	\item Where feasible, project-specific hackathons can be organized to engage the community (especially students and new users) in using the products, and developers in contributing new features. 
	\end{enumerate}

Forming cross-disciplinary teams for developing innovative products and user communities is critical for the advancement of the national CI and is already encouraged by the CSSI program. However, some of the workshop participants (particularly, early-career) mentioned that they have challenges in finding collaborators for forming such teams. These participants appreciated the format of the CSSI workshop and recommended that additional workshops/meetings that can provide opportunities to develop cross-disciplinary collaborations should be organized. While the CSSI program PIs can request for a supplement to their existing CSSI awards to offset the cost of organizing such events, a Research Coordination Network (RCN) award has the potential of conducting such events and providing additional in-person and online services to the community in a sustainable manner. Hence, the CSSI program could consider funding such RCNs. Additional strategies for encouraging or developing cross-disciplinary collaborations are as follows:
	\begin{enumerate}[label=(\roman*)]
	\item NSF and/or NSF-funded PIs could continue to organize Birds of a Feather (BoF) sessions at major conferences like Supercomputing (\emph{e.g.}, \href{https://sc19.supercomputing.org/proceedings/bof/bof_pages/bof157.html}{NSF BoF at SC19}). NSF and/or NSF-funded PIs could explore opportunities to organize such sessions at the scientific symposia in domain science focused national meetings (\emph{e.g.}, ACS, APS, MRS) to foster conversations around software and data products that brings together domain-scientists and CISE experts. 
	\item NSF could continue funding the organization of cross-disciplinary workshops/meetings/conferences, and encourage the PIs to consider networking with the organizers of domain science meetings (\emph{e.g.}, ACS, APS, MRS) to foster conversations around software and data products with the domain scientists (and other potential end-users of the products).
	\item The sessions at the annual CSSI PI meeting could be organized in a way that promotes forming new collaborations - for example, theme-based sessions, and brain-storming sessions.
	\item It may perhaps be useful to have NSF funded workshops and conferences in partnership with the widely known publishers such as ACM/IEEE/Springer that can help in disseminating the results of the projects. These publications can be peer-reviewed and the workshop/conference venue can be in the same city as another major conference and just before or after it. The scope of the annual PI meeting could perhaps also be extended/modified to enable this.
	\end{enumerate}

\subsection{Best Practices and Policies}
There is a clear need for creating and sharing the best practices and processes for scientific software development, data management, and dissemination of the products. Such practices and processes can lead to high-quality products that are easy to use/reuse, discover, maintain, and sustain. Therefore, the CSSI project PIs could be encouraged to:
	\begin{enumerate}[label=(\roman*)]
	\item Make the products convenient to use/reuse by adopting:
		\begin{itemize}
		\item Software-as-a-service model, where appropriate, instead of just the download model
		\item Containerization of the products, where appropriate
		\item Software change management processes that ensure backward and forward compatibility with the published interfaces
		\item Portable and scalable file-formats for computational data (such as HDF5 and netCDF)
		\item Provide clear instructions/user-guides/videos on product use
		\end{itemize}
	\item Consider adopting the ``continuous integration/continuous delivery'' model
	\item Consider adopting modern programming standards/models/tools  
	\item Conduct a thorough assessment of the existing products that could be a part of the tool-chain, and integrate or build upon those existing products that have a sizeable user community as doing so will help in preventing the software and data products from decaying due to the lack of availability/support/need/utility of the internal components used in the products’ composition
	\item Consider the life-cycle of the hardware on which they are developing and deploying the products
		\begin{itemize}
		\item The PIs should be encouraged to invest effort in the preservation of their software and data products through their institutional repositories and libraries
		\item The PIs could ensure that their products continue to function on modern hardware equipped with latest software stack
		\item Consider incorporating the best practices recommended by the \href{https://www.softwarepreservationnetwork.org/}{software preservation network}
		\end{itemize}
	\end{enumerate}

The CSSI program could also consider defining policies (or setting expectations) on making the software and data products accessible, discoverable, and citable by the community. Such policies have the potential of bringing uniformity to the process of dissemination of the products funded through the CSSI program and in tracking their use in the community. It is recommended that the products funded through the CSSI program be encouraged to have DOI numbers or some other citation mechanism. Tools are becoming common for DOI creation for public software releases and other artifacts (\emph{e.g.}, Zenodo can be configured to tag every release of a public GitHub repo with DOI automatically). This recommendation also ties back to the evaluation metrics and can help in assessing the impact of the project on a broad community. PIs should consider publishing peer-reviewed papers on their products to illustrate new features in new product releases. Some potential venues that encourage publication of papers and artifacts associated with software and data products are:
	\begin{enumerate}[label=(\roman*)]
	\item Gigascience - open source software for big data analytics
	\item Springer Nature Computer Science (SNCS) journal on Software Challenges to Exascale Computing
	\item Patterns - new journal ``The science of data - Patterns'' is a premium open access journal from Cell Press, publishing ground-breaking original research across the full breadth of data science. Data are the foundation of all research, and all data are in scope, regardless of original domain.
	\item WIREs Computational Molecular Science
	\end{enumerate}

\subsection{Sustainability}
Sustaining the products and their evolution after the award period can be challenging for many PIs even if there is a clearly demonstrable need in the community for their products. Majority of the workshop participants agreed that it is impractical to seek or expect continuous support through the CSSI program for sustaining and maintaining the products in the long run. Some recommendations for sustaining and maintaining the products after the CSSI award period are as follows:
	\begin{enumerate}[label=(\roman*)]
	\item When a software or data product matures, the PIs could consider seeking industrial partnerships, patents, or commercializing their products to raise additional funds. The PIs could also consider additional funding opportunities such as NSF SBIR and NSF Icorps. Hence, while writing their CSSI proposal, the PIs should carefully consider their choice of license and software delivery model. If feasible, the CSSI program could also consider encouraging the PIs to seek patents for original and transformative work by explicitly mentioning such a possibility in the solicitation.
	\item NSF could consider providing new funding opportunities for supporting the evolution, and maintenance of existing products and services that are popular in the community and are innovative.
	\item The PIs can consider engaging with Software Institutes, and if applicable, seeking support through virtual collaborations such as eXtreme Science and Engineering Discovery Environment (XSEDE) in deploying their products on the resources that are a part of the national CyberInfrastructure (CI).
	\end{enumerate}

\subsection{Staffing}
Majority of the workshop participants agreed that staffing is a big challenge in the successful completion of the projects in a timely manner. Finding talented developers and keeping them in academia can be challenging. Some workshop participants mentioned that training short-term staff so that they are productive on the projects is difficult and costly, and there are insufficient funds to do long-term hiring. For mitigating the staffing related challenges to some extent:
	\begin{enumerate}[label=(\roman*)]
	\item The CSSI program could consider requiring that head of the PIs' departments provide a letter stating that the PIs will be ensured adequate staff support to execute their projects
	\item The PIs could be asked to include their hiring/recruitment plans, and also provide a risk mitigation strategy for addressing the change of staff or managing the effect of delays in hiring
	\item Engaging consultants to complete the implementation of the projects can be considered as an option, and such consultancy services could also be offered by the large software institutes funded by the CSSI program
	\item Where viable, PIs could consider outsourcing product development to industrial partners or collaborators who have the required expertise. The PIs should provide sufficient rationale to support such an effort.
	\item The PIs should have the required expertise to take the project to completion with the help of students or temporary staff
	\item The projects should budget time for onboarding and training staff/students 
	\item The software engineering processes adopted by the projects should ensure that the products and the related artifacts are of high-quality so that the learning curve of the new staff on the project is not too steep
	\end{enumerate}

\section{Organization of the Workshop}
Well before the workshop, the organizers discussed with program managers about suggested brainstorm themes and questions about future directions for the CSSI program. These were posted in advance in a pre-workshop survey. Each participant was asked to prepare their thoughts about the brainstorm questions and suggest topics that they thought should be discussed at the workshop.

The workshop's morning program on the first day featured two lightning talk sessions and an extempore session. There were two breakout sessions in the afternoon on the first day, followed by summary presentations from the breakout groups. The breakout sessions were organized as per specific themes: each breakout group focused on one of the themes. To help ensure productive discussions, each breakout group had two steering committee members to guide the discussions and a worksheet with questions to address during the sessions and in the summary presentations. The first day event ended with another lightning talk session. On the second day, an invited talk on quantum computing was featured, followed by two breakout sessions and summary presentations from the breakout groups.

\section{Impact of the CSSI Program}
The CSSI program and its precursors (SI2 and DIBBS programs) have fostered a growing scientific software community. It is a unique program that has impacted the career growth and development of many domain and computer scientists. The CSSI program has opened new opportunities for interdisciplinary collaborations and dissemination of products to the broader scientific community. The success of the CSSI program has led to many innovative products driven by the need of the scientific community. The program has raised the awareness of the importance for best practices in developing software and data products such as thoughtfully choosing licenses, defining the software engineering process, and considering the sustainability of the projects in the long-run. 

Software and data products are instrumental in powering discoveries and maintaining the nation’s competitiveness globally. Hence, the CSSI program and its precursors, SI2 and DIBBS programs, have played a vital role in the sustainable growth of the scientific community and in supporting continuous scientific innovations.

\section{Themes for Brainstorming Sessions}
\subsection{Software and/or Data Management Processes and Sustainability}
\noindent{\it\ul{How can sustainability be supported while encouraging creativity?}}

The word `Sustainability’ has multiple meanings and the answers may depend on multiple factors, such as, developers vs. users, and design vs. coding. Some of the following meanings were raised and discussed: Does your software adhere to the standards? Does it have a wide user base? Can it survive without NSF funds for any period of time? Does it keep up with standard technologies (such as, OpenACC, OpenMP, and MPI)?

There are always tug-of-war issues between these angles where sustainability and creativity can be at odds. Research questions encourage creativity, but limit sustainability. Sustainability can support creativity, but it is harder to make a reverse connection. 

The conflict between these two also arises because sustainability includes tasks such as maintaining, documentation - there is no prize for that. It takes away from creativity since all these mundane work takes away time. Panels (especially in domains) think science is the most important, so tend to prioritize creativity. Domains fund most of the software development, NOT computer science. Software development is not credited in tenure evaluations (not even in computer science).

The following points were recommended:
\begin{itemize}
\item Software development and coding should be an integral part of scientific training.
\item Software may need to be updated to reflect the advances made in the domain-sciences - for example, to use a new type of mathematical model. In such situations, one may need to re-write software in order to support innovative features - such activities should probably be supported, not discarded, especially if there are potentially promising improvements of initial versions of the products.
\item NSF does emphasize the importance of the software aspects of the proposal, and need to continue to do so.
\item The CSSI program should consider encouraging in-person interactions like in the CSSI 2019 workshop to promote cross-disciplinary interactions.
\item Who are you asking of ``creativity'' and what do you mean by ``sustainability'' - it is important to define these up front.
\item It may be useful to set the expectations for sustainability on projects that are of the level of ``Frameworks''. ``Elements'' proposals (typically small projects - for hardening the prototypes) could be rated leniently with respect to the sustainability criteria. 
\end{itemize}

\noindent{\it\ul{Is it realistic to expect high-quality software or data product that is perpetually free, and for how long should the NSF funding support the development life-cycle?}}

Expectation for ``free'' software varies strongly across disciplines from ``definite yes'' to ``definite no''. There were questions regarding whether it is free for users or for commercial usages. It was also indicated that it is difficult to make scientific software free forever. The following alternative suggestions were made:
\begin{itemize}
\item Once bridge from innovation to application is made, NSF should stop funding. Community should support it, somehow. Bridges made to other funding organizations (\emph{e.g.}, NIH) via joint calls, unsolicited proposals, \emph{etc.} can be explored.
\item The PI can continue to support the project via other grants that innovate, with overhead of maintenance.
\item Gaps need to be filled between rapid prototyping and sustainable software. For example, a multi-stage approach (prototyping to hardening software to user community) can be explored. 
\item Long term funding may be warranted to drive the project, if there is a large community behind it.
\item NSF could consider supporting projects that define best practices and processes for software development and data management.
\item Some attendees indicated that it is not an NSF requirements that all things need to be free forever - sometimes these tools need to be licensed for money or developed commercially.
\end{itemize}

\noindent{\it\ul{For sustainability and cost-effectiveness, should the PIs be encouraged to buy software and services from industry instead of spending effort in developing open-source and free versions of the commercial software, and when is it appropriate to compete with commercial solutions and build from scratch?}}

This question had mixed answers. 
Some attendees argued positively to purchase software with the following reasons: 1) Good commercial software is built based on the point of view of a domain scientist. 2) Open-source and free versions of the commercial software can be of poor feature coverage and poor interoperability. 3) It is unrealistic to expect monolithic software packages to cover all research done in a group. 4) ``Competing'' can be risky --- Cease \& Desist.

Several attendees argued negatively to purchase software and preferred to focus on effort in developing open-source and free versions of the commercial software with the following reasons: 1) A community may not be served by commercial structure. 2) Severe licensing restrictions (\emph{e.g.}, per-core licenses) can make commercial solutions unfeasible. 3) Vendor lock-in can cause extra unforeseen expenses — \emph{e.g.}, the dependence on the MKL library requires purchasing Intel processors. 4) Commercial versions can limit reproducibility, or encourage ``black box'' mentality of data analysis. Even those in industry are concerned by ``black boxes''. 5) If there is no competition, vendors can take advantage of the users. 6) To avoid undue influence of the industrial vendors on research agendas. 7) Some domain software may or may not have commercial value, depending on the applications/science. Activities such as methodology testing, and understanding the fundamentals of science may not be of huge interest to the industry. However, the applied aspects of the scientific domains, such as binding energy calculations, could potentially be of interest to industry and could potentially be commercialized. 8) Testing new methods and theories is not of large commercial interest, therefore, the support from the NSF is essential for such activities.

Some attendees also indicated that NSF should not be taking a stance one way or the other. Since there are too many variations on what you need, this should be grant-specific and any/all strategies need to be well-justified. The cost of extending open source software should be considered and carefully evaluated.

There was also some opinion that funding agency/industry partnership may be useful for reducing costs. The cloud-access subsidy for the PIs, or providing credits for compute and storage on the cloud may help increase sustainability of the projects. There was also some opinion that the cost-benefit can vary widely, \emph{e.g.}, computing on commercial clouds can be expensive as compared to the access to the large-scale HPC systems and cloud computing systems funded by NSF. However, the commercial clouds do provide the advantage of on-demand scalability at competitive pricing, and several commercial cloud providers offer free computing credits to the PIs for educational and research purposes.

\noindent{\it\ul{Industry maintains the lead over the academia in establishing software development processes, especially for web-portals, and has crystallized the best practices. Are such practices and processes easy to adopt in academia? Should the PIs be encouraged to subcontract the development of web applications and portals to colleagues from industry? What are the challenges and advantages of doing this?}}

Academic environments are not well-suited to handle strict adherence to coding guidelines/processes, as the needs of graduate students and PIs can be quite different from the colleagues in the industry.

For those types of software that are likely to be easily outsourced, costs are an issue. Other types of software are too ill-defined or ``special'' to be outsourced, and may not even be able to leverage high-quality software engineering practices for this reason. In fact, some argue that academic software may need its own specialized processes to support development scenarios that are unique to academia, and do not commonly occur in industry.

One strategy for buying software engineering expertise is performing a ``group buy'' of software engineering services. This can help to reduce costs -- sharing a single engineering contract among several small projects reduces overhead. This could work well for consultants that are dedicated to helping teams improve the quality of their existing codebases.

In spite of the unique nature of academic software development, the lack of proficiency in industrial software engineering practice is harmful. Since academic and industrial software engineers tend to be on separate career tracks, training tends to be different, so academic software engineers use fewer sustainable practices. Instituting something like software carpentry in the academic environment would be a good way to get people trained in these practices, as well as to create the next generation of trainers, providing ongoing sustainability.

Some of the workshop participants also mentioned that it would be nice if the future CSSI solicitations encourage investments from industry during the funding process. One of the participants from the industry mentioned that instead of the community spending a lot of time in developing custom versions of certain popular functions/libraries/software components, it would be nice if the community tells them (industry) about their needs, and they (colleagues in the industry) expand on the functionality of the software at their end using their own investments.

Academia  could collaborate with industries to build prototypes, and test solutions. For the community to propose changes/updates to existing vendor functions/libraries/software, it would be necessary for the newer functions/libraries/software to be tested by the contributors (or developers) in the community before they can be incorporated into the vendor software and tools. For any such suggestion, a prototype is almost the first important step. Partnering and working closely with the colleagues from the industry can make this plausible. To that end, collaboration with industry in the form of internships, or short-term funded projects can be encouraged as part of the solicitation.

Some of the workshop participants mentioned that Where viable, development can be outsourced to industrial partners or collaborators who have the required expertise.

\noindent{\it\ul{What if the NSF funding is used to develop a minimum viable product that can be made free and open-source, and extended features and services are offered on top of the free versions at a cost? Does this look like a reasonable approach for sustainability?}}

It is valuable to move software maintenance costs from CSSI/other innovation-focused funding to domain scientists that use the software. However, attaching a cost to software incurs risk that PIs will choose to not support software, and instead will create forks/competing projects to avoid monetary costs. Software support costs are often underestimated by academics, so this strategy will rarely lead to a more sustainable solution, and instead fragments a user base.

The ``open core'' model described in this question has the potential downside where companies release barely usable or severely restricted base versions in order to maximize the likelihood that a customer must buy the software to get good use of it. This can extend beyond missing features into poor documentation. The classical way to guard against this is to insist that base capabilities developed via grant funding must remain open and available.

\noindent{\it\ul{How should the software be maintained in the long run? When a new software compiler and hardware infrastructure come out, if needed, who should update the software accordingly - developers and/or the community for whom the software was developed?}}

Community support for software is not reasonable until the community is quite large (\emph{e.g.,} the Linux kernel project), as software should reach a community that is much broader than its developer community. Until a project reaches that point, the core developers are uniquely qualified to make modifications to their own software -- context, undocumented design decisions, previous experience in its construction, and cognizance of high-level concerns enable efficient introduction of features, hardware support, and so on.

Centralized funding of maintenance is an issue, as grants are usually directed at innovation instead of support. There appears to be a wide-spread sentiment that important, impactful software is hard to support long-term via NSF, even though such support is considered important.

There are business models targeted at providing support for legacy projects. These include support contracts, fee-for-service, and pay-as-you-go arrangements. Companies following these models can provide sustainable ongoing maintenance for open source projects.

\noindent{\it\ul{Do you use your institution's infrastructure or the national cyberinfrastructure for the long-term retention of your software and related artifacts?}}

Many PIs use institutional or commercial solutions for artifact retention. GitHub, BOX, Google Drive, \emph{etc.} are popular choices. Unfortunately, many are unaware of the national cyberinfrastructure that can be used for these purposes.

\subsection{Software/Data Reusability and Quality}
\noindent{\it\ul{How do you measure the quality of the software/data product that you produce for public release?}}

``Quality'' of software/data product has multiple meanings/attributes. First, there is
technical quality. Second, there is the impact that the software has on the community.
The former is a prerequisite to the latter, but is not sufficient.

Software/data product has high technical quality if it has the following attributes:
	\begin{itemize}
	\item It comes with a thorough set of tests (unit, validation, performance)
	\item Test coverage is measured and reported
	\item Tests are invoked automatically (\emph{e.g.,} via Continuous Integration infrastructure)
	\item If applicable, software is deployed automatically via Continuous Integration
	\item Complete user and reference documentation
	\item Clear license terms
	\item In case of data products, it is important to have precise curation/provisioning of the data, support for data retrieval in all industry-standard formats
	\end{itemize}

Some technical quality attributes are more subjective, difficult to define, and are inter-related; these generally correlate with the ``hard'' attributes above:
	\begin{itemize}
	\item Easy to use (usability, UI/UX) and easy to modify/extend
	\item Well-designed
	\item Feature set is complete
	\item Maintainability
	\end{itemize}

It should be noted that the technical quality metrics used by the industry may not be as
useful in academia due to differences in product objectives and life cycles. Thus it is not sufficient to simply copy the best practices from the industry.

There are a number of metrics for gauging the software impact:
	\begin{itemize}
	\item Number of downloads/releases
	\item Number of users
	\item Number of external contributors and developers reusing the software
	\item Number of publications citing and acknowledging the use of software
		\begin{itemize}
		\item This metric may not be easy to use for middleware developers if they cannot track how many people are reusing their products because their products may not receive as many citations in publications as an end-user application
		\end{itemize}
	\end{itemize}

\noindent{\it\ul{What is the software engineering process or data management process that you (and your group) follow currently?}}

The workshop participants reported that they follow standard best practices that ensure high-quality of software, and some of these are:
	\begin{itemize}
	\item Automated testing of various kinds (such as, correctness and performance testing)
	\item Pull-request-centric development workflow, with peer review of code
	\item Modular design
	\item Structured software evolution:
		\begin{itemize}
		\item Regular code review and refactoring
		\item Release planning
		\end{itemize}
	\end{itemize}
Standard best practices for data management that are followed by the workshop participants include:
	\begin{itemize}
	\item Adoption of automated processes for data management, and ontologies
	\item Making the data Findable, Accessible, Interoperable, Reproducible (FAIR) 
	\item Adoption of data governance practices
	\item Compliance with regulatory requirements, government approvals, and following certification processes for different regulations such as FISMA, FERPA, \emph {etc.}
	\end{itemize}
Some of the workshop participants mentioned that developing software/data repositories while following best practices is expensive.
	\begin{itemize}
	\item Small labs can’t do ``all'' the best practices, but some are easier to implement than others such as version control, maintaining usable documentation, and using modern tools and technologies.
 
	\end{itemize}

\noindent{\it\ul{How often do you favor using existing software components (either by buying or using for free) and, if needed, customizing it for your needs instead of writing from scratch?  What are the challenges, if any, that you typically face in reusing existing software or data products?}}

	\begin{itemize}
	\item The opinions range widely, from ``should always reuse'' to ``long term it’s best to not reuse''.
	\item For some foundational pieces of software (BLAS/LAPACK, FFT, MPI) reuse is a no-brainer: the cost of not reusing them is extreme.
	\item There are pros and cons for reusing existing software and for writing software from scratch
		\begin{itemize}
		\item Quality and reusability of existing software in science varies greatly; these traits can only be discovered by trying to reuse the software.
		\item Writing software has educational value, especially for junior developers, but junior developers are unlikely to produce something as robust and reusable as an established software component.
		\item There is a chance that the software component will stop being  maintained.
		\item Extra dependencies add to the complexity of software configuration and building; using packaged components is relatively easy.
		\item The cost of learning someone else’s code may not outweigh the benefit of using it.
		\item Software evolution can introduce non-backward-compatible changes. 
		\item IP/licensing issues can create bottlenecks. Need to ensure correct attribution for tools (especially, when wrapped into larger frameworks).
		\item For large data sets, data sharing and accessibility can be a challenge. Additionally, data privacy and compliance can bar one from reusing data meaningfully.
			\begin{itemize}
			\item Raw data sets can be easier to reuse in multiple scenarios than the finished data products. Not only the finished data products, but also the raw datasets and all the changes applied on the datasets to obtain the finished products (such as algorithms for noise reduction, calibration, and other processing) should have long-term retention value.
			\end{itemize}
		\end{itemize}
	\end{itemize}

\noindent{\it\ul{How do I get/find the software components to compose the software framework/system?}}

Whereas experts within a domain are aware of most/all established pieces of software, a more systematic cataloging/search capability is necessary to enable novice and other non-expert users to find software (whether supported by CSSI or not). Participants listed a number of ways for how they look for software/data repos:
	\begin{itemize}
	\item Internet search engines (\emph{e.g.,} Google). Google data search is in development. However, quality of search engine results can be affected by SEO factors. It can be difficult to know what term to search for and/or how to filter out irrelevant results.
	\item Google Scholar and citation databases are used but they are biased towards publications rather than software/data.
	\item Some scientific publication include links to supplemental data, GitHub, data repositories. 
	\item Word of mouth. This also provides feedback on ease of use, availability of tutorials, existence of local/online community support, \emph{etc.}
	\item Code repositories (GitHub, Gitlab, BitBucket, SourceForge, \emph{etc.})
	\item Some projects have web portals that provide information on the project software/data.
	\item Specialized software catalogues, \emph{e.g.,} Netlib, are available for browsing.
	\end{itemize}

It is suggested that a catalogue of software or data products developed using the funds from the CSSI program is maintained. This catalog should have tags or metadata for describing the different types of software and data products. PIs should be able to add predefined tags (keywords) to their project proposals and the proposal could be compared against the related previously funded projects.  

\subsection{Project Staffing}
\noindent{\it\ul{If software engineers are available for contract jobs at industry rate, is it more cost-effective to hire their services for short-term to complete the project on time than have full-time staff hired for the job?}}

	\begin{itemize}
	\item If the user base is small, then contracting may not be cost-effective.
	\item Funding/hiring long-term staff can help with the sustainability of the software. 
	\item Using students/postdocs would train the next -generation workforce.
	\item Hired staff needs to go through a learning curve to understand the domain science.
	\item For production code, having contracted resources can help with the overall architecture of the product, design, testing, and documentation.  
	\end{itemize}

\noindent{\it\ul{Does new hiring introduce delays on the project, and do you plan for that in your project timeline?}}

	\begin{itemize}
	\item New hiring definitely introduces delays, but the amount depends on the specific domain and/or scientific focus.
	\item It is important to build a pipeline of talent with postdoc, graduate, and undergraduate students.
	\item Having scientific computing added to the undergrad curriculum can help build the foundation earlier with classes, and organizing tutorials, and hackathons can be helpful for skill-development. 
	\item The following ideas are suggested:  
		\begin{itemize}
		\item Universities could help match-make between domain scientists and software engineers with certain skill sets. 
		\item Universities or professional societies could setup professional degrees or ``professional certification for software engineering'' to help incentivize students to pursue a career path in academic research computing/software engineering. 
		\end{itemize}
	\end{itemize}

\noindent{\it\ul{When the senior personnel on the project who are in charge of the development work get promoted or take a lucrative job elsewhere, or when the students graduate, how do you address the loss in progress on the project?}}

	\begin{itemize}
	\item Large groups can accommodate the loss easily.
	\item The PIs should be qualified to complete the project on their own to minimize the loss in progress.  
	\item The PIs should have continuous training/replacement activities in the group - planning ahead for the structural changes that might occur - PIs should have a staffing/risk-management plan specified in the proposal.
	\item PIs should setup best practices in the group. They should utilize standard resources (for example GitHub) for sharing code, have solid documentation, cross-train students, and encourage code reuse.
	\end{itemize}

\noindent{\it\ul{What are the tradeoffs in having a full-time, senior-level staff, learn the basics of software engineering and deliver on the project, versus hiring trained or trainable graduate/undergraduate students?}}

	\begin{itemize}
	\item It is helpful for the sake of continuity to have a full-time staff, but funding is difficult, making it difficult to hire full-time staff in a sustainable manner. 
	\item For PhD students, the experience helps understand the project/code and provides continuity.  
	\item It is important that the personnel have a strong domain knowledge for the success of the project. For example, the HPC Center at the University of Houston conduct tutorials during which scientists train CS students on how to use their tools, so that CS students can contribute to their projects.
	\item Engaging trained or trainable students in the development of software and data products can contribute towards the workforce development efforts. On some projects, there can be a learning curve that the students would need to climb before they are productive. However, once they are productive, the senior staff members can have time freed up for focusing on quality and other critical activities on the project.  
	\end{itemize} 

\noindent{\it\ul{Should we put PhD students in the project for non-research oriented tasks (\emph{e.g.,} software maintenance)? How about recruiting students pursuing Master's or undergraduate degrees?}}

	\begin{itemize}
	\item For masters and undergrad students, it is a good experience for their career. 
	\item PhDs could initially use the experience to be familiar with the project and provide continuity.  The task should be related to their research focus. 
	\item Software engineering is an important and necessary skill for CS and domain scientists.  Students can be trained via programs like Software Carpentry. 
	\end{itemize}

It is suggested that the CSSI program could fund ``software institutes'' that provide software development services to domain scientists on the lines of the Extended Developer Support model offered by the Science Gateway Community Institute (SGCI) or the Extended Collaborative Support and Service (ECSS) model of the XSEDE. Such institutes could perhaps maintain staff at competitive salaries. The staff could be having expertise in software development (\emph{e.g.}, in C, C++, Fortran, Python, and parallel programming), cloud computing, data lifecycle management, and Artificial Intelligence (AI). The domain scientists could consider requesting some percentage of the time of the full-time staff at such institutes and accomplish their code development and maintenance objectives without feeling burdened by the need to hire full-time staff for developing production-quality code. As an experiment, independent software providers (or companies) could be invited to bid for small-duration code development and maintenance projects through a central agency.

\subsection{Inter-Disciplinary Collaborations}
\noindent{\it\ul{Should the domain-science projects - funded through the CSSI program - be required to have the CISE collaborators engaged at a substantial level of effort to ensure that best practices for software engineering and data management are followed? Should the CISE projects - funded through CSSI - be required to ensure accountability from domain-scientists (who could have promised to provide test-cases in their letters of collaboration)?}}

	\begin{itemize}
	\item The CSSI program funds projects at different phases and from a diverse background, \emph{i.e.,} physics, materials science, chemistry, geography, biophysics, biology, and fluid mechanics. As different projects may have distinct demands, it would be ``too strong'' to make it a requirement to substantially engage CISE collaborators. However, such collaborations should be encouraged, especially for larger proposals that involve team efforts. To engage CISE collaborators is helpful to take advantage of rapid changes of technologies. One suggestion is to engage social sciences to figure out how to do better team sciences.
	\item For element proposals, where grants are limited in size, it may not be feasible to involve CISE collaborators at substantial levels. Making a rigid requirement to have CISE collaborators would cut out a large percentage of proposals in the program. If such a requirement is necessary, the NSF solicitation could also have language to encourage collaboration with library science for preparing the data management plan, campus HPC infrastructure where applicable, and including trainees (students/postdocs) for product development.
	\item Some of the workshop participants suggested that NSF could consider helping in pairing up domain projects with potential collaborators, \emph{e.g.,} by allowing domain scientists to describe requirements. This is probably important to pick up CISE collaborator across the whole CSSI domain. Opportunities for speed networking could also be helpful. 
	\item Ensuring long-term commitment from all collaborators who provide the letters of collaboration can be challenging. As an industrial practice, outsourcing software engineering requires rigorous specifications, but it is rare in domain science. Aligning interests of domain scientists with computer scientists is possible but challenging. Budget support for postdocs and graduate students to attend collaborative meetings is a useful mechanism for enabling deeper synergy in the long-term.
\item For addressing the accountability issue, a collaboration should be more synergistic than just meant for getting help from the CISE counterpart on software engineering. Having joint research publications may help address this issue partially. As computer scientists can derive interesting problems when they work with domain scientists, both parties can benefit in the end. 
	\item The unfunded collaborators could be requested to briefly describe the nature of their contribution to the project in their ``letter of collaboration'' and their two-page resume should be included in the proposal to help the reviewers in objectively evaluating the nature of the collaboration. 
	\end{itemize}

\subsection{New/Emerging CSSI Domain Science}
\noindent{\it\ul{How is heterogeneity in the hardware considered in the modern software development and data management practices?}}

Software:
	\begin{itemize}
	\item In general, many domain scientists narrow their focus onto platforms they deal with and/or access routinely (mostly CPU). State-of-the-art in programming accelerators is still fluid/immature; domain scientists do not expect to have to deal with this. While some libraries like the Message Passing Interface (MPI) consider different hardware architecture during their development phase, there are many software products that do not work out-of-the-box on  different hardware architectures and may need adaptation to get installed, and work correctly and efficiently. 
	\item Materials science: Computational scientists spend significant effort in porting and optimizing to new architectures. The main driving force is to reach realistic physical conditions. With the development of quantum co-processors, people would be willing to redesign their entire stack to use them due to its revolutionary capability.
	\item Fluid mechanics: Computational scientists also spend significant effort in porting and optimizing to new architectures.
	\item Chemistry: The field is moving slowly, need future CSSI support to improve the performance of legacy software. In some cases, platform concerns are often secondary to the research problems.
	\item Biophysics: Some applications already use GPUs, and in some cases, heterogeneity is not really an issue.
	\item Biology: It is not an issue for smaller labs (they use whatever code runs on their laptop/HPC/cloud), but biologists tend to write scripts/pipelines instead of proper new algorithms.
	\item GIS and spatial analysis: There are huge challenges to transform convention sequential algorithms to high-performance and scalable algorithms that can harness the power of heterogeneous hardware.
	\item Image processing: Large in-memory systems have been used for streaming data analytics in applications such as image processing.
	\end{itemize}
Professional services for code adaptation and optimization for different hardware platforms are needed. Such services could be available locally through the research computing centers at the PIs' institutions, or through NSF funded institutes and facilities. 

\noindent Data:
	\begin{itemize}
	\item Domain knowledge behind data is critical to meaningful applications of machine learning.
	\item Domain scientists lag behind in awareness and proper best practices (and/or lack desire) for data management; it also takes time and funding support. Some of the workshop participants noted that the importance of data management, in some cases, is even greater than the simulation itself.
	\item Library/Information Science staff are important for data management practices (and may offer know-how for using campus hardware and best practices for data management) - input from librarians and research computing staff should be encouraged/promoted to ensure that PIs follow best practices.
	\end{itemize}

\noindent{\it\ul{Are artificial intelligence techniques such as deep learning disrupting or impacting the future directions of your discipline? If yes, then how?}}

	\begin{itemize}
	\item Yes or No, depending on discipline. We are at the early stages for deep learning in some research areas. Funding agencies could facilitate these efforts. Machine learning (ML) may need to be broadened to data-driven and/or hypothesis-driven science, where same concerns apply.
	\item Importance of education for using ML effectively and safely. Modern tools not only make using ML easy but also too easy.
	\item Successful stories: Computer scientists have been trying to use deep learning to figure out certain trends in communication patterns, and use this to design better algorithms.
	\item Successful stories: Physicists have been using deep learning in spaces where they previously used low-fidelity simulations to look for interesting phenomena (\emph{e.g.,} neural networks to run faux simulations at 1000x faster in shock physics). This can now be done on desktops, so more people can easily adopt deep learning approaches.
	\item Unsuccessful stories: Biophysicists have been analyzing large amount of synchrotron data, but there are so many things that can go wrong. It is not ready to use deep learning right now because people do not trust the data due to noise and experimental variability.
	\item Unsuccessful stories: There are too many other problems in biology to solve first and too little baseline knowledge in most biological disciplines. 
	\end{itemize}

\subsection{Software/Data Dissemination Plans}
\noindent{\it\ul{What are some of the effective approaches for software dissemination?}}

	\begin{itemize}
	\item Using containers on large compute systems to address challenges of reimplementing software and dependencies 
	\item Using central repositories to load data and code (\emph{e.g.,} Code Ocean)
	\item The manuscript publishers could consider making software dissemination mandatory by requiring authors to add software as a part of their publications 
	\item Where feasible, the PIs could consider using popular tools of interest to the community so that the information about their products can be relayed through the communication channels that are already established for these tools 
	\end{itemize}
		
\noindent{\it\ul{How can software and data products be made discoverable, accessible, and usable by the community?}}

	\begin{itemize}
	\item Disseminate results and information through conferences, and symposiums
	\item Consider creating documentation, tutorials, and YouTube videos for wide dissemination 
	\item Choose the right keywords for tagging the products while disseminating information 
	\item The PIs could define policies on how the data must be used and shared 
	\item Funding agencies could make it mandatory that the PIs consider making their data discoverable, accessible, and usable 
	\end{itemize}
	
\noindent{\it\ul{Hosting software or data on GitHub may not be sufficient for meeting the community engagement needs of the project. What are some of the approaches to engage the community?}}

	\begin{itemize}
	\item Hosting summer schools and providing hands-on experiences is another way to disseminate data and engage with the community
		\begin{itemize}
		\item A success story being summer school for astrophysics where the participants quoting the summer school experience have led up to over 180 publications
		\end{itemize}
	\item Engage not just with users but also developers 
	\item Engage local participants via in-person workshops but also engage with remote participants via online materials and tutorials
	\end{itemize}
	
\noindent{\it\ul{How can software be enabled on XSEDE systems?}}
	\begin{itemize}
	\item First up understand users’ wants and needs 
	\item It is not practical to ask developers to maintain software across XSEDE and other systems
    \item It is not practical to have XSEDE maintain software across all the systems affiliated with XSEDE
	\item The PIs could consider talking to the XSEDE campus champions who could be aware of the need for software and packages in their local communities
	\item The PIs could consider having their teams and/or campus champions attend programs such as the
Argonne Training Program on Extreme-Scale Computing (ATPESC)
	\item The PIs could consider using containers on XSEDE systems 
	\end{itemize}

\noindent{\it\ul{Other topics discussed}}

	\begin{itemize}
	\item Reproducibility of research results from both software and data perspectives were discussed. The discussants mentioned that community-wide efforts are required to enable reproducible science. Some changes on this front have started happening (\emph{e.g.,} some journals mandate open-access of software/data or extra documents for reproducibility). 
	\item Challenges with reproducibility: Hardware and software ecosystem changes can prevent legacy code from compiling or running on modern systems. This is a key challenge in doing reproducible science. A possible solution is to deposit containers as paper artifacts to be able to reproduce data in perpetuity.
	\end{itemize}
	
\subsection{Feedback on the Award Process, Scope, and Impact of the CSSI Program}
\noindent{\it\ul{What are the metrics for evaluating the impact of a project? Number of users? Number of downloads? Number of citations?}}
Some of the workshop participants commented that as the CSSI program has a broad range of domains and projects of interest, it is hard to have a general rubric to measure the impact of the projects. Both short-term and long-term metrics should be required to set a mindset for good project management and for considering community adoption/use in product evaluations.

	\begin{itemize}
	\item Short-term metrics:
Number of users, frequency and mode of usage, downloads, and site visits should be considered as recommended software and data product evaluation metrics. Evaluation metrics should be tracked over time. Evaluation metrics can also change or include additional indicators as time goes on. Additional indicators, such as bug reports, support forum posts, number of contributing external developers, and case studies or new sciences enabled, can also be used.
	\item Long-term metrics:
As software and data products mature, additional evaluation metrics should be included. This can be in the form of self-conducted user-studies, highlighted success stories, and the number of software citations. Journals/reviewers need to enforce citation, attribution, and acknowledgement to the software. Metrics should favor speed and ease of use. Additionally, community activities and interactions must be evaluated as an important part of building a user community around the software.
	\end{itemize}
	
An open question is how to track the long-tailed impact after a project ends. Some participants discussed the idea of developing a method for tracking the use of products developed using CSSI funds. If a tracking mechanism is in place, perhaps the CSSI projects funded in future may be encouraged to use such a mechanism for reporting statistics on the long-term impact of the software/data products. \href{http://blogs.exeter.ac.uk/rmas/about/}{The Research Management Administration System} implemented at some UK universities is a good example of a system for tracking the projects funded by the Higher Education Funding Council for England (HEFCE) over a course of multiple years.

\noindent{\it\ul{How can the magnitude of the impact of the CSSI program be amplified?}}
	\begin{itemize}
	\item NSF supported facilities or organizations or institutes could support the dissemination of the information about the products. This could be a good way to help users discover new software and CSSI projects. 
	\item Solicitations should encourage the use of NSF-funded cyberinfrastructure resources and services where feasible - the PIs could be encouraged to do feasibility studies for technology integration \emph{a priori}. 
	\item NSF could consider making NSF-funded resources and services easy to find.
	\item The investigators who teach courses at universities can incorporate relevant/applicable software packages with tutorials/resources to teach the use of software packages to their students, and help grow the user communities of their products.
	\item Investigators could explore funding opportunities to facilitate incorporating CSSI projects into their courses or in porting them on NSF supported facilities (\emph{viz.}, XSEDE).
	\item Project-specific hackathons can be implemented to get CSSI PIs together and have students/trainees test out and learn how to use their code. 
	\item NSF, the public relations team at the universities, or project-specific communication and outreach personnel may be helpful for communicating the information on research products to the community.
	\item The PIs from the CISE domain may need to reach out to broader audience and raise awareness in domain-specific conferences. Software Institutes can help provide the human resources or training for storytelling or scientific communication between computer and domain scientists. Computer scientists and domain scientists should also reach out to each other locally. The CSSI PIs could consider offering webinars related to their products.
	\item Integration between use cases and cyberinfrastructure needs to be done well across an entire project lifecycle.
	\item NSF and CSSI PIs could consider taking an active role in forming collaborations, fostering linkages across different projects in the program, and coordinating with other agencies (such as NSF-DOE joint effort) and national laboratories. This can be organized at the CSSI PI meeting or in the form of an NSF sponsored workshop.
	\item The PIs could consider organizing Birds of a Feather (BOF) sessions at major conferences like Supercomputing
	\item The PIs could consider engaging with professional societies to reach out to scientists in different domains
	\end{itemize}
	
\noindent{\it\ul{NSF has invested millions of dollars - across multiple directorates - for building web-portals that support computation, data analysis, and data management from the convenience of the web-browser. Are you aware of such investments, and if needed, will you be able to comfortably leverage these investments in your current or future projects?}}

	\begin{itemize}
	\item NSF has invested heavily in science gateways (\emph{i.e.,} web portals that serve as computational and/or educational workbenches or provide data management services) through different programs. There are numerous gateways out there but the impact of many of these gateways on research/development/education is not very clear. While some gateways have hundreds of users, others support the number of users in double digits only. It may be useful to assess the long-term research and education outcomes supported by such gateways while making funding decisions. Usability and data discovery are some of the key challenges associated with several gateways in the community.
	\item Most people are aware of such investments, however, many people are unable to take advantage of these investments. Awareness is higher in some cases (such as XSEDE portals and nanoHUB) than others. However, there are too many out there. Awareness and visibility still needs to be improved. 
	\item nanoHUB is useful for education and classroom teaching for engineers without a computational background, but it is very difficult for software development or implementation (non-standard code). Some workshop participants have also tried to use XSEDE and Jetstream for teaching purposes. Neuroscience gateway has many highly cited publications. XSEDE and SGCI provide a list of web portals that can be useful in research and education, such as:
		\begin{itemize}
		\item Data.gov, fema.gov, archeology data portal, digital rocks portal, ICPSR, harvard dataverse
		\item HydroShare, Datacenter share, nanoHUB
		\item Harvard dataverse will take any type of data from anyone (need to be small)
		\end{itemize}
	\item Some participants indicated that many portals have performance issues and are not user-friendly. Portals can make simple tasks very complex (\emph{e.g.,} downloading a dataset) and do not stay updated with changing technologies (\emph{e.g.,} browsers and different hardware such as tablets). Reproducibility can also be an issue but is supported by some portals. Technical support for portals (via email) is challenging because staff don’t know about users’ data/problem. It is often easier to use campus HPC resources and troubleshoot face-to-face.
	\item Searchability and organization of data need to be improved. A single online place could be developed to find out what resources and services are available. Documentation for these resources is important: to crosslink different tools in the portal. PIs need to advertise these support at different conferences.
	\item The diversity of CSSI funded projects is challenging to build portals accessible to users across domains. Portal developers do not always know exactly what researchers need because domain users driving their own science (\emph{e.g.,} life science). Workflows that are up and running may not be innovative in terms of the science. Probably it is worth comparing/partnering between agencies - NIH has many portals and initiatives that are relevant and complementary/overlapping. As NSF has already funded a software institute for supporting web-portal based activities - SGCI - the PIs proposing to develop new web portals could be guided to take advantage of such institutes as much as possible.
	\end{itemize}

\noindent{\it\ul{How to advance the CSSI program in the future?}}

	\begin{itemize}
	\item The CSSI solicitation should lucidly communicate what the CSSI program aims to accomplish and its vision, and how this vision fits into the broad cyberinfrastructure ecosystem and science communities.
	\item The CSSI program could promote a new culture that recognizes the importance and value of research software and data capabilities.
	\item NSF could ensure that the CSSI panelists equally weigh the science, software engineering practices, and community engagement aspects of the proposal. Panelists could be instructed specifically on how to gauge the sustainability of the proposed projects. SBIR metrics of evaluation could be leveraged (perhaps not entirely).
	\end{itemize}

It is recommended that the CSSI program's solicitation clarifies the degree of novelty expected in the projects. A clear statement on what is considered ``novel'' would greatly help the PIs put their proposals together. 
 
If the novelty is with respect to the science:
	\begin{itemize}
	\item A novel algorithm will need to be proposed
	\item A prototype of the algorithm will need to be developed
	\item The prototype will need to be extensively evaluated with different scenarios/test cases
	\item Results from the tests acting as feedback will need to be incorporated into the algorithm devised
	\item Further experiments would need to be performed before the novel algorithm is considered a success
	\end{itemize}
The above steps can by itself take 12-18 months at the least depending on the algorithm and its complexity. This could also largely vary from application to application. 

If the novelty in science is a component of CSSI, the proposal would be expected to have the above steps in place, followed by the steps below as the goal is to create sustainable software. Some of those steps may entail: 
	\begin{itemize}
	\item Assuming a preliminary version of the corresponding software already exists
	\item An ecosystem will need to be created and prototyped that can be accessible by small, medium, or large communities
	\item The ecosystem would also need to be able to run on small to large-scale computing platforms, potentially including cloud resources
	\item Setting up partnerships with relevant stakeholders 
	\item Developing relevant education materials to train the next-generation workforce as we are aiming to build a sustainable cyberinfrastructure 
	\item If the scope of the project involves domain scientists in the mix, appropriate best practices will need to be in place such that the domain scientists can also seamlessly use the proposed cyberinfrastructure 
	\end{itemize}
	
Executing these above steps among others to create a sustainable software in the remaining 18 months or so of a project may not be feasible. Hence some degree of clarification is required to the novelty component in the proposal. What are the expectations? What is the rubric that this bullet will be evaluated against? 

\section{Closing Thoughts}
The CSSI program aims to foster a culture of innovation across multiple disciplines of science and technology. At the center of achieving this objective are best practices for software development and data management, sustained efforts for community-building, and initiatives for workforce development. The software and data products funded through the CSSI program should be well-designed and well-documented to promote reusability, composability, and maintainability. An ecosystem of innovative and well-maintained products funded by the CSSI program has the potential of impacting the rate of advancement in computational and data sciences.

\section{Acknowledgement}
This workshop was made possible with NSF Award \#2034617. We are grateful to NSF for the same.

\clearpage

\appendix

\section{ APPENDIX}
\subsection{Workshop Overview}

The goal of the National Science Foundation (NSF) Workshop on Future Directions for Cyberinfrastructure for Sustained Scientific Innovation (CSSI) is to elicit input from the community to guide the future directions and funding scope of the CSSI program. This 1.5-day workshop brings together researchers and practitioners from different fields of sciences to identify the current and future cyberinfrastructure needs of the community for enabling innovation in the various disciplines.

This  workshop comprises of presentations and brainstorming sessions on some of the key thrust areas related to the cyberinfrastructure. A key outcome of the workshop is a report capturing the feedback/input from the community. The report is submitted to NSF and will be made available to the community to inspire future cyberinfrastructure research and development activities. Additionally, the workshop will broaden its participants' perspective on the various topics related to the cyberinfrastructure and will serve as a venue to form new collaborations.

\subsection{Brainstorm Themes and Breakout Group Worksheet}
\subsubsection*{A.2.1 Software and/or Data Management Processes and Sustainability}
\begin{enumerate}[label=(\alph*)]
\item How can sustainability be supported while encouraging creativity?
\item Is it realistic to expect high-quality software or data product that is perpetually free, and for how long should the NSF funding support the development life-cycle?
\item For sustainability and cost-effectiveness, should the PIs be encouraged to buy software and services from industry instead of spending effort in developing open-source and free versions of the commercial software, and when is it appropriate to compete with commercial solutions and build from scratch?
\item Industry maintains the lead over the academia in establishing software development processes, especially for web-portals, and has crystallized the best practices. Are such practices and processes easy to adopt in academia? Should the PIs be encouraged to subcontract the development of web applications and portals to colleagues from industry? What are the challenges and advantages of doing this?
\item What if the NSF funding is used to develop a minimum viable product that can be made free and open-source, and extended features and services are offered on top of the free versions at a cost? Does this look like a reasonable approach for sustainability?
\item  How should the software be maintained in the long run? When a new software compiler and hardware infrastructure come out, if needed, who should update the software accordingly - developers and/or the community for whom the software was developed?
\item  Do you use your institution’s infrastructure or the national cyberinfrastructure for the long-term retention of your software and related artifacts?
\end{enumerate}

\subsubsection*{A.2.2 Software/Data Reusability and Quality}
\begin{enumerate}[label=(\alph*)]
\item     How do you measure the quality of the software/data product that you produce for public release?
\item     What is the software engineering process or data management process that you (and your group) follow currently?
\item     How often do you favor using existing software components (either by buying or using for free) and, if needed, customizing it for your needs instead of writing from scratch?
\item     What are the challenges, if any, that you typically face in reusing existing software or data products?
\item    How do I get/find the software components to compose the software framework/system?
\end{enumerate}

\subsubsection*{A.2.3 Project Staffing}
\begin{enumerate}[label=(\alph*)]
\item   If software engineers are available for contract jobs at industry rate, is it more cost-effective to hire their services for short-term to complete the project on time than have full-time staff hired for the job?
\item    Does new hiring introduce delays on the project, and do you plan for that in your project timeline?
\item    When the senior personnel on the project who are in charge of the development work get promoted or take a lucrative job elsewhere, or when the students graduate, how do you address the loss in progress on the project?
\item   What are the tradeoffs in having a full-time, senior-level staff, learn the basics of software engineering and deliver on the project, versus hiring trained or trainable graduate/undergraduate students?
\item   Should we put PhD students in the project for non-research oriented tasks (\emph{e.g.,} software maintenance)? How about recruiting students pursuing Master’s or undergraduate degrees?
\end{enumerate}

\subsubsection*{A.2.4 New/Emerging CSSI Domain Science}
\begin{enumerate}[label=(\alph*)]
\item    Should the domain-science projects - funded through the CSSI program - be required to have the CISE collaborators engaged at a substantial level of effort to ensure that best practices for software engineering and data management are followed? Should the CISE projects - funded through CSSI - be required to ensure accountability from domain-scientists (who could have promised to provide test-cases in their letters of collaboration)?
\item    How is heterogeneity in the hardware considered in the modern software development and data management practices?
\item    Are deep learning techniques disrupting or impacting the future directions of your discipline? If yes, then how?
\end{enumerate}

\subsubsection*{A.2.5 Software/Data Dissemination Plans}
\begin{enumerate}[label=(\alph*)]
\item     What are some of the effective approaches for software dissemination?
\item     How can software and data products be made discoverable, accessible, and usable by the community?
\item    Hosting software or data on GitHub may not be sufficient for meeting the community engagement needs of the project. What are some of the approaches to engage the community?
\item    How can software be enabled on XSEDE systems?
\end{enumerate}

\subsubsection*{A.2.6  Feedback on the Award Process, Scope, and Impact of the CSSI program}
\begin{enumerate}[label=(\alph*)]
\item    What are the metrics for evaluating the impact of a project? Number of users? Number of downloads? Number of citations?
\item     How can the magnitude of the impact of the CSSI program be amplified?
\item   NSF has invested millions of dollars - across multiple directorates - for building web-portals that support computation, data analyses, and data management from the convenience of the web-browser. Are you aware of such investments, and if needed, will you be able to comfortably leverage these investments in your current or future projects?
\end{enumerate}

\subsection{Participants}
In addition to the participants listed below in the alphabetical order of their last names, the following NSF officers also attended the meeting: Dr. Alan Sussman, Dr. Alexis Lewis, Dr. Amy Walton, Dr. Micah Beck, Dr. Joseph Whitmeyer, Dr. Stefan Robila, and Dr. Vipin Chaudhary.\\

\noindent
Alexey Akimov, University at Buffalo\\
Jay Alameda, NCSA, University of Illinois at Urbana-Champaign\\
Ilkay Altintas, San Diego Supercomputer Center, UCSD\\
David Anderson, University of California, Berkeley\\
Ritu Arora, Texas Advanced Computing Center, UT Austin\\
Bill Barth, Texas Advanced Computing Center\\
Wolfgang Bangerth, Colorado State University\\
Ira Baxter, Semantic Design\\
Timothy Berkelbach, Columbia University\\
Sanjukta Bhowmick, University of North Texas\\
Emre Brookes, University of Montana\\
Holly Bik, University of California Riverside\\
Ting Cao, University of Washington\\
Umit Catalyurek, Georgia Institute of Technology\\
Sunita Chandrasekharan, University of Delaware\\
Yong Chen, Texas Tech University\\
Margaret Cheung, University of Houston\\
Aurora Clark, Washington State University\\
Davide Cureli, University of Illinois at Urbana-Champaign\\
Matthew Curry, Sandia National Laboratories\\
Diego Donzis, Texas A\&M University\\
Rudi Eigenmann, University of Delaware\\
Peter Elmer, Princeton University\\
Maria Esteva, Texas Advanced Computing Center\\
Dan Fay, Microsoft\\
Ian Foster, University of Chicago\\
Geof Hannigan, Merck\\
Hrant Hratchian, University of California, Merced\\
Howie Huang, George Washington University\\
Tanzima Islam, Texas State University\\
Shantenu Jha, Rutgers University\\
Ananth Kalyanraman, Washington State University\\
Kate Keahey, Argonne National Laboratory\\
Gerald Knizia, Penn State University\\
Antia Lamas-Linares, Spectral Quantum Technologies\\
Elliot Lefkowitz, University of Alabama at Birmingham\\
Ben Levine, Michigan State University\\
Xiaosong Li, University of Washington\\
Si Liu, Texas Advanced Computing Center\\
Zhenfei Liu, Wayne State University\\
Andrew Lumsdaine, University of Washington, PNNL\\
Paul Macklin, Indiana University\\
Amit Majumdar, San Diego Supercomputer Center, UCSD\\
Madhav Marathe, University of Virginia\\
Devin Matthews, Southern Methodist University\\
Robert McLay, Texas Advanced Computing Center\\
Bronson Messer, Oak Ridge National Laboratory, University of Tennessee\\
Barbara Minsker, Southern Methodist University\\
Shirley Moore, Oak Ridge National Laboratory\\
Catherine Olschanowsky, Boise State University\\
DK Panda, Ohio State University\\
Abani Patra, Tufts University\\
Joshua Patterson, NVIDIA\\
George Percivall, Open Geospatial Consortium\\
Sushil Prasad, University of Texas at San Antonio\\
Xiaofeng Qian, Texas A\&M University\\
Diana Qiu, Yale University\\
Ravi Radhakrishnan, University of Pennsylvania\\
Rajiv Ramnath, Ohio State University\\
Bhanu Rekipalli, BioTeam\\
Natalia Rosales, University of Texas at El Paso\\
Andre Schleife, University of Illinois at Urbana-Champaign\\
Loren Schwiebert, Wayne State University\\
Sahar Sharifzadeh, Boston University\\
Sameer Shende, University of Oregon\\
Subhashini Sivagnanam, San Diego Supercomputer Center, UCSD\\
Alexander Sokolov, Ohio State University\\
Dan Stanzione, Texas Advanced Computing Center, UT Austin\\
Hari Subramoni, Ohio State University\\
Gary Trucks, Gaussian Inc\\
Ed Valeev, VirginiaTech\\
St\'ephanie Valleau, University of Washington\\
Robert van de Geijn, University of Texas at Austin\\
Sudharshan Vazhkudai, Oak Ridge National Laboratory\\
Shaowen Wang, University of Illinois at Urbana-Champaign\\
Nathan Wiebe, Pacific Northwest National Laboratory, University of Washington\\
Bryan Wong, University of California, Riverside\\
Elisha Wood-Charlson, Lawrence Berkeley National Lab\\
Michael Zentner, San Diego Supercomputer Center, UCSD\\
Xiao Zhu, Purdue University

\subsection{Workshop Program}

\noindent {\bf October 29, 2019}
\begin{table}[ht]
\begin{tabular}{ll}
8:30 AM & REGISTRATION, TEA/COFFEE/BREAKFAST, Location: near Primrose A/B  \\
&\\
\multicolumn{2}{l}{MORNING SESSION I:( Location: Primrose A/B)}\\
&\\
9:30 AM & Opening Remarks, 10 minutes, Workshop Chairs \\
9:40 AM &Welcome from TACC, 5 minutes,  Dan Stanzione, TACC  \\
9:45 AM &  NSF CSSI Program Overview, 15 minutes + Q\&A, Vipin Chaudhary, NSF \\
10:05 AM & Overview of the data-focused CI, 5 minutes,  Stefan Robila, NSF\\
10:10 AM & Lightning Talks, 8 minutes each, no Q\&A\\
&1) ``Thoughts on the CSSI/SI2 program''\\
&\qquad Wolfgang Bangerth, Colorado State University\\
&2) ``The Perils and Promises of Sustainable Data, a Cyberinfrastructure Perspective''\\
&\qquad Maria Esteva, TACC \\
&3) ``Keeping Software Infrastructure on Life Support Past SI2 or CSSI Funding''\\
&\qquad Robert Van De Geijn, UT Austin \\
&4) ``To Interface Or Not To Interface Or: What is an Interface Anyways?''\\
&\qquad Devin Matthews, Southern Methodist University \\
&5) ``A Science Gateways Perspective''\\
&\qquad Amit Majumdar and Subhashini Sivagnanam, SDSC  \\
10:55 AM & BREAK - 15 minutes \\
11:10 AM &Lightning Talks, 8 minutes each, no Q\&A \\
 &6) ``Building Collaborative User Communities''\\
 &\qquad Elisha M Wood-Charlson, Lawrence Berkeley National Lab \\
 &7) ``Network Analysis in the Era of Inexactness''\\
 &\qquad Sanjukta Bhowmick, University of North Texas \\
 &8) ``Speeding Up Science in Environmental -Omics ...''\\
 &\qquad Holly Bik, UC Riverside \\
 &9) ``Supporting Modern Programming Environments for High Performance Computing''\\
 &\qquad Jay Alameda, NCSA \\
11:45 AM &Extempore, 45 minutes \\
12:30 PM &LUNCH (Food \& Beverage Service)  - 60 minutes, Location: near Primrose A/B \\
\end{tabular}
\end{table}

\begin{table}[ht]
\begin{tabular}{lllll}
\multicolumn{2}{l}{AFTERNOON SESSION:} \\
&\\
1:30 PM& Brainstorming Session-1, 60 minutes  \\
2:30 PM & Brainstorming Session-2, 60 minutes  \\
3:30 PM & BREAK - 15 minutes \\
3:45 PM & Presentations, 5 minutes each, Session Moderators  \\
4:35 PM & GROUP PHOTO  \\
4:45 PM & BREAK (Food \& Beverage Service), 60 minutes \\
5:45 PM &Lightning Talks, 8 minutes each, no Q\&A   \\
&10) ``CSSI Workshop: Challenges and Gaps''\\
&\qquad Mike Zentner, SDSC \\
&11) ``Design Capture as a Paradigm''\\
&\qquad Ira Baxter, Semantic Designs \\
&12) ``Next-Generation Frameworks for Irregular Data-Intensive Applications''\\
&\qquad Howie Huang, George Washington University  \\
& 13) ``Software Challenges at TACC, CSSI Thoughts, and the LCCF''\\
&\qquad Dan Stanzione, TACC  \\
&14) ``Clowder: An Open Source Customizable Research Data Management Framework''\\
&\qquad Barbara Minsker, SMU  \\
& 15) ``Improving Reproducibility with Clouds and Notebooks''\\
&\qquad Kate Keahey, Argonne National Laboratory \\
&16) ``Software Challenges for ML driven HPC Simulations''\\
&\qquad Shantenu Jha, Rutgers University \\
6:50 PM&Self-paid dinner at a nearby restaurant \\
8:00 PM &Report writing discussion (for only those having report-writing assignments) 
\end{tabular}
\end{table}

\clearpage
\noindent {\bf October 30, 2019}
\begin{table}[ht]
\begin{tabular}{lllll}
8:00 AM & TEA/COFFEE/BREAKFAST (Location: Primrose A/B) \\
&\\
\multicolumn{2}{l}{MORNING SESSION (Location: Primrose A/B)}:\\
&\\
9:00 AM & Opening Remarks/Agenda of the Day,  5 minutes, Workshop Chairs \\
9:05 AM &Invited Talk, 20 minutes\\
&``Quantum Computing's Killer Application'', Nathan Wiebe, PNNL/UW\\
9:25 AM & Brainstorming Session-3, 60 minutes \\
10:30 AM & Brainstorming Session-4, 60 minutes  \\
11:30 AM & BREAK - 15 minutes \\
11:45 AM & Presentations from Different Groups, 60 minutes \\
12:45 PM &LUNCH (Food \& Beverage Service)  - 60 minutes (Location: near Primrose A/B)  \\
&\\
\multicolumn{2}{l}{AFTERNOON SESSION:} \\
&\\
2:00 PM& NSF Report Writing Session, 5 hours, Steering Committee \\
\end{tabular}
\end{table}

\pagebreak
\subsection{Pre-Workshop Questions and Responses}
\subsubsection*{A.5.1 Is it realistic to expect high-quality software or data product that is perpetually free, and for how long should the NSF funding support the development life-cycle?}
\begin{itemize}
\item Yes, the Linux operating system is perhaps one of the best examples. As far as scientific software goes, there are many high-quality open-source solutions to electronic structure and molecular mechanics/dynamics available nowadays (\emph{e.g.}, Quantum Espresso, CP2K, NWChem, GAMESS US, Psi4, LAMMPS, NAMD, VMD, \emph{etc.}) These solutions are widely used by the research community, if not to say more - they have become de facto the standards for some groups. The NSF should continue supporting the future developments of the free software at every stages, as far as such developments are justified by the community or scientific needs, even is the community may not seem to be very large. The commercial entities may not always be interested in such developments since they won't be able to gain much profit, either because the community may be relatively small or because the developments would focus on the fundamental scientific aspects which may not have an immediate fast return. Without the NSF support, some important problems may remain unaddressed for long time.
\item 1) For some widely used infrastructure, perhaps yes; witness Linux, GCC, PostGres and OpenOffice. For infrastructure with limited audiences, absolutely not; the economics don't support it. 2) ``Development lifecycle'' needs a definition; when does development ``stop''? The problem with useful artifacts is the the supporting structures rot, and so the software has to be modified, and demands for improvement require the software to be modified. An unchanging software artifact is a dead artifact. I think NSF has to decide which artifacts have limited audiences but need ongoing support (\emph{e.g.,} Automatic Differentiation tools?) and plan on long term support as investment in national infrastructure.
\item Yes. Open source software has shown that it can stay relevant for a long time, both inside and out of academia, and so I see no reason why high quality open source software can stay free forever. As for funding: NSF supports other core infrastructure -- say, telescopes. As long as the community believes (through periodic peer review) that a particular infrastructure piece is useful for the broader community as opposed to just a small and potentially diminishing part of it, then NSF ought to continue supporting its continued development and its community. It is on the PIs to describe the size of its user community and the impact a piece of software has.
\item It is never free, someone has to pay for it. Either it is paid up front, for quality development and support, or on the backend, by license and support fees. I think there is a gradation of software, a small amount that would benefit by direct, long-term support because of its global importance, and others that would be better with an initial investment and followed by a diversification of support mechanisms. The mechanisms are critical, as without the support, the software dies quickly. I can't put a single number on how long NSF should support the development lifecycle, the answer is far more nuanced than that.
\item Potentially - Yes. A very good example is the MVAPICH2 MPI library that has been/is being supported by NSF over the last 18 years and has been free throughout its lifetime will continue to be. However, the NSF should not support a pure development/maintenance effort. NSFs role is to promote innovation and research. Thus, as long as the software is innovating and creating new and novel solutions, NSF can and should continue to support it. It is up to the creators of the software to then find a way to balance development/maintenance with innovation.
\item The current model is good, but building a user base with sufficient users who are motivated enough to maintain and develop the software for free for their use takes a lot of time and effort. For new projects, building a user community happens only after the software has sufficient functionality to meet user needs. So, NSF funding is required for longer than some might consider ideal. For instance, a new software project of reasonably large scope isn't going to be self-sustaining after only 3-5 years of funding.
\item It depends on the functionality of the software. Does it need to ``grow'' by adding new functionality or new architectures? If so, then NSF funding may be needed periodically for a new generation of the software. If the functionality is relatively fixed and the software is layered so that changes in architecture are handled by other libraries (\emph{e.g.,} the BLAS), then software could be perpetually free and may not need much further maintenance.
\item It is not realistic because sustainability and maintenance of software is a really difficult problem to solve - especially if the user base grows exponentially and the software needs continual web resources, support, and training in order to stay relevant to research. However, perpetually free (open source) software is the ideal for scientific research, and NSF should find long-term ways to keep funding such software maintenance and development.
\item This is only possible if the software never gets updated. However, due to the changes in computer availability and growing scientific developments, scientific computing software is frequently updated. The person who works on updating the software will need to be funded, whether that is a graduate student, postdoc or specialized staff. Therefore continued support for the development of the code is necessary.
\item Research groups should focus on developing a stable version of the software or making significant changes to the existing version. There should be a different mechanism/entity for maintaining software. Instead of competing for funding for maintaining software, groups should compete for inclusion of their software into that entity. Or have institutional grants for supporting maintenance
\item no, software should be built into modular containers where they can be repurposed and used across different systems and projects. Data should be built into frictionless data packages using community standards for metadata so that they too can be reused in a modular fashion across sites. Data storage and compute resources need to be democratized, however, to promote access.
\item This requires sustained effort and personnel to maintain existing software which seems at odds with the academic goal of cutting edge research. I see no educational benefit attached to maintaining existing software which precludes the use of students and post doctoral fellows. It is also a difficult task for a PI who typically would be supporting new research.
\item We cannot expect software to be perpetually free, in general: The math doesn't work. Software sustainability is expensive. With few exceptions (\emph{e.g.,} industry wants to pay for open source, the user community also happens to be a passionate developer community,) software developers and maintainers need to be paid, and NSF can't pay for them all.
\item No. Sustainability of the software or data product has to transition away from NSF and be sustained by the community. NSF can fund innovation, but the community is responsible for sustenance. ``After how long'' will vary by situation, but there should be transition path that begins as soon as possible once the software starts getting adopted.
\item Data and software stewardship is one of the key problems to be addressed, so far communities can sustain it through different funding strategies but there is never a warranty that software or data will be available after a few years that the project ends and new resources and funding is needed to sometimes ``reinvent the wheel''.
\item Yes. It needs to be. Free access to software is an important step for leveling the field so that small institutions with little to no research funding can provide research experience to their students. NSF should fund productive projects as long as they are producing research results and educational experiences for students.
\item No it is not realistic. I can say that two cycles of NSF funding can be justified if after the initial funding there is actually some usage and/or impact to the community. Even in this case, the second funding should be for a proposal where further improvement, new feature (based on initial usage) is proposed.
\item I think it is. NSF needs to proactively identify software that is likely to stand the test of time (function of utility and timeliness), and invest in early development that makes it portable. The funding model can be a recurrent short bursts model - 1 year grants every $\sim$3-4 years of the software.
\item No, that wouldn't be a realistic expectation. It will depend on the product and its evolution. I believe via active discussions with the PI, a time stamp on how long should NSF support support the development life-cycle can be answered as different products have different expectations and goals.
\item In computational chemistry, there are high quality, free software packages available. However, they often benefit from perpetual support, typically from DOE (\emph{e.g.,} GAMESS, LAMMPS, NWCHEM) or NIH (NAMD, VMD). NSF might consider these cases as models for free, high-quality software.
\item Yes. Many software packages have a free version with basic features and the option to subscribe for a more fully featured version. This model seems to be sustainable. NSF needs to support the software until it's developed enough for industry or a non-profit to take over.
\item It would be helpful if after the major development, there is minimal funding from NSF just enough to support its regular maintenance. And when a major upgrade is needed, there are funding mechanisms such that the PI could apply for some larger grants to do the upgrade.
\item Yes. There are plenty of examples: \emph{e.g.,} Linux kernel. In science there are examples too: \emph{e.g.,} BLAS/LAPACK/MPI/\emph{etc.} NSF funding should support the development as long as the case can be made that supporting the development advances the science, which is the goal of NSF.
\item It is realistic to release a project and leave it free, however, it is not reasonable to expect it to be updated. NSF should also consider work with other agencies (\emph{e.g.,} NIH, USDA) to support the further development of software in a specific scientific domain.
\item It may be realistic to expect high-quality free software, while the application remains an area of active research interest, but if this is the case, there must be funding for software development and maintenance throughout its entire lifetime.
\item NSF should support the entirety of the development cycle, perhaps with multiple renewal opportunities. It takes at least 5 years to develop a good quality software, so NSF should support such a project for at least 5 years, if not more.
\item Yes, I believe the big data community has shown this is possible. Dask is another example of a NSF funded project that's picking up use as well as Numba. I believe 3-5 years is a reasonable time for projects to become self-sufficient.
\item I personally am an ``open-source believer'', but high-quality free software requires both funding and long-term personal emotional involvement. The role of commercial software should be considered where appropriate.
\item NSF funding should support development of the software product. As long as more features are developed or the software is updated, funding should be provided, to keep the software perpetually free.
\item Well, nothing is free, someone pays the salaries of the people who develop it. But open source is a good model in my opinion. How long NSF should support depends on the program and the objectives.
\item It is not realistic to expect a high quality software or a data product for free. I think NSF funding should support 2-3 development life-cycle for products that are used by the community
\item It's realistic to expect so for the software or data generated during the project period. The NSF funding should support the development for the project period.
\item It is realistic, but it depends on the user community. For the domains of relatively small user space, it will need much longer support from NSF.
\item Yes, there are such software available in my field of electronic structure, such as Quantum Espresso, that are highly used and very good.
\item Perpetually free is not realistic. Both code and data require maintenance. As long as they are used they will need to be maintained.
\item yes to the extent that it enables new science in ways not otherwise possible. The length would be determined by the same criterion.
\item Yes, because this is critical national science infrastructure that is much cheaper to maintain than it is to create.
\item Yes. it is not possible for a single lab to do this without consistent support from the NSF.
\item Not realistic. NSF should consider funding the ones that have shown value to the community.
\item Free, no. Subsidized, yes. Funding period will vary greatly depending on the product.
\item I think it is ideal but not realistic in perpetuity
\item Yes, it is a major gap right now.
\item Somewhat, 5 years or more.
\item no; 5-6 years
\item sure, 3 years
\end{itemize}

\subsubsection*{A.5.2 How can sustainability be supported while encouraging creativity?}
\begin{itemize}
\item In my experience, creativity in scientific software is $\sim$100\% driven by scientific inquiry. To some extent, commercial software provides sustainability while allowing creative input via feature requests by users. However, this approach tends to disfavor high-risk high-reward features. In academia, sustainability is usually sacrificed due to time and funding limitations, and a reduced focus on software quality (because it is not directly incentivized). Expecting individual scientists (even with additional funding) to produce high-quality sustainable software is generally not realistic, but funding interactions between groups of scientists, between domain and computer scientists, and perhaps between scientists and industry in order to collaboratively produce larger frameworks and community codes that spread out the burden of sustainability might be a better solution. On-going maintenance roles and funding is a major issue.
\item The creativity and the ``new'' aspects of research have been emphasized both by NSF and nearly all peer-reviewed scientific journals. However, this philosophy is not supportive of the idea of sustainability at all. As a result, we see a large number of ``new'' solutions being proposed all the time, whereas the continued efforts on ``established'' (not fully developed though - that's why the quotes) approaches are not typically considered popular or fundable. Perhaps, NSF could revise (or extend) the definitions of the merit/impact criteria to reflect the need for ``sustainability'', at least at the scale of individual programs, \emph{e.g.,} within CISE.
\item I assume that this is a question about what CSSI should fund. A first question is what is CSSI's goal. The name says that it is about building cyberinfratructure, but funding decisions seems to emphasize cyberinfrastructure research, \emph{i.e.}, creativity with no attention paid to sustainability. I'd argue that for CSSI to have an impact, it needs to be serious about evaluating sustainability plans, and to recognize that open source is NOT a sustainability plan in itself.
\item I think in order to be sustainable one needs to be creative and break out of conventional ways of solving a problem. So they both should go hand in hand IMHO. For example - producing green energy - Wind solar geothermal ways to extract energy. Need creative techniques. Evolution has to be democratic. Another example - How to tackle climate change? Creative ways such as smart fabrics using sensors that can sense temperatures and adjust temperature. That will determine heating/cooling requirements.
\item To be honest, I'm not sure. I think many research codes run into a problem where the initial implementation is done in a quick and dirty way to get results and then the code dies after the graduate student or postdoc who developed it leaves the group. One way to avoid this might be to offer some sort of technical support for code refactoring and optimization and to make collaborations with this technical support a requirement for using in-house codes on NSF supercomputers.
\item I think the answer lies in who gets the funding to sustain/perform continued development of a given software. Current models encourage same team of developers or centers. But if we have to ensure creativity and wider community contribution for large-scale software, I think preference should be given to new developing team of contributors each funding cycle. This would not only promote creativity but also would facilitate decentralized management (ownership) of software.
\item Sustainability is what makes our research usable. I think there should be a separate opportunity, perhaps a small amount of money or allocation grant on XSEDE or the cloud storage, for making data and results produced accessible to the community. I think building a storage cloud and interfaces for scientists to share and access their artifacts long after the 3 or 5 year development cycle will go a long way in making research sustainable.
\item The challenge is to get developers to invest the time needed into making each feature robust and flexible for a wide range of users before moving on to develop the next feature. This requires some discipline within the project team that is difficult to enforce from outside the project. It isn't clear what metrics could be used to determine how well a software development project is being developed for sustainability.
\item Innovation and stability is achieved in software development by having specifications for the concepts and implementation specs. The conceptual specifications capture the concepts which evolve slower than technology. Implementation specs show how the concepts are implemented in various technologies. The conceptual specifications form a sustainable architecture that enables creativity of implementations.
\item Sustainability is hard because artifacts are becoming more complex, and we insist on forgetting how and why they are built the way they are. We literally forget the design, and then we pay highly to recover enough to do ``sustaininabilty'' work. We should push methods to capture and retain design information as one key way to reduce sustainability let yet people be more creative. 
\item I think ``usage'' of a software is key to sustainability. NSF shouldn't fund something because it is exciting; realistic usage potential has to be evaluated and demanded explicitly in the proposal call and carefully reviewed during the review process. ``Usage'' will also lead to creativity as new needs from users will give rise to requirement for new features and which will lead to creativity.
\item This is exactly where federal funding is likely required. Once software becomes a commercial product, changes to the software will likely become incremental since new funding is dependent on the success of funded changes. More radical change requires longer term funding, which for the kind of software supported by CSSI means, well, CSSI-like funding.
\item Every project could be a mix of innovation and sustenance. Or else innovation and sustainability can be the primary responsibilities of different communities with different funding/support streams. In addition, there is creativity in making something sustainable as well. This should be recognized.
\item More early prototyping grants for ``risky'' software, especially those that encourage collaborations between interdisciplinary researchers (\emph{e.g.,} Biologists and Computer Scientists, and/or industry partnerships especially amongst the freelance data visualization community)
\item I don't think they are mutually exclusive. I think it is best to sustain software that is evolving and developing in new and wonderful ways. Simply trying to keep the lights on is the opposite of a dynamic software ecosystem.
\item This is a difficult question. One method could be by supporting long-standing projects that are 1) being used by the community 2) continually innovating, and 3) have an impact on the scientific community.
\item That's a question for software design. Most of the successful projects are led by experienced software developers who can come up with designs that allow for extensibility in view of new requirements.
\item Some crucial elements of scientific software infrastructure are so important that it may make sense to invest in their maintenance. So NSF should balance these two (sometimes exclusive) objectives.
This could be the down side because there is a lack of motivation to create something new if funding is stable. Perhaps the community can create a committee to constantly create new ideas.
\item Encouraging establishing a community of developers, some of which are responsible for sustaining the code and others are responsible for building new (creative) features.
\item These two don't have to be mutually exclusive, but they have to be viable. There has to be some end user that wants to use these or the output will inspire new research.
\item Different thrusts of the program can accommodate the early stage high risk creative project, while the established ones can focus on the sustainability.
\item I don't see why these two goals should be at odds with one another. NSF should fund CSSI proposals that demonstrate both sustainability and creativity.
\item Sustainability can drive innovation by introducing new design constraints that shape how key resources are used in products and processes.
\item If the thrust of this question is related to the question above then they are two different tasks albeit equally important.
\item Encourage the formation of open source consortia with production software and continued open source innovation.
\item Funds need to be allocated that value both. Furthermore reviews should accommodate these two views as well.
\item I don't think there is a contradiction between the two. A creative work could also be sustainable.
\item A software project can serve as a firm foundation for research into different aspects in an area.
\item good question. i guess creativity is involved in the beginning and will be always evolving
\item Through sites like Dock Store that promote community sharing and ongoing innovation.
\item Allow more non-traditional funding, including crowdsourcing, donations, endowments.
\item need strong oversight and need to change peer review standards for software papers
\item Federal funding and reducing redundancy between the funding organizations
\item on a case-by-case basis. General principles would be worth a discussion.
\item Easy. Provide a separate source of funds to support sustainability.
\item Gaining invested support in a life-cycle plan for the project.
\item Have continuous funding to support the software development
\end{itemize}

\subsubsection*{A.5.3 For sustainability and cost-effectiveness, should the PIs be encouraged to buy software and services from industry instead of spending effort in developing open-source and free versions of the commercial software, and when is it appropriate to compete with commercial solutions and build from scratch?}
\begin{itemize}
\item Perhaps, sometimes it makes sense to encourage the use of commercially-available software. But the PIs should not be forced to obtain such software and services from the commercial entities. By definition, the primary goal of the commercial entities is to make profit, which may inhibit the transparency and the ideas' sharing vital to the scientific progress. Scientific software should be open-source and freely available to the community, so the community could help the software's future grow and development. This is especially true for the developing areas, where the solutions are not well established yet. For more-or-less standard solutions (\emph{e.g.,} a range of ``standard'' electronic structure or molecular mechanics methods) the companies that involve a wide scientific community may be reasonably positioned to provide reliable software solutions to the broader community. In this cases, it would make sense to encourage the PI to use such solutions, provided the PI's goal is in the applications of the software and not in the development of new approaches, unless there is a way for the PI to obtain access to the code base, free of charge.
\item I strongly believe that products out of academic research is key to creativity. To that end, I don't believe PIs must be encouraged to use existing software and services from industry. Competing can be encouraged once it is clear that the PI has spent enough effort to refute existing solutions (via experiments) and that there is a good/high demand for newer solutions. This does put a heavy burden on the PI to do substantial homework before proposing a solution but I believe that is what research is all about. Many a times we resort to an existing commercial solution as it is one of a kind and probably the best out there and cost-wise it might take a lot more \$\$\$ to build something similar which is only going to be marginally creative and useful. So I believe estimating the reward and the demand ahead of time is key to demonstrating the need for new open-source software.
\item In general this can be problematic for a few reasons. If the source code is not available it will be very difficult to understand how to add any additional modules if needed. On the other hand, even if the source code is available it can be very time consuming to learn how a code was written to be able to add a new part of code to that existing code. Lastly, software from industry often comes at a high cost which adds to the expenses of PIs in academia. Open source code on the other hand solves the above problems.
\item You frame the problem incorrectly. The choice is NOT just between commercial and free. 1) An important alternative, as illustrated by, \emph{e.g.,} Globus, HUBzero, InCommon, is support from stakeholders for software developed by not-for-profit entities aligned with needs of academic research. 2) Many widely used open source solutions (\emph{e.g.,} Linux, Mysql) are successful because of substantial commercial contributions. 3) Software-as-a-service as a delivery vehicle for research software can change cost calculations.
\item Software and services from industry make sense only if they are offered at a competitive price, which doesn't often seem to be the case. Building competing systems makes sense when commercial solutions are either prohibitively expensive or lacking in the required functionality or flexibility for users to accomplish their research. The dividing line depends also on the size of the community. The larger the user community, the lower the requirements for deciding to developing a competing project.
\item Perhaps. It is appropriate to ``compete'' with commercial interests to ensure open, portable, reproducible research that avoids proprietary lock-in, and to avoid the risks of corporations unexpectedly exiting the market (bankruptcy, deciding a service isn't profitable, \emph{etc.}) Open source, portable infrastructure should be hosted on low-cost commercial platforms. Software / data maintained by open source and scientific community, easily ported from one commercial host to another for safety.
\item At least in my field, transformative and innovative new theoretical methods require modifications to the program; some commercial packages allow/encourage this, but open source is beneficial in this case. In more extreme cases, an entirely new program is the most pragmatic way to move forward, but definitely is a burden on sustainability. Re cost-effectiveness: cost-effectiveness in terms of PI and student time is generally inversely correlated with sustainability.
\item Yes, this definitely makes sense. One way PIs could be encouraged to do something like that is by allowing them to allocate money from the budget to buy license for the duration of their development. This will allow them to figure out the kinks, and address all issues with the vendors before their grant expires, so that afterward they feel confident about spending money from their pocket to sustain this investment.
\item I feel if something is well developed and serviced by industry and where there are competition among industries, it is OK to do that. Academia will do new CI development and to some extent truthfully (\emph{i.e.,} CI should be usage driven) propose new CI/software ideas and this is something industry will not do. And hence new software/CI development proposals from industry an be built on something obtained by industry.
I might be biased as I lead an open source project, but I believe value is in data, not in the SW. Opensource also seems to bridge more communities than divide them, so I believe PIs should be encourage to opensource AND if they do contribute to other existing ecosystems and adopt standards. The goal should be to advance the space, and not create more close to but slightly different libraries.
\item There is no publishable work connected with reinventing the wheel. Doing so is a disservice to both students, postdocs and the PI. Software efforts should fill in gaps and push the state of the art. Otherwise, this is as ridiculous as writing your own word processor. Competition with commercial solutions include ongoing support and maintenance requiring mundane yet time intensive duties.
\item I think that developing software will always be necessary for cutting edge computational research, as frequently the most interesting applications are new areas of research where software does not yet exist. The problem with working exclusively with industry software and services is that they are usually proprietary, making modifications or new development challenging.
\item Developing open-source versions has the advantage of greater expandability, and it is relatively easy for the developers to modify such versions by adding new features, compared to the industry-level software. However, this practice should only be supported when there is absolutely such a need, \emph{e.g.,} when commercial software lacks a feature that is well desired.
\item PI's specific requirements for using the software and the amount of funding available will determine whether they need to buy software or develop open source version. It is beneficial to develop open source solutions when the desired features of the software are community driven and the community can make recommendations on how the product matures over time.
\item No - open source should be the main emphasis. I have philosophical and practical concerns with closed source (proprietary) and fee-for-service workflows that are opaque and non-reproducible. Open source alternatives to commercial software would be highly valuable to the scientific community, even if it means building this software from scratch.
\item In general, license fees for software developed by the industry can be expensive. In the absence of open source competitors, these license fees can be jacked up quickly. However, doing everything from scratch is also not viable for very complicated packages like Fluent. Thus, one needs to have a good balance the two.
\item I don't see a reason why PIs should not be allowed to buy commercially available tools from industry. However, ``encouraged'' is a strong word in this context. I think commercial tool usage as part of research should be allowed when a strong justification in terms of productivity and cost-effectiveness can be made.
\item When a small group of well-paid private companies control research software they also control the research agenda. The methods used to produce the data can't be explored in the same way. Now, when the software handles a decided part of the workflow that isn't an issue, but I am not sure that is common.
\item Depends on the capabilities in a given lab or institution. This is often true, most lab rely on graduate students and postdocs whose skills are fleeting, and therefore cannot be depended on for robust bullet proof code. Labs with developers, however, may be able to produce robust production-level code.
\item Wow. Where to start on this. It really depends on the goals of the development; industry can be far outside of the places where research software needs to go. The answer on this is basically - it depends, on a lot of complicated factors. And, don't forget there are costs literally everywhere.
\item The PI should be given both options. While it is difficult and few can do it, there are some PIs who can develop and sustain high quality codes that are highly used --they should be allowed to do this and supported by NSF because it benefits the scientific community.
\item It should be in the mission of the NSF to support Open Source software as a common good for the US and it should be considered vital to the support of scientific research to do so. Certain types of services should be outsourced, including web-portal development.
\item There is a continuum of software options in my field, ranging from 100\% publicly funded to largely commercially supported. The publicly funded projects are competitive with the commercial ones. NSF should continue to fund both free and commercial packages.
\item Most software/algorithms available commercially are not suitable for research. In particular, research has many intricate questions for which knowing how to modify the software or collaborating with people developing the software is essential for success.
\item There are two reasons (that I can think of) for developing community versions of software. One is if the community versions are advance innovations and the second is when relying completely on industry versions is a risk to the scientific community.
\item There's a third answer: NSF could outsource support for software and services on a commercial contract basis to ensure continuity, with the proviso that the software resulting is free to at least research institutions if not outright open source.
\item I think the PIs should definitely leverage the resources from industry, if the commercial solution is solid. The PIs could then spend more efforts on building tools that do not exist yet, and that is where creativity occurs.
\item Partnership with industry should be encouraged, some companies have open-source versions with possible limited features but new ideas or exploring features that are not priority for industry can be explored by academics.
\item Depends on the availability of commercial software. If high quality software is available, that might be preferred to build-from-scratch. If software is very specialized, build from scratch might be the only alternative.
\item I don't think NSF should encourage. But if there is no clear need to develop software from scratch it should be proposal reviewers and panels who should make this judgement and recommend industrial/commercial solutions.
\item Depends. There are cases when the answer is a clear yes or a clear no. In general industry solution need not be the best (nor should be ruled out). It depends on how mature the systems have become.
\item No! PIs should be encouraged to support open source sources of software and contribute to its development. Industry software is fine if it is affordable and fills the need.
\item This is a complex business problem. There is probably some middle ground around developing free software, but providing services like support through a commercial model.
\item If industry software is available at a reasonable cost and can be integrated with other software then it should be used. Otherwise open source development is justified.
\item From my experience with Gaussian that prohibits benchmarking, it is stifling to the progress of research when the test of new ideas needs tinkering and modifying codes.
\item PIs should be encouraged to embrace the most cost-effective solution, keeping in mind that there is cyber-training value in developing open-source and free versions.
\item Neither should be discouraged at least, to foster the diverse approaches. However, the focus shall be placed on the intellectual merits and impacts.
\item no point reinventing if open source tools and packages are around. Key is open source. It is appropriate to create when new methods are developed
\item No; it is appropriate when the developed software has some additional capability that the commercial software does not
\item If the goal is science and discovery. Yes, rather researchers do good science then spend time on developing software
\item If this functionality is readily available commercially, I see no reason not to go with commercial products.
\item No single answer. Sometimes industry solutions are a perfect fit, some times NSF requirements are unique.
\item Commercial software is often not scalable for HPC applications and systems.
\item when intellectual merit and broader impact justify it.
\item only use free open sources, no need to buy.
\end{itemize}

\subsubsection*{A.5.4 Industry maintains the lead over the academia in establishing software development processes, especially for web-portals, and has crystallized the best practices. Are such practices and processes easy to adopt in academia? Should the PIs be encouraged to subcontract the development of web applications and portals to colleagues from industry? What are the challenges and advantages of doing this?}

\begin{itemize}
\item Industry maintains the lead because of a need for an iterative process in software over many years. Until the NSF begins to create funding opportunities that are many years long, this is not needed. It is up to anyone who picks up the open source software after the fact to deal with it. The PI should be encouraged to subcontract the portion such as the web applications and portals development to industry. The main challenge would stay the same, which is the communication between the researchers and software developers, in particular on usability. In fact, it could be improved considering business mission of the industrial partner. For sure, the subcontract should be kept short term with renewable option. In addition, NSF has funded the project like SGCI and PIs should be strongly encouraged to work with SGCI for related science gateway development.
\item Challenge: Discourage early career academics and put one more hurdle in their pathway that will discourage them to pursue funding from NSF. In turn, this might discourage academics altogether to pursue this career option. I am an early academician, and I feel intimidated by the fact that there is one more duck I have to get in the row, a rather large and non-cooperative duck at times, before I can pursue NSF funding. This option will only work IFF NSF arranges for such collaboration. Advantage: The potential benefit is obvious. Following software engineering practices and leveraging web-portals from the professionals, if subcontracted, can only improve the quality of software that is output from a project. I am just worried, that such requirement will make these funding opportunities less accessible to early career people.
\item The primary challenge here is that an academic's funding depends on his/her ability to publish papers. Adopting software best practices is time consuming (at least at the outset), and that time does not result in publications. If researchers can expect sustained funding for software development, it will become easier for academics to adopt best practices. It is my understanding that this is CSSI's goal. Perhaps the sticking point is that it's not obvious to potential PIs that they should expect ``sustained'' funding for software development projects the way that they can for fundamental science.
\item I don't get the emphasis on web portals. They're surely not the most important part of research software? Some relevant perspectives: 1) Building high-quality software needs specialized skills and teams; it is unusual that these can be established in a university setting, but can be done so, \emph{e.g.,} Condor, Globus, HUBzero. 2) Research software should make much heavier use of software-as-a-service methods than it does, these can semi-automate a lot of the required development. 3) See https://docs.globus.org/modern-research-data-portal/ for how the right platform can simplify things.
\item With proper support, I believe they can be. While industry can lower the cost of the development and maintenance, it increases the risk of proprietary lock-in, unexpected cost increases beyond our control (\emph{e.g.,} Oracle-style business practices), unexpected loss of services (\emph{e.g.,} Google wakes up and decides to discontinue a service), \emph{etc.} I'd encourage PIs to use commercial low-level services to reduce data hosting and other costs, but to build the higher-level software \emph{etc.} on the top that can be easily moved from one commercial vendor to another.
\item Industry best practices are efficient and end-product-driven, which I think is a good thing. This is not always easy to adopt in academia however because of time demands on PIs and the difficulty of finding good people to employ (\emph{e.g.,} brain drain to industry, much lower salaries that make these positions less attractive). Subcontracting out to industry is fine, as long as the development focuses on open source tools and there is no long-term commercial/for-profit agreement that would produce negative consequences for scientific end users.
\item No, it is not easy to adopt such practices and processes because academia has reward systems that generally do not reward industry-scale software development. Subcontracting development of web applications and portals would be great if the budgets can be sufficient to support this. Advantage would be working with professionals who know how to do software design well. Challenges would be managing the additional cost.
\item I would say academia will be saving time if this component is off-loaded to already well established and existing solutions. The advantage is that the academia will not be re-inventing the wheel as I believe there are many such solutions out there and the downside is lack of possibility to customize according to needs. So this would depend on one's flexibility and adaptability to existing solutions.
\item The practices followed in industry, while good, can be too cumbersome for academia to follow due to lack of software developer time. In my view, one can compare software development in academia to a startup company --- A startup follows some set of best practices while ensuring that the process does not impose too rigid of a restriction on the pace of development.
\item The primary challenge is disseminating this information on best practices to those working on the software development and encouraging adoption of these practices. The actual processes are not as difficult as getting them to be used. Subcontracting these services is a viable option, depending on costs including long-term maintenance.
\item I think productization and maintenance could be handed over to entities more capable of doing that kind of work. These entities could be industry organizations or they could also be professional research infrastructure support organizations like super-computing centers or research computing entities within the university.
\item PIs can be encouraged to subcontract the development of portals to industry if the existing open source solutions are not a good fit based on the requirements. The advantage is outsourcing of the development work and the challenge is cost, additional cost for new features, and continued maintenance
\item I think tasks such as web applications are best left to industry. The effort of academicians and researchers are best spent on research and generating and testing new ideas. NSF can partner with industry to help facilitate academic-industry partnerships.
\item Not sure. Are the web apps and portals mainly for outreach? If it is not specialized to an academic subfield, it may be advantageous to subcontract to industry. Otherwise, I think academic software development processes may be very subfield dependent.
\item It depends on the partners and the goals. In general this has become relatively mature and industry does provide very good alternatives (cost is a different matter). There is of course the broader question of training the next generation scientists.
\item It will be great to partner with industry. The main advantage is the mutual benefits to both academia and industry as the former can concentrate on scientific development. The main challenge will be to standardize and speed up the process.
\item I think that industry members would be interested in collaborating with academics who can build the tool but lack the web-portal capabilities. A big challenge would be linking the academic groups with the industry members who could benefit.
\item Subcontracting for more ``standardized'' implementations is likely preferable, because it ensures cutting-edge web portals that are safe, and the scientist can focus on putting creativity and effort into building backend functionality.
\item Does industry make its processes and products available? I am not aware of solutions that would readily support our academic needs. As for subcontracting, this should only be done on a case by case basis dependent on need.
\item Depending on the software, some are easy to adopt and some need significant training and education. But having uniform applications with enough competing commercial tools will help in long run for academia and industry.
\item The question is why should academia need to develop these? It is indeed possible and, yet again, requires manpower very different from most PIs, students and postdocs areas of expertise. Is this a worthwhile endeavor?
\item I think that is totally valid. However, there are also open-source equivalents for these well established software, and it might make more sense to subcontract the development of the open-source ones if available.
\item Challenges: the process is expensive if collaborating with industry, also the requirements of industry and academia are very different Advantages: More accessible software, outsourcing of ``non-research'' component
\item Sometimes it may make sense, but I am afraid that there is just not enough money in most PI budgets to afford industry-grade solutions. I think this should be done on a case by case basis.
\item Perhaps the NSF can allow the budget for a research staff who can do this thankless work. No postdoc or grad students I have worked with wanted to spend time on this.
\item I believe these processes are easy to adopt. GitHub for instance is how RAPIDS plans and tracks all roadmaps and issues. Many best practices are low cost.
\item Depends on the nature of the job - some jobs require a sufficient understanding of the science, which industry may not work as well as people in academia.
\item It is not easy to adopt this in academia since student-training time is so long, whereas industry workers are typically trained from the start.
\item I don't understand the emphasis on ``web portals''. Surely that isn't the most important topic in the scientific sustainability space?
\item Software for web-portals have become commodity products. Populating web portals with NSF content requires PI guidance.
\item Yes, these best practices need to be adopted and used in academia, often times labs are not infrastructure savvy
\item college students are creative and they can do great development, given resources, examples and experiences
\item Yes, providing funding for web application development and user interface is very important.
\item Yes, processes can be adopted as long as a savvy software engineer can support the project.
\item Probably yes, unless they can pay undergraduates or support non-research staff to do it.
\item Web design seems like a good task for students. They will learn and earn.
\item need to educate or change software engineering education
\item I don't agree with the premise.
\end{itemize}

\subsubsection*{A.5.5 What if the NSF funding is used to develop a minimum viable product that can be made free and open-source, and extended features and services are offered on top of the free versions at a cost? Does this look like a reasonable approach for sustainability?}
\begin{itemize}
\item It sounds reasonable at first and could potentially work. However, there is still a risk that a wide range of versions of the ``extended features'' may be provided by different vendors, in which case the sustainability may be compromised since it will not be clear which ``camp'' to choose. As I indicated in another answer, it may be good to rely on the commercially-available solutions with lower variability - that is for more-or-less established methods (\emph{e.g.,} Hartree Fock could be done by various commercial solutions more-or-less identically). Then, NSF could support further efforts on finding consensus among various new directions and solutions to important unresolved problems. For instance, there are tons of approaches for nonadiabatic and quantum dynamics, but it is really hard to know how they all relate to each other and what their limits of applicability. In such situations, some comprehensive analysis across various methods is needed to make further developments sustainable. This is an example of the area with a greater variability in the methodological approaches. Although it is important to foster this ideas' diversity, the sustainability would require finding common grounds among them and finding the ways these methods to be interoperable with each other mutually reinforcing their strengths.
\item Is this cheaper for NSF? I recognize that it will impose the cost of the software on the NSF-funded projects of those people who actually use it, rather than the whole community. But in the end, NSF will have to pay. My concern would rather be that the unavailability of the real functionality ``for free'' will encourage PIs to develop their own software because that is ``for free'' -- it just costs PIs and NSF-funded grad students time they could spend on more relevant research, and it will almost certainly lead to worse software. I believe that the end result will simply be less and worse research.
\item I suspect that the main driver to develop new features is to explore new science. Thus, making these new features a source of income might not offer much additional incentive, especially since the target audience may be very small. For open-source codes, I think a requirement that any branches resulting in publications also need to be made freely available within a year would be a good incentive for the community to contribute actively to code development and maintenance.
\item I like this idea and I think mixing up NSF funding and topping it up with features and services sounds like a win-win. Again this would largely depend on how much in agreement both parties are and will require several dialogues and compromises from both ends. But worth a try. This way academia is able to brainstorm one piece of the puzzle and come up with novel ideas and lean on others to complete the rest of the puzzle.
\item I'm not convinced that, at least in many cases, this would be attractive option to the academic community. What is the value added and who has budget to pay for these additional costs? Users often underestimate the time/cost of developing a system from scratch and the drawbacks, including debugging and reproducibility of results are often overlooked in these calculations.
\item This is possible. I think if anybody wanted customer service, they can pay extra for this. However, I don't think it is realistic for a research lab to offer customer service. Perhaps we can see ourselves as the R\&D arm in a company for an example, and leave productive development to other specialists.
\item The success of this model would depend on the size of the market for extended features. Would it be profitable for companies to take on? and would scientists have budget to spend on those types of add-ons? Also, the add-ons are often where the interesting research questions are.
\item Again, we need to determine what is the goal of NSF/OCI funding. I think that it should be to create high-quality, sustainable research software, not fund research in software. The question of whether the software is free or commercial seems to me to be a red herring.
\item Possibly. The problem is that the kind of people who like to develop software in academia are not necessarily interested in creating a company. Shouldn't we encourage such people to stay in academia by funding the valuable software they develop?
\item Who would offer the commercial version? What would the extended features be? Services (I guess support?)? What would the effect be on sustainability of innovation and creativity (at least as important as sustainability of the software itself)?
\item yes. If the added feature are big science enablers, the equation changes a little. Whether OAC pays for development/sustainability and what domain directorates would pay for services should be considered and balanced across directorates.
\item This question is confusing. Development of MVPs that transition in to sustained baseline is a good plan but it needs an architecture to enable the transition and an infrastructure to promote the open-source community, \emph{e.g.,} Apache.
\item It would depend on the size of the market for a particular project. If a PI proposed and justified this model for sustainability in a specific proposal, I would definitely support it. It's not a one-size-fits-all solution, though.
\item That is a great approach, a free and open source core, and derivative ``products'' offered for a cost. The weak copyleft license used by the Eclipse foundation, plus the processes the foundation employs, is one way to handle this .
\item Yes, this is a viable approach. However, we have seen a lot of questions (especially in NSF reviews and as feedback from the community) asking why an NSF funded project is not made available free of cost.
\item This appears to be moving in the arena of commercialization. It can be done; however, it requires ongoing funding and stable personnel. It seems counter to the academic mission of education and research.
\item I think there is good reason to believe anything that originates from NSF funding should be free to the broader community until commercialization of the tool (via startups or adoption in industry).
\item This needs significant effort and funding from NSF to keep developing these tools. For some it might be reasonable but for other it might not be realistic to maintain for long time.
\item With the current funding model, it seems like this is all that can be done and a reasonable approach for sustainability while encouraging diverse prototypes.
\item Yes, NSF should help researchers to develop a Minimal Viable Product (MVP) that can then be hardened by industry. Academia is good at R\&D and innovation.
\item Maybe, it would depend on implementation and how effectively it is managed. I think it would require a dedicated group just to manage this and promote it.
\item Sounds like a good idea. Then there is the huge question of what features should be paid-features and what features should be kept in the free version.
\item This is possible, of course the hard part is to decide on MVP. But the original developers have a continued role in systems that are good and improve.
\item Yes, but it depends on the nature of the product. I don't think some of the research software could fit in this model.
\item Yes, but the path to sustainability depends on the situation. Freemium services could certainly be one path.
\item I don't think people would be interested in a minimal product so I don't see the point of doing this.
\item Yes, but the NSF should ensure that previously free features are not moved to the paid version.
\item it is a good idea and I can see myself being excited about this model.
\item Yes. Matches my ``third answer'' point above.
\item Yes, this seems like a reasonable approach.
\item Yes, this could be one potential solution.
\item Yes this is very reasonable.
\item Yes - see my response above
\item Yes that seems reasonable.
\item This is a terrific idea!
\item this sounds a good idea.
\item Yes, possibly
\item Possibly yes.
\item Perhaps ...
\item Yes.
\end{itemize}

\subsubsection*{A.5.6 How should the software be maintained in the long run? When a new software compiler and hardware infrastructure come out, if needed, who should update the software accordingly - developers and/or the community for whom the software was developed?}
\begin{itemize}
\item Perhaps, there could be a ``cyber-infrastructure updates'' category of the NSF grants, similar to the instrumentation grants to experimentalists. The rationale is very similar - to help advance science via getting better instruments/infrastructure, so perhaps this new grants category could re-use the existing structure of the instrumentation grants applications. These grants could support solely the software updates due to major changes in the compilers and hardware infrastructure. The proposals would then address the needs for such transitions/updates, potential benefits from it, and how the transition will be made. The project could probably be rather short-termed – 1 to max 2 years and could support hiring professional software engineers.
\item I think both are equally responsible to keep the software up to date. The person who created the software is naturally going to know all the hoops and jumps and hence completely offloading the software to the community to update will not sustain and might break the software. Instead I believe the onus is on the developer to find those handful from the community that could help him/her update the software in a systematic way. This effort could grow in the right direction with inputs from the developer. This is also needed as the developer can maintain the software only to a certain point, sooner or later he/she is going to need help esp. if the software is being widely popular and used.
\item I think there is a case to be made that with every NSF research project that gets funded, some part of the money should be allocated to software development in the budget. Currently there is no incentive in terms of publication for such development exercises. That is why academia doesn't do it. Creating an added incentive in funding (\emph{i.e.,} research expenditure) can help mitigate this issue. All universities have qualified MS and UG students who can possibly be supported and this kind of added budget component (in the form of a SW Dev. Supplement) can make PIs think of creative ways of engaging with the available student talent at a university across all levels.
\item Depends on the software. Some developers are willing to do this type of work, some are not. What is crucial here is the mechanism for translating the ``demand'' for software development from users into financial support. This works differently for each case. Some developers use CSSI to do this translation. Some do this directly by raising money by selling software. The latter is more direct and probably ``honest'' in a sense that it is clear what the funder pays for. But we should not have to become business people if we want to provide software to the community.
\item Ideally, this should be a shared responsibility across the user community. Assuming the developers are still using the software, they should have motivation for doing this, but so should others. When there is no longer funding for the software, it is not realistic to have a single point of support. When there is funding for research that relies on the software, this motivates someone to maintain the software. This may be the best solution in the long run.
\item The real question is whether there should be a ``gate keeper'' to ensure continued quality control as new compilers and hardware infrastructure (and/or functionality) comes along. We would like to think of such a person more as a facilitator, who supports a community that contributes while safeguarding the quality of the software. This person (or group of people) is of extreme importance to the software and is often expensive to fund.
\item It depends on the skill set of the community. Some software is meant to be used by researchers that would not know where to begin to maintain the software. Other software is more of a collaborative development, and the development and maintenance burden can somewhat realistically be shifted to the ``user'' community that is using it. In that case, the developer community and user community lines really blur.
\item In most cases, maintenance of software for new compilers and/or small updates in the hardware platform is something the successful projects can do through their community: If projects have a successful structure of incorporating community contributions, then small updates will come from the community. (Of course, this will not apply for shifts such as from CPUs to GPUs.)
\item In the long run academia is not the best place to do this, they rely on students often as the students graduate the work is neglected. It is also the case that funds need to be provided to do the work. At this point, the level of funding given out to build good systems is small in my opinion and when compared to funds from NIH for \emph{e.g.}
\item It is OK for the software to charge small amount from its users. This will automatically prove if the software is needed. Sometimes the basic CI development can be funded by NSF but afterwards some specific user oriented organization (such as NIH) may pick up funding it; this is for a CI which is applied in biomedical field.
\item This depends on the application and scope. It can be extremely time consuming so who has the incentive, manpower and funding to perform such tasks? Outside of computer science, this is a distraction from new research and can be very far afield from concentrated areas of study (\emph{i.e.,} chemistry, physics, mathematics, \emph{etc.}).
\item It can be a combination of both. However, the original developers of the software or someone who has in-depth knowledge of the software must review the changes and ensure 1) it is taken through appropriate thorough testing, 2) there is no performance regression and 3) there are no corner cases at the application level.
\item The NSF should fund any modifications to keep the software robust if the software is critical to a specific research. Research grants should outline the software that is critical and needed and funding should be given accordingly, ideally after reviews or some types of assessment.
\item This seems like a great educational opportunity for CS students. Many CS students are required to work for a company for a semester before graduation -- I wonder if there could be software institutes that train undergrads in software engineering using these codes.
\item I think we need a solution for the ``customer service'' aspect of the software development. It is not reasonable to expect research labs to do this. Perhaps people running NSF funded national computing centers (like TACC or SDSU) can take on this job.
\item There needs to be a sustainable model for updates, whether it's transition to industry, a non-profit, or an open source consortium. Any of these models work, but knowing how to get it done can be difficult.
\item I think adapting to new infrastructure should be done as a collaboration between the vendor and the developers, in order to update things most efficiently. This can be done through workshops and hackathons.
\item The developers shall maintain in the long run together with the community. It will be nice to have all the updates maintained by the developers, however it should not be explicitly or strictly required.
\item *Funded* developers. See third point above. [One might have ``funded'' development accept contributions and vet/improve them, as happens with many open source projects.
\item Developers will likely do it more economically. While to cost to NSF may be similar, expenses across directorates should be balanced.
\item Again, it depends on the situation. There could be a core development team, or the community could contribute resources to do this.
\item Community - but this is probably unrealistic. We need better metrics and incentives for community-sustained software projects.
\item Communities are the champions for the resources they need, engaging a community and involving them in the development is key.
\item Promote containerized code, with descriptive CWL, with a eye towards optimization and modularization where appropriate.
\item Foundational ideas like FLAME should be extended to more domains to allow for algorithmic and hardware flexibility.
\item Software should be available and maintained to some level as long as it is used where there is no alternative.
\item Probably a mix of both. Developer if possible, but trying to build a community to help is essential.
\item I think it can be led by the developers but definitely the community should chime in.
\item It should be a combination of both. The SW should have stewards on both sides.
\item good questions. but whoever the community/company who is using it should do it
\item The developer should update the software based on the community input.
\item Again, this is a question best answered on a case-by-case basis.
\item If you open source, you can get community contributing.
\item That needs to be part of the sustainability plan.
\item Both, depending on capability and need.
\item Ideally, it should be the community.
\item responsibility lies on both
\item Community and industry
\item Maybe both?
\item Either.
\item Both
\end{itemize}

\subsubsection*{A.5.7 Do you use your institution’s infrastructure or the national cyberinfrastructure for the long-term retention of your software and related artifacts?}
\begin{itemize}
\item Yes, I use national infrastructure (\emph{e.g.,} nanoHUB) for retention, and industry-standard places (GitHub, sourceforge) for long-term software archival. I use institutional archives and pubMED for long-term archival of research papers, as well as bioRxiv (public+philanthropic subsidized) I fear the long-term, and truly believe it is in the national interest to long-term archive software created by public money. Should there be a national GitHub mirror?
\item There is no national CI for this, that I know of. For the project I recently completed, we used foundation provided infrastructure for everything about the software.
\item Both. The software and artifacts are available on a project web site within the University's domain, but the software is also available on GitHub.
\item I use NSF XSEDE resources and infrastructure, we also have a web-server, and small compute server for small jobs that take too long to stage.
\item We use GitHub and/or bitbucket for most of our software. If a similarly useful tool were available through NSF, we would consider using it.
\item For long term support we rely on mature open-source managed in a community, \emph{e.g.,} Apache Software Foundation, and on industry.
\item I have used my institution's infrastructure. I don't know if XSEDE has any option for long-term retention.
\item At this point, my institution's infrastructure, or ``free'' infrastructure like GitHub.
\item I would rely on national cyberinfrastructure for the long term retention.
\item We use local infrastructure, but we are not your typical ``science'' shop.
\item I use my institution's cyberinfrastructure for both of these purposes.
\item Yes, we store all of our software on our own and national resources.
\item I don't understand this question. We use GitHub to store software.
\item We have not used this as much as we would like.
\item I'm at ORNL, so I use ORNL infrastructure.
\item I use GitHub, paid for by my university.
\item We use our institutions infrastructure.
\item My own/institutional infrastructure.
\item using institution's infrastructure.
\item We have everything in-house.
\item Institution. This is policy.
\item Yes - national CI for data.
\item Do Box and GitHub count?
\item When appropriate, yes.
\item institution resource
\item no. I use GitHub.
\item We use GitHub
\item National CI
\item Sometimes
\item national
\item GitHub
\item Both
\item yes
\end{itemize}

\subsubsection*{A.5.8 How do you measure the quality of the software/data product that you produce for public release?}
\begin{itemize}
\item To date, the main focus has been on its validation by reproducing known test cases. The metric here is the number of model problems that can be handled well by the program. This metric also includes the number of methods that can handle particular model problems. The secondary metrics are the amount of educational materials (quantified as the \% coverage of the use cases possible by the software) and the amount of documentation provided (both as the comments in the code and via stand-alone website content, although the two are related via the auto-doc generation). Finally, an important metric will be the number of unit tests that work as supposed, although this has been the area we paid a bit less attention due to the limits of time/human resources available and due to the funding climate being somewhat non-rewarding these efforts.
\item The documentation should be clearly readable, code needs to be annotated with comments, people can follow them to setup, install, and use, they should be able to explore any and all options that software promises to deliver ---- when all these can be ensured, I deem this software as a quality product that can be released to the public. Any software that is made public under MIT or such license, does not need a whole lot to be out there. Also, I am all for the community to pitch in and make pull requests to augment a software and extend it to make it their own. I think the most important part is for the software to work and the documentations to allow for easy extendability.
\item Metrics! First up standards/specification need to be set that will determine the quality of software/data product. Aggregate experts on the topic (industry, labs, academia) and come up with a specification/standards document on the metrics to be achieved and reasonings on what defines those metrics and document them. Look at the state of the art on the particular software/data product that others have released. Set expectations. Metrics, expectations, standard, specifications and thresholds would help to measure the quality of the software/data product. Other indicators include reliability, testability, security, maintainability, efficiency and usability.
\item Primarily through testing, especially of the new features. As well as regression testing based on prior results. We maintain a set of tests that can be run against the code along with being used as examples for new users. In the case of features that have been added based on community feedback, by requesting feedback on a pre-release version.
\item High quality is the result of mature software development processes which is the result of people caring about quality. So checking whether a project has a test suite, uses GitHub, does peer review for each patch, and has a continuous integration suite is a very good indicator for a project's quality.
\item Have I introduced any new bugs that affect accuracy or efficiency? Have I broken existing models? Does it perform tests as expected? Can it make reasonable scientific predictions? Can end users create models as easily or more easily in this release than in the last?
\item Testing, testing, testing, against a well-defined release cycle. Since the software is interactive, this involves putting release candidates through usage testing that typifies paths a developer would take using the software.
\item 1. Number of downloads/releases 2. Number of organizations using the software stack 3. Number of external contributors 4. Number of publications arising out of the project to measure the innovation component.
\item Does it have a testsuite? Does it pass? Is there complete documentation? Can users actually use it (more difficult to measure)? Are there field-specific performance criteria that can be measured?
\item Two methods: 1) Certification tests against adopted open international standards, 2) Software sprints that involve developers and users.
\item Software performs its expected functions, is easy to use, and fills an important societal need that other software does not.
\item Partners in industry have told us that the software we produce is of higher quality than they can produce in house.
\item We run internal benchmarking. Groups related to RStudio are also building methods to quantify package quality.
\item ease of use and documentation, optimized and fast computation, can be combined into user-defined workflows
\item New features, range of applicability for a variety of users, reliability, speed, ease of use and support.
\item Not sure. There are several metrics for several aspects (from point of view of user, developer, ...).
\item Hard question: We do a lot of internal testing and acceptability tests and release things in stages.
\item We don't, measuring is a huge undertaking and measuring quality is not currently a metric needed.
\item I run a commercial software tools company. We build (we think) thorough tests of of products.
\item We don't directly develop software but use other CI software for specific domain sciences.
\item Rigorous testing. Continuous monitoring of usage and errors. External security reviews.
\item Easy to compile, run, good readme files and moderately easy to read the code to change
\item Publications arising from the software. It takes years before these arrive, though.
\item One way we measure the quality of software is by performing rigorous testing.
\item Benchmarking of scalability, number of individual downloads, citations
\item documentation, examples, tutorials, user comments and feedbacks
\item By testing and comparing to existing published software/data.
\item Code reviews by my group and the software engineering group.
\item Industry and government best practices and recommendations.
\item User interfaces, API for extensions, Simplicity in usage
\item CI/CD and lots of unit test plus reviews from peers.
\item End user growth, scientific publications that result
\item don't have really good metrics. Worth a discussion.
\item I basically archive them or upload them to GitHub.
\item Haven't done much recently, but in the past, yes.
\item By how much it is used by the public.
\item The number of issues users create :)
\item Number of citations/downloads/users.
\item user feedback and analytics
\item By the extent of its use
\item Number of citations.
\end{itemize}

\subsubsection*{A.5.9 What are the software engineering processes or data management process that you (and your group) follow currently?}
\begin{itemize}
\item The modular approach, interoperability, and hybrid programming. Through modular approach, we try to improve the reusability and maintainability of various computational methods and facilitate their interfacing with each other in a simple, intuitive way. Such interoperability can help boost the advantages of various approaches when combined together. In addition, it makes the systematic benchmarking of various approaches possible. Our programs are being periodically refactored as we find more systematic and general ways/strategies to put the broad range of methods on a similar footing. The refactoring ensures the backward-compatibility until the older use scenarios become obsolete, at which time the old functionality is dropped. We also utilize the methodology prototyping approach which is facilitated by the hybrid C++/Python programming. In this approach, new methods are being implemented and tested using Python (with the help of the C++ components of our code already exposed to Python). Such developments are significantly faster and easier than they could be in plain C++. Once the functionality is established, the Python functions and data types are translated into C++ and re-exposed to Python. At this point, the newly added methodologies become interoperable both with other C++ methods and with the Python codebase. The original Python code is then moved into the ``obsolete'' directory, which can be still useful for other purposes – as the ready code blocks for other developments or as the educational examples. We preserve and share the most critical data used or resulted from our research. In fact, for every publication, I request my students to create a GitHub repository and deposit such critical data. I request the data to be narrated enough for an external person to repeat our ``computational experiments''. In addition, for some of the most important use cases, our code would automatically save all the input configuration used to run the calculations. This is important when one utilizes on-the-fly computations (\emph{e.g.,} via Jupyter notebook), when one can easily change the input parameters directly in the script. In the common ``black-box'' approaches the input configuration is always predefined and is manually created by the user. That’s why in this type of calculations there is no much need for such a bookkeeping of the input configurations. In the ``on-the-fly'' approaches with larger flexibility, such automated bookkeeping of the input configuration is important.
\item We have a Strong Procedure for Design, Development and Release: Research is done for exploring new designs; Designs are first presented to conference/journal publications; Best performing designs are incorporated into the codebase; Rigorous Q\&A procedure before making a release; Exhaustive unit testing - Various test procedures on diverse range of platforms and interconnects Test 19 different benchmarks and applications including, but not limited to - OMB, IMB, MPICH Test Suite, Intel Test Suite, NAS, ScaLAPACK, and SPEC - Spend about 18,000 core hours per commit - Performance regression and tuning - Applications-based evaluation - Evaluation on large-scale systems (Lassen, Frontera, Summit etc); All versions (alpha, beta, RC1 and RC2) go through the above testing.
\item (1) Git repository and version control (2) Put emphasis on modular designs, develop a component architecture, and ensure simple interfaces between the modules and components (3) Develop iteratively (4) Make sure to develop unit testing for each module and component (5) Comment code, and document a component before moving onto developing another component. For data management, my team has completely moved to HDF5 data format, all data that we collect are through the HDF5 interface or we use our own converter to convert data first to HDF5 format. HDF5 provides a scalable method for extreme-scale I/O and provides the only way we can survive the data deluge -- portability and reusability. We use GitHub for hosting our data, but also looking into cloud storage for long term storage.
\item Software engineering processes: Use of version-control software and a git repository shared by all developers. Pull-Requests at GitHub automatically spawn tests of a test suite in Amazon Web Services with a Plethora of Different Compiler and Runtime Options. Automated builbot prevents commits from breaking code. Established coding conventions help bring uniformity to code. Dedicated slack workspace improves communication between developers. New applications are validated against analytical results and compared to other independent implementations of the same method. Data management processes: input and output files uploaded to public repository
\item My students build machine learning algorithms and we publish them as Brown Dog/Clowder extractors, as well as making them available on GitHub. We share and archive our datasets on Clowder. We partner with NCSA on projects where we are one use case for cyberinfrastructure development. My students work closely with their research programmers to identify requirements that they can use to develop/improve the software. We use Slack for communication among the team, a wiki for notes and documents, and a software repository for sharing code.
\item Dev and master branches, numbered releases, representative qualitative / behavioral tests on multiple platforms (it is stochastic agent-based software that wasn't originally engineered for bitwise reproducibility when run multithreaded), update user documentation, write change log and release notes.
\item We have tailored an Evolutionary Software Development process to meet our needs. The processes combines use case development, agile sprints in the context of an abstract/conceptual architecture, and informed by a technology forecasting activity to foster innovation.
\item For the recently completed project, we used code reviews facilitated by gerrit, and we practiced source management through git. Additionally, all features or issues that needed to be addressed were managed through bugzilla.
\item Our software engineering process aligns with the crystal agile development model. We focus on iterative and incremental development, active user involvement with project stakeholders and on-time deliveries
\item We work with an industrial partner (data viz studio); our biology lab are end users and guide the devleopment process and also provide feedback on the software features and UI/UX
\item Have followed fairly standard and accepted processes. Spiral processes are followed along with processes for documentation etc that typical industries follow.
\item Doesn't directly apply to us. We are starting to get into data management and will need to gain more experience before I can answer confidently.
\item Peer review of each patch, use of CI suites, a very large test suite, periodic automated regression tests on a large number of platforms.
\item For a recent project, we used prototyping to check the requirements and feasibility of some design decisions.
\item Combination of agile and kanban with a focus on data driven design and test driven development.
\item Something like agile with continuous sprints, supporting a longer waterfall model.
\item agile development, focus on prototypes and ``early failures'' for long term success
\item We use a modified scrum cycle. All of our work is tracked on GitHub using zenhub.
\item We currently do not produce software but do have multiple ways of storing data.
\item We use revision control software. We are implementing automated unit testing.
\item Ad hoc with some mixture of industry-grade elements (code styles, tools, \emph{etc.})
\item theory development to code development to test to application to maintenance 
\item Unit tests, regression tests, valgrind, testing with multiple compilers.
\item We use version control and document the code and follow C++ guidelines.
\item CI testing. Code versioning. Modular and/or layered software design.
\item GitHub or bitbucket are used a common space to share group code.
\item Try to apply best practices from industry and literature.
\item I am not the right person in our group to answer this.
\item Nothing unique. We use generally accepted practices.
\item follow the requirement from the funding agencies
\item we use GitHub for release and version control.
\item We follow a small set of Agile practices.
\item Using GitHub for version control.
\item It is all ad hoc at this point.
\item Version control, documentation
\item code review and unit testing
\item We have in-house standards.
\item We strive for AGILE.
\item We use GitHub
\end{itemize}

\subsubsection*{A.5.10 How often do you favor using existing software components (either by buying or using for free) and, if needed, customizing it for your needs instead of writing from scratch?}
\begin{itemize}
\item Whenever the functionality and methodology is well-established and is sufficient for our research needed, I would use the available solutions. There is no need to invest time and efforts into something other people have already done and have done it well. I would always give the first preference to the open-source software. Another aspect on existing software components – sometimes they are in a rather poor condition (not documented, not following best software development practices). In this cases, it would again be reasonable to write such things from nearly scratch (refactor, translate, redesign), but perhaps follow closely the ideas expressed in those prototype codes blocks.
\item This depends on the topic. If it is a well-explored area, for example DNA sequence alignment, there are many tools out there. Trying to use existing software components would be my first to-do. And then the goal would be to (a) determine if the tool meets expectations of my ideas? (b) how usable (buggy?) the tool is (c) has the tool been maintained and being maintained, can I create PRs? (d) Readability (e) Portability (f) Are others using the tool? and other questions. If the answer is yes to most to all of the questions, I would use someone else's tool else I would build from scratch.
\item Default is to compare innovative directions against the existing baseline. If a parallel development is taken to an existing capability, then help for the community is developed: guides on how to choose between the options, software tools (bridges, facades) to enable transition.
\item Being a commercial vendor and distributor, we have to evaluate this differently from an academic group. We own all of the rights to our software. Having said that, we do not extend this to tools used internally. We concentrate on our area of expertise.
\item It depends on the requirement. \emph{e.g.,} We use a software called HWLoc from Inria, France to detect hardware topology of a system instead of writing our own. However, we customize some portions of it to match our requirements.
\item I prefer using existing software components when available. However, customizing others' code can be difficult unless its been designed in a modular form that enables us to insert a module for the needed customizations.
\item As much as possible, especially components with licensing terms that are compatible with the licensing of the overall package. Not so much by buying, but rather using components with an appropriate open-source license.
\item If it does the job effectively, you should use what exists. a) you can't replace it for any reasonable cost, b) you can't discover all the gotchas needed to make it work reasonably on your specific project budget.
\item Always. I am a scientist by training, and we like to spend our cycles in more creative and forward looking problems than reinventing the wheel. This should be the only way rationally we should approach projects.
\item We always try to favor open source projects. The NSF should strongly encourage, if not force, researchers to look for open source alternatives and contribute to open source projects.
\item It depends on the situation. BTW, there are publicly available ``build vs. buy'' decision making frameworks that can be used to figure this out for each particular situation.
\item Existing software components are used if they are open-source and license compatible. Maybe 25-50\% for more generic library/infrastructure (\emph{e.g.,} Google Test) components.
\item Rarely. The software complexity and dependency costs (particularly for cross-platform support) often exceed the benefits of gaining one or two minor features.
\item Whenever possible. It's had to put a number on it, but my research program depends heavily on both free and commercial software developed outside our group.
\item If the software is available and meets ``production-level'' robustness we always use existing code. We only write what is needed to amplify existing efforts.
\item Always review the ecosystem and see where SW can be extended rather than building from scratch. RAPIDS tries to build as little as possible form scratch.
\item We use existing free software whenever possible. In particular, we have adopted standard file formats instead of creating our own.
\item About half of my research focuses on developing theory and codes, and the other half focuses on using other codes/software.
\item I strongly prefer customizing existing open-source packages with my feature, and never write things up from scratch.
\item Whenever they are available, my team is always looking for best practices and validated tools available.
\item Almost always for production-level code. Analysis/viz code can be better from scratch, but it depends.
\item Almost all the time. Nowadays, it almost makes no sense to start from scratch for any projects.
\item Being able to call existing libraries is almost always better than building from scratch.
\item We write everything from scratch, building on a decade of our own development.
\item Yes--but there are not many to suit our needs and it is difficult to integrate
\item Whenever we can use existing, we do. For example, commercial cloud services.
\item If we have to use a third party tool, we use only open source/free versions.
\item I am in favor of this, if the code is a ``standard'' and easy-to-use code
\item The first step is to always do an assessment of what already exists.
\item Always, as long as they do the job and the solution is sustainable.
\item I favor modifying/adapting/expanding existing software products
\item In my field, there is not much software for purchase.
\item every time. We don't write software from scratch.
\item more using existing, unless i need to develop new
\item We try to do it in almost all the cases.
\item this should be done as much as possible.
\item Always when we are aware of it.
\item Whenever it makes sense.
\item Most of the time
\item Almost always.
\item Quite often.
\item Quite often
\item Always.
\end{itemize}

\subsubsection*{A.5.11 What are the challenges, if any, that you typically face in reusing existing software or data products?}
\begin{itemize}
\item One issue is evolving capabilities or interface evolution, both of those can prove to be difficult to manage. Even worse is if the software stops to be supported, that can prove to be a big issue. Finally, for integration of large components, IP provenance is a big deal - trying to ascertain the licenses and IP-cleanliness of the pieces can become a showstopper.
\item Poor documentation or no documentation at all, or that the required environment is too outdated, so the software fails. We were successful in some cases to leverage the inner-workings of the software to develop our own, but that is a waste of cycle. Oh well.
\item Open source projects are not really ready to use out of the box, so small modifications are often needed to match a specific part of research and the NSF should fund that as a part of a research project.
\item My edits may depend on a specific version of the package. Then when the package is updated to new versions, I might have to redo my edits for my feature to work in the new version. That takes time.
\item Not having the relevant features and the code not being commented properly so that it can be easily modified. I've found this to be the case with free and for-purchase software.
\item Often you only have access to the executable of the program. Therefore without seeing the source code it is difficult to establish the reason for eventual errors in the code.
\item Lots of last mile bugs, where you're almost done but can't put a definitive timeline on hitting the quality you want. That said, it helps the next person along.
\item We are worried about the software developers being funded to keep the software active and supported if the software is a NSF or other govt org funded software.
\item On some occasions, this software is not sufficiently well-documented, so getting a correct implementation can be a challenge on some obscure or subtle points.
\item An existing product tends to continue being valuable until the user requirements change or the underlying software technology moves to a new version.
\item Sustainability (with free/open-source but also with commercial software), software quality is often not up to par. the code is not modifiable, \emph{etc.}
\item Difficulty getting them to build and run; they are often out-of-date because they were a research artifact that was not maintained and updated.
\item (1) Figuring out to make the software work. (2) Getting bugs fixed (3) being able to rely on the software in the long term, where appropriate.
\item It is typically easier than writing from scratch. The lack of general APIs for interfacing to electronic structure codes is a bottleneck.
\item the user documentation is poor. the sample scripts are buggy. and people who published using the common software do not share input files.
\item Cross-platform support (\emph{e.g.,} works on Linux only). Ease of installation and maintenance for non-CS end users. License compatibility.
\item Poor documentation of interfaces. Poor configuration management (Makefile fragments, autoconf, homegrown configuration scripts).
\item I've found that software is poorly maintained, unit testing is lacking, and understanding the code takes longer than rewriting.
\item a) Doesn't work on my platform. b) doesn't integrate with the partial problem solution I already have.
\item learning cost. Complexity. Unwanted and unneeded features. Mismatch of functionality and exact need.
\item The context of this is important. There are questions of ownership, portability, reliability, \emph{etc.}
\item Documentation. Functionality. Ease of extension. How best to contribute to the existing product.
\item Difficult to install, rely on old/outdated libraries, sparse documentation on how to use
\item Difficulty in integrating with our code. Not knowing the error bounds of the codes
\item Interface mismatch. Performance/stability problems. Build system integration.
\item Investing time for training and effort to go through documentation
\item Understanding the code, debugging issues in the software product.
\item To make them fit into the specific tasks (\emph{i.e.,} data formats)
\item different format of input/output, parallel vs serial, \emph{etc.}
\item Lack of documentation, support and outdated features.
\item porting to new platforms for testing and comparison.
\item some time do not completely meet my specific needs
\item lack of documentation and inability to scale.
\item Bugs, unclear documentation, not maintained.
\item Becoming familiar with someone else's code.
\item Deprecation \& poor documentation.
\item Understanding how to use it
\item documentation and training
\item Adding new features
\item Poor documentation
\end{itemize}

\subsubsection*{A.5.12 How can others get/find your software components to compose their software framework/system?}
\begin{itemize}
\item As a commercial company, we have a standard offering of software components. We are findable on the web. We offer the components commercially with a special option for research. We offer training in using our software, and we offer custom services. I don't see how research groups that are focused on doing science can do this very well; its not in their mainstream goals.
\item Our software is available for free for download and has an extensive userguide (over 400 pages). We also have a vibrant online discussion group where users can post questions to the developers and the community.
\item Through GitHub, dedicated software websites, through media (periodic Twitter and Facebook updates and highlights), through mentions in the scientific publications and trough presentations and seminars.
\item These are all available through the Eclipse foundation - the process to check out the components is well-documented (though not trivial, neither is the learning curve for developing with them).
\item Google, literature review. Freely available on GitHub and sourceforge. Alas, nothing equivalent to R and Python package managers in the C/C++ world. (But GitHub is close.)
\item 1. Conform to standards 2. Publicize the code 3. Provide support 4. Provide initial support to get people over the hump with regard to adopting the software.
\item Mostly through repositories and conferences. We have found that the most efficient method is word of mouth within the community it serves.
\item I would say we have not figured the best way out so far. We dot he standard things and they have only worked some.
\item We are usually aware of NSF (or other) projects that funds software component that we need.
\item container repositories, GitHub, and from recent publications via google scholar alerts
\item Mostly GitHub or thru my research group website or in footnotes of my publications
\item We maintain a registry of software that implements the OGC standards.
\item cloud sharing repositories, journal online supplementary material
\item There are both in-house and outside collaborations.
\item GitHub or contact me for file transfers via ftp.
\item Open source project available from GitHub
\item GitHub, description in scientific papers
\item Word of mouth; internet search engines
\item We put it out on GitHub or bitbucket.
\item GitHub - https://GitHub.com/rapidsai
\item through online access, or email
\item Most of our codes are in GitHub
\item They can see the group webpage
\item We keep repository on GitHub.
\item ORNL has a software portal.
\item Google. Stack overflow.
\item Buy or download them.
\item Public repositories.
\item from our web page.
\item Not sure.
\item GitHub.
\end{itemize}

\subsubsection*{A.5.13 If software engineers are available for contract jobs at industry rate, is it more cost-effective to hire their services for short-term to complete the project on time than have full-time staff hired for the job?}
\begin{itemize}
\item Depends on the task. In the software industry the affordable people will usually write code for a spec. Producing such a spec can be as much (if not more) job than doing this using students and/or yourself. For a well-defined task it should be possible to hire someone. For tasks that are more open-ended it will be difficult to expect a favorable outcome from a short-term involvement of an outsider, unless they are truly an expert.
\item Software engineers for contract jobs might be better if there is someone to maintain a critical software. Students come and go, they are never going to be the source that provides sustainability. So, if it is a project that needs sustainability in the long-term, then hiring a full-time staff for the job is a better option.
\item Industry rates for software engineers is likely to be significantly higher than the compensation that full-time often receive. There is also the issue of dealing with any future bugs or software enhancements. It also reduces opportunities/budget for hiring post-docs and Ph.D. students.
\item Of course, we use this commercially at times, but there are concerns with long term support and familiarity. Most software is not simply modular and can not be dropped in and maintained without in-depth knowledge. Proper documentation can help.
\item Contracting with industry really is not something that is supported at my institution. Rather we try to have a group of developers that can work on a multiplicity of projects, and we try to juggle this talent amongst projects.
\item In my experience the kinds of software engineering challenges encountered in scientific software are generally not suitable for short-term contract work. For smaller contributions that are this might be an effective strategy.
\item Even if industry resources are hired to complete the project, someone has to maintain it long term. That has to happen with internal staff or by the community.
\item We maintain a staff and augment it with short-term to meet temporary needs triggered by more work than the staff can handle or to bring in new expertise.
\item I believe short-term would be more cost effective as you can get the exact talent needed for as long as needed, but I don't have empirical evidence.
\item Possibly; especially if you can hire in cycles of short-term contracts that allow for end user testing in between periods of intensive development.
\item In the scientific computing community, I didn't think you can hire ``contract programmers'' (by definition for limited periods) and be effective.
\item It could be. Some universities HPC centers are beginning to essentially sell a fraction of a developer's time to support software development.
\item Yes, but this takes a very special software engineer. I need the code to be written ``my way'' so that I can continue to grow it in my lab.
\item Yes, sometimes -- we sometimes do this with the Research Software Engineering group at ORNL rather than hiring our own programmers.
\item Yes if it is in the initial phase of the project. For maintenance and continuing support full or part time staff is required
\item Continuity and establishing an institutional memory is important. You don't get that from hiring contract workers.
\item Yes. Although if a project is doing to continue for a very long time, it might be good to have project's own staff
\item No. There is a steep learning to understand complex scientific software. Short-term staff are rarely useful.
\item Probably. It would depend on the length of the project and more importantly funding availability.
\item Yes, unless the project is long term and will likely continue with other projects in the future.
\item Rarely. We don't need guns for hire. We need long-term support for our work and our end users.
\item Most time, it depends on the availability of software engineers, but we prefer short-term.
\item Unlikely, but depends on the problem and the frequency of needing this service.
\item full time staff is better. the high turn over rates impedes research process.
\item It depends on the specific projects, some may benefit a lot, some may not.
\item Yes, retaining full time staff is not viable for software development
\item depends. for short term the former, for long term the latter
\item Yes, but the quality of final product might not be optimal.
\item Rarely. What is ``on time'' in academic software projects?
\item Varies depending on need, cost, and time frame.
\item At current funding levels, this will not work
\item I don't have experience related to this topic
\item yes, if they have the right expertise
\item It depends on the project.
\item Sometimes, yes.
\item That could work
\item Probably not
\item No comments.
\item Yes
\end{itemize}

\subsubsection*{A.5.14 Does new hiring introduce delays on the project, and do you plan for that in your project timeline?}
\begin{itemize}
\item Yes--we get a grant, then have to advertise for a developer or postdoc. We can't plan for this because the funding cycles and both funding and university bureaucracy make it impossible. Unless I have startup funds, I can't advertise for a position until I have a funded grant to pay for it. Funding agencies make this worse when they say ``congratulations, your funding is retroactive starting 2 weeks ago, and your first milestones are due soon.'' This inevitably contributes to projects that do not keep pace with milestones when they start faster than students and staff can be hired.
\item New hiring can introduce a significant delay on a project. We try to manage this by a) maintaining a group of developers that are working on multiple projects; we actively manage their project workload so that we can make milestones amongst the multiple projects, we also have developed a practice where we have a standing search (for about a year at a time) that we keep open, so that we can reduce the lead time at the beginning of a search if we need to hire more staff to handle additional project load.
\item Yes, new hiring does cost cycles! One way to mitigate that is to be actively involved in the first few months of the project in training that staff. I always plan for delays in my project timeline. I budget my own time into the project. I am always confident about my ability to accomplish what I propose.
\item Yes and yes; I have trouble finding skilled people to work on software development projects as a new PI in a biology department. My professional network does not include the type of people I need to work on software development, hence why we contract out to industry
\item Yes it does coz every project has a steep timeline. One cannot avoid taking this into consideration so yes, I plan for that in my project timeline. It is often a hard problem as when funding begins, deliverables begin too. So this is the hard part.
\item We have an established team, so this does not apply. We are very fortunate to have a very dedicated core team that intends to stay with our group as long as funding continues to flow. If funding dries up, the project will be terminated.
\item Yes it does. In soft money organization staff are fully funded (otherwise they are not in the organization) and so if new funding comes in and at the same time another funding doesn't end, it is hard to match staff to funded projects.
\item Yes. Hiring is a problem. I'm not sure if there is a good solution other than constantly training competent people within your group to ensure that the pipeline is full.
\item New hires need training in technology on which they are going to work. That adds significant delay. We try to hire people long term to amortize the training costs.
\item Yes, there are always delays that should be taken into account although sometimes reality surpasses all the possible planning one can anticipate.
\item Yes and we do our best to plan for this in the project timeline, although this more often gets factored into a no-cost extension.
\item It can, but I think this is not different from fundamental science. I don't see this as a major obstacle.
\item There is very little we are able to plan in an academic setting because there is so much uncertainty.
\item Yes new hiring introduces delays as a new hire requires training before they can become productive.
\item We tend to bring in part time support to existing staff rather than delay start.
\item Yes, and the delay is usually unpredictable, so it is difficult to plan.
\item Typically, we find new capable people to continue writing the software.
\item Yes. We usually get extensions to the project to deal with this.
\item Yes, it introduces delays because it is hard to find people.
\item yes, there could be some delays in the hiring process.
\item Yes, it does introduce delays. Yes, we plan for these.
\item I desperately try to maintain talent to avoid this.
\item Yes, this often is a longer lead than expected.
\item Hiring is fundamental and must be factored in.
\item Yes, there is always an administrative delay.
\item sometimes is does cause delays.
\item sometimes yes
\item No comments.
\item Yes and yes
\item definitely.
\end{itemize}

\subsubsection*{A.5.15 When the senior personnel on the project who are in charge of the development work get promoted or take a lucrative job elsewhere, or when the students graduate, how do you address the loss in progress on the project?}
\begin{itemize}
\item I try to build a pipeline long before a student graduates. It requires meticulous planning. For \emph{e.g.,} I always pair up a Masters student with 2 undergraduate students in their junior years. An MS student graduates after 1 year of research, and by that time the 2 undergrads are trained. Then either they continue or we get another MS student or 2 more undergrads to train for the project. This is a great way to build a pipeline and goes a long way in sustaining a project. So far, I have never had any project that fell through, while providing research exposure to the undergraduate students.
\item I try to find another student to pick up from where it was left. It is not a complete match always but works most of the time. Having said there are times when there is a long pause between a student leaving and a new student jumping into the game. The development of the software takes a hit and a back seat. Sometimes I have trained an undergrad student along with a grad student, so if the grad student is leaving then the undergrad student picks up the steam until a newer grad student joins - vice-versa, so this way the gap is somewhat addressed.
\item Hire as quickly as possible, transition other students into leadership. We work to make sure software is as well documented as possible to mitigate these risks. Usually documentation takes far more time than coding, but we are aware of this. We also are developing ``computational apprenticeship'' STEM education practices to make sure more senior students are training more junior students and passing on their knowledge.
\item Two ways - one is we try to have more than one person on a particular project, with exposure to a broad range of what is going on in the project, so that the risk associated with the loss is mitigated. Secondly, we are always mentoring new developers, to have them come up to speed with the specific skills needed for the projects.
\item (Commercial company... we don't have students). Loss of serious talent is always a setback. It is hard to control. We work to ensure we have the software and appropriate design documentation; we try to have some overlap in work so that somebody has a chance of picking up the pieces with some background in place.
\item Each area of the project has two people working on it. Thus, when one person leaves, the other person takes off. We also have an extensive knowledge transfer session where we the employee leaving the organization does a series of presentations to hand over items they're working on.
\item Students should not be writing production software. A good organization should have redundancy in personnel, and reward staff in ways that discourage them from leaving. These issues relate to the overarching need for approaches designed to provide for sustainability and quality.
\item By having multiple people on the development and bringing on new students before the current students graduate. Having the PIs and CoPIs working with the code makes it easier to ensure there is adequate documentation and less progress is lost when group members depart.
\item There must be continuity. That requires that more than one person be versed in the details of a project. Frequent meetings and documentation can aid in this process. There must be stable management of the project whether that is long term staff or the PI.
\item We have an established team, so this does not apply. We are very fortunate to have a very dedicated core team that intends to stay with our group as long as funding continues to flow. If funding dries up, the project will be terminated.
\item Understand the potential loss from the very beginning and have a well-designed process in place, such as documentation. Do not let the senior person or student dictate the progress.
\item We hope to bring in new people whom the promoted staff can oversee and guide. I can see that for some other projects student/postdoc leaving can cause problem.
\item We try to make sure the software is maintainable to begin with. After that we support the software in ad hoc ways until we can get to a more stable situation.
\item Senior personnel leaving should hopefully be anticipated and planned for. With students, knowledge transfer to more junior students is generally relied on.
\item I'm from industry, but we typically have people shadow to avoid key person dependencies. In addition, we release very often to have regular checkpoints.
\item can be a substantial problem due to small group size. I try to get all group members involved in infrastructure activities to mitigate the effect.
\item Good project leadership builds personnel with redundant skills. That includes leadership: Good leaders plan to become replaceable themselves.
\item Sometimes this ends up making the software languish and/or become obsolete for years, unfortunately. This is a hard (but common) problem.
\item By hiring staff (part time temporarily) who has similar skills and who might be looking for funding. Promoting junior members.
\item Interesting this would not happen if you contract experts and once the project is done you dont have to deal with them.
\item Do a technology transfer, so that the student shows how things work to the next designated student in the group.
\item Try to move the project to someone in-house who is already familiar with it and can get up and running quickly.
\item Make sure there is sufficient time overlap between the person leaving and the next person who will do this job.
\item I try my best to start training the new person before the old person leaves. This is a hit and miss strategy.
\item I try to have a pipeline of students, but sometimes, I spend my weekends coding. That is not a solution.
\item Both from promotion from within and from new highers; often new highers come from our temporary staff.
\item Try to transfer knowledge at least a few months before they leave, but some delay has to be expected
\item Everyone loves their job. No one ever leaves. (We also hire replacements as quick as possible.)
\item Tough question. No easy answer but a real concern. We have been lucky and stable thus far.
\item I find another student to have some transfer of knowledge from the departing person
\item We avoid loss by keeping all the group code and software on a shared repository.
\item It often hurts the project and the position is often filled with a new employee.
\item Documenting and training a new person before the other person leaves.
\item I try to hire new students who can be trained by those graduating.
\item oh yeah. all the time. it's terrible when it happens.
\item We don't really have this problem at a national lab.
\item I have to be up-to-date myself on the code.
\item We do cross training to avoid braindrain
\item need to find replacement before that
\item Train a new grad student/postdoc.
\item Nice documentations and backups
\item need overlap and handover
\item No comments
\end{itemize}

\subsubsection*{A.5.16 What are the tradeoffs in having a full-time, senior-level staff, learn the basics of software engineering and deliver on the project, versus hiring trained or trainable graduate/undergraduate students?}
\begin{itemize}
\item Full time developers, while expensive, are very valuable as they can easily lead development efforts, and do not require lots of close management to make sure that goals are achieved. Graduate and undergraduate students: mileage varies considerably, it is possible to have a simply amazing student working on a project, and for every one of those, you may work with 2-3 less than amazing students. Also, graduate students are not ``cheap'', they really are an investment, and one needs to factor in the mentoring needed, to achieve the success you desire. So, this is where some additional risk can be introduced into your project - I think it is important to think of what the goal is for the work (a production software product, or training the next generation) when trying to set the balance on the project.
\item The advantage of a full-time staff is long-term sustainability. Any project such as establishing a cyber-infrastructure would require at least one such full-time staff. Hiring students, on the other hand, has obvious benefit of training the workforce, some might continue as the full-time staff. Since students are cheaper, we can get more students to engage in the project, and in the process get diverse and creative minds to enhance the quality of the overall deliverable. Providing long and sustained service though is a challenge with students.
\item We all have to keep learning and adapting :-) The constant in our field is - ability to pick up new stuff. Full time, senior-level staff were once upon a time undergraduate/graduate students and to that end I strongly believe in training the next-generation workforce. Yes, if there is a major deliverable else the project would fall apart, I will need to hire a senior staff - but that also means I need more funding to hire talented people so it is not an easy problem to solve.
\item The graduate student also understands the physics and so will be able to make more informed choices about how to develop the software based on context. Additionally, the graduate student can take that knowledge and continue to apply it for the benefit of their scientific community while a software engineer without the background may not see connections between different physical problems that require similar software tools.
\item Software engineer costs a lot of money and I have to constantly think about grants about supporting their position into the future. But they bring continuity and sustained quality to the project. I'll take one good senior-level staff over 4 graduate students any day of the week! Undergrad and grad students do play a vital role, though. Often, they bring immense creativity and nucleate new projects / ideas within the lab.
\item It requires a tremendous amount of work to mentor a student who knows basic programming to use software engineering best practices. Additionally, senior level staff cost too much, so we end up with students who don't know software best practices writing all the software because the actual cost to write good and maintainable software is too high and makes proposals noncompetitive.
\item Staff can often make more progress quickly since they are full time. However they don't always have a research mentality for innovation and publication, which doesn't advance the field the way a student would. Plus there is the advantage of educating the student, but they tend to be slower and sometimes their code is not well written.
\item For reasons of continuity and/or competent management it is necessary for the senior (including senior PI) level people to be involved in software. But in an ideal universe where people that the PI can hire are better in a short period of time than the PI, sure it makes sense to outsource the software activities.
\item It depends. While full time staff members allow us to make faster progress, good Graduate/Undergraduate students are equally productive. Taking the MVAPICH2 MPI library as an example - Most of the designs in MVAPICH has been primarily (~70\%)been done by graduate students with full time staff members (~30\%).
\item Full-time staff will generally make faster progress, although they may be more resistant to adopting new SE methodologies. Of course, getting students involved is part of broader impacts and educating the next generation. Students may also be on the project longer or are at least less expensive.
\item Senior level staff is clearly expensive so the deployment of their resources is best done in their field of experience. Training students for such tasks should only be done when it is beneficial to their future careers and never when it distracts from mastering skills in their field.
\item Salary is a BIG tradeoff, as is competition with industry. Pros of grad student are longevity for ~5 years whilst in the program, and the fact that students are more affordable. But students can work slow and produce fewer outputs compared to senior/experienced staff.
\item Some senior staff are capable and willing to learn in these areas, but those that aren't cannot be persuaded to change in my experience. Targeting students (in an educational way--not as code monkeys) will likely be more effective in the long run.
\item For small to medium project the former is often cost prohibitive. For large projects, having a senior staff person may be desirable, even though post docs may be good for that role and fit the university mission better.
\item Senior staff (such as postdocs) can often be more productive earlier in a project. However, graduate students offer more certainty in some cases. They are less likely to leave unexpectedly mid-project.
\item Having a full-time senior-level staff person be the project manager for the PAPI project, back when I used to run it worked great because it provided continuity and we also hired students.
\item Students learn fast but might also leave to take a lucrative job elsewhere with the skills learned. Senior level staff can be mature and methodical in the learning process.
\item Graduate students have more expertise in the science and might be more driven to develop new applications, but basic software engineering could delay their research career
\item The full-time staff have to be considered educators as well. The real goal of my lab is to train students. I can't do that if students aren't allowed to touch the code.
\item When a ``production level'' tool is needed, a full-time is needed, still Master students can assist in several parts of the project and learn from the development phase.
\item Having a full-time staff is better as they will stay longer in the organization and don't have to balance between research and work as would be the case for students.
\item Staff members are critical. Students simply are not the ones one can rely on to build and maintain the systems (they of course play an important role)
\item Experienced full time staff will speedup research. Time invested in training students could be detrimental if you have limited budget and resources
\item Graduate students are not reliable, you need real developers for infrastructure. Grad students help with innovation and piloting new ideas.
\item If your software needs to survive and be sustained long term, there has to be a mix of professional staff and students in the team.
\item Students should not be writing production software. Meeting needs for sustainability and quality requires a professional approach.
\item I can't speak to senior research staff learning basic software engineering. Student labor is guaranteed to be ephemeral.
\item Students have no long-term interest in software quality, and they have other priorities (namely graduating).
\item Graduate/undergrads can do wonderful things, but its difficult to rely on them for the core development.
\item We don't have this issue. Our full-time, senior-level staff is well-versed in software science.
\item For faculty, this will be good, but the cost is high which shall be mitigated by the fund.
\item The former will typically need a higher salary than graduate/undergraduate students
\item long term development of methods does require a full time person for the long term
\item Having a pyramid of talent is typically more cost effective in my experience.
\item Sustainability. Extended time needed for training. Cost.
\item accuracy. can't trust graduate students all the time.
\item We mostly work with grad/undergrad students.
\item Hiring full-time staff is expensive
\item Both can work.
\item No comments
\item depends
\end{itemize}

\subsubsection*{A.5.17 Should we put PhD students in the project for non-research oriented tasks (\emph{e.g.,} software maintenance)? How about recruiting students pursuing Master’s or undergraduate degrees?}
\begin{itemize}
\item I think, it makes sense to involve PhD students in non-research tasks related to software development, but only partially. Here are some reasons why: a) these efforts may help students develop their software development/maintenance skills, which may eventually be valuable to them if they decide to go to non-academic positions; b) the PhD students assigned these tasks would have good knowledge of the scientific aspects of the software anyways, so it should not be a big overhead to them, as long as this is not a really-time-consuming effort; c) through tasks such as software maintenance, students may also increase their own visibility and grow their network, which may eventually benefit their career pathways. Master’s students may or may not be suitable for such tasks. Typically, to be able to maintain software, one needs to know not only the bare implementation details, but also some underlying scientific basis. Within 1 or 2 years of Masters training, students may not reach this level. Yes, they may contribute to somewhat more trivial tasks though that requires more guidance – \emph{e.g.,} to running tests and analyzing some extensive data. The tests should be designed or pre-created for them, either by the PhD students or by senior staff members. Analogously, the undergraduates are less suitable for tasks such as software maintenance but may be suitable for massive testing and analysis efforts.
\item Software engineering is an important concept. Lately we are hearing a lot about Research Software Engineers (RSE). These are personnel technically expected to be equipped with excellent software engineering skills. RSEs could be PhDs which means they need to be trained with software engineers. and maintenance skills. So I wouldn't totally separate both the aspects but balance it out. I would definitely put more focus on such skills at the MS or undergraduate level.
\item Although PhD students shouldn't be spending a significant fraction of their time doing this, performing software maintenance is a good way for new students to learn the code. And, it is a good learning experience for students to have to maintain and improve their code. Outstanding Master's students or UG students could perform these functions, but they may not be invested in the project enough or around long enough to handle all these requirements.
\item The PhD student who was responsible for the innovation/research should be involved portions of the software maintenance. This is valuable experience for the students in how to develop good software and will definitely help them in the future when the join the workforce in industry and are called upon to architect/design/develop large software frameworks. Same goes for MS/Undergrads.
\item They should all pitch in, but software is usually a means to an end, rather than the primary goal. I work regularly with students at all stages who pitch into these efforts. Sometimes, we cycle between growing / building software infrastructure, then applying them to new scientific problems, and then back to infrastructure / training materials motivated by the last cycle.
\item We should not put PhD students in the position of writing software unless it is directly tied to their current research. Undergraduate students are much better candidates as their availability and interest is much higher. However, the work to mentor an undergraduate (or graduate) student to use the best software practices is large and cannot be understated.
\item Definitely MS or undergrad students do an excellent job in maintaining software. I think Ph.D. students should never be put in that position, unless that earns them an assistantship to cover their tuition. Ph.D. is a special training that takes away a good chunk of lifetime. One should only do it to fulfill one's dreams or make a change they believe in.
\item Not all Ph.D. students wish to go on to careers in fundamental research. Those who wish to go to industry benefit greatly from involvement in such tasks. In particular, students earning their Ph.D.s in application areas (\emph{e.g.,} Chemistry) can make a stronger case for future employment in computational areas if they have this experience.
\item Certainly MS and UG students will gain valuable experience even from non-research oriented tasks. PhD students also learn a lot about the business of software from non-research oriented tasks, so they could certainly work on such tasks during the initial part of their program, before transitioning to primarily research.
\item This seems exploitative. We have joint postdocs with academia and never assign projects that are not publishable. How does it further a student's education to assign difficult tasks outside of the field of study? Is such deployment worthy of NSF funding or, put another way, the best use of research dollars?
\item I think it is OK for PhD (and even MS or undergrads) students to gain some experience in non-research oriented software work as both in academia (where they themselves may get software related funding or their research may involve some software) and industry they may deal with software.
\item I think in technical fields many ``non-research-oriented'' tasks are part of the research. \emph{e.g.,} is learning how to operate NMR research-oriented task? Is preparing samples for shipping to a synchrotron? Our students will need software skills that go beyond pure-research activities
\item Putting PhD talent on projects (again from an industry perspective) sometimes inspires them to do things slightly differently or find new bottlenecks to research and improve. I think some time in software maintenance makes them appreciate trade offs of development.
\item I think this is not fair to PhD students to be put on too much of the non-research oriented tasks, it is far too easy to put them at risk of not achieving their desired degree. I think it is less problematic with Masters graduate students or undergraduates.
\item No. It is wrong to use PhD students for such purposes. The Ph.D. students should be the innovators rather than the code monkeys. Master's students don't stick around long enough. Ditto for undergraduates.
\item Sometimes they need the support and if they can still do something innovative for their thesis, it can be helpful, although slower than when the project is more directly applicable to their research.
\item Yes, but like I noted earlier, I think if given a SW Development Supplement as part of a grant, the PIs will find creative ways to deploy the services from talented MS and UG students as well.
\item Only in a very limited way that supports their research and career goals. I think students joining a research group expect to make progress on research, not just learn software maintenance.
\item PhD students will suffer by having task that do not help them in their future careers. MS and undergraduate students may gain valuable experience doing this but may not be qualified.
\item Generally speaking for PhD students in science I would say no, unless they request this explicitly and / or aim to pursue a career in software engineering after their PhD.
\item It depends upon the PhD students interest. It is important to have a person in the position that understands professional software development and maintenance.
\item PhD students can benefit from working on these tasks depending on their primary area of research. Masters and undergraduate students definitely benefit.
\item That's a question of how much time they spend on it. I think that's part of learning how to work with software. Students need to learn that as well.
\item All students should have some experience in software development and maintenance, but no matter their level, they should not focus solely on that
\item No. just let us hire research staffs. perhaps several groups can share one research staff who get paid for a few months from each of our grants.
\item This isn't a yes/no answer: the question is driven by the cost to train for the task vs the amount of work delivered for that training.
\item Students should not be writing production software. Meeting needs for sustainability and quality requires a professional approach.
\item I think it is better to use full-time staff for this purpose if possible -- master's or undergraduate students can help.
\item NO! They should learn critical skills in these areas but their primary job is to learn and to research.
\item Minor maintenance tasks can be done by students. This is also a potential for cross-disciplinary work.
\item They all need training on this topic and we need to change the culture in WRT software sustainability
\item It will be fine if the PhD students are ok to cover both non-research and research oriented tasks.
\item No, graduate students efforts are best doing research. Maybe undergraduates for training purposes
\item PhD students can do it, but it is always helpful to train under and MS students
\item If their degree focus is on software development, then that could work.
\item Master's have proven to benefit the most in performing those tasks.
\item Maybe for a short time to give them the needed experience.
\item This is common practice and some students even prefer it.
\item PhD will probably perform the best for such a task.
\item Yes, but their role needs to be carefully defined.
\item Both sounds like a good idea.
\item No, very bad idea.
\item Yes and yes
\item Yes
\end{itemize}

\subsubsection*{A.5.18 Should the domain-science projects - funded through the CSSI program - be required to have the CISE collaborators engaged at a substantial level of effort to ensure that best practices for software engineering and data management are followed? Should the CISE projects - funded through CSSI - be required to ensure accountability from domain-scientists (who could have promised to provide test-cases in their letters of collaboration)?}
\begin{itemize}
\item First part of the question: probably not. Not all domain-science projects (\emph{e.g.,} in Chemistry/Material Sciences) require substantial software developments. In this case, collaborations with CISE awardees may not be needed. Another aspect – sometimes, the domain-science projects’ PIs may already be sufficiently versed in the software best practices or can obtain such a training (\emph{e.g.,} via MolSSI’s workshops in Chemistry). Nonetheless, the CISE collaborations should be encouraged – this looks like a good idea in general. The second part of the question: due to the NSF’s policies on the format of the collaboration letter, it is unlikely that the domain-scientists who provided letters of collaboration would have an opportunity to explicitly state the scope of their work. To date, this remains up to the PI to describe the nature of the collaboration. I think, it sounds like a good idea to incentivize some accountability from domain-scientists, but it may be harder to do if they are not funded (\emph{e.g.,} unfunded collaborations).
\item This may be a good step, but should be in an opposite way. It may be more appropriate to require ``software maintainability plan'' along with a data management plan, etc for domain-science projects - funded through the CSSI program. CISE collaborators are not necessary. This could be done through consulting service. On the other side, CISE projects-funded through CSSI should have the domain-science collaborators engaged as collaborators at a substantial level to provide and test the software/data and ensure the usability and quality. It is NSF’s mission to promote the progress of science. In addition, the domain-science collaborators are required to help recruit more researchers using the software. Most often, it only stays within the research groups of the domain-science collaborators.
\item The domain-science projects need to show how they will ensure best practices, but that doesn't necessarily have to involve CISE collaborators. Perhaps they have advisors from industry or their students are doing or have done internships in industry where they learn good practices. They could also send their students to courses where they will learn good practices. CISE CSSI projects should have accountability for good test cases from domain scientists if their software is to be used by researchers.
\item All proposals need to be judged holistically. It is not realistic to expect all domain-science proposals to achieve best practices at the same level as a CISE proposal. On the other hand, it is not realistic to expect CISE projects to achieve the same level of domain impact as domain-science projects. I think it is fair to judge proposals on both accounts, with the understanding that some projects will be stronger in one aspect and some will be stronger in the other.
\item Projects should be able to demonstrate that (i) they know about and follow best practices, and (ii) are relevant. That can be done in many different ways -- using some kind of CISE- or domain-affiliation is one of these, but ultimately I would leave the method by which they demonstrate these points to the PIs: There are excellent software engineers also in the domain sciences.
\item To the extent that it helps projects without stifling creativity, yes. Especially if funded by CSSI. I like the potential of this ``match maker'' role: CISE projects should include domain scientists, at least on an advisory board, hopefully funded on some level. Domain projects should include some CISE staff. Hopefully the CSSI grants can fund these.
\item I don't think there should be a set rule. Academics are most innovative if processes are not prescribed. Indeed, the SI2 program, which has morphed into CSSI now, is much too controlling. The special requirements that seem to now be part of the rubric by which projects are awarded or evaluated do not help innovation. They stymie innovation.
\item In general, yes. There may be cases where non-CISE researchers have strong software engineering skills, but this is the exception. Yes, in general to the second question. Without working with domain scientists, it is unlikely that the code will provide the necessary functionality for anything beyond trivial/obvious problems.
\item A big YES! Interdisciplinary science cannot win without both parties involved equally. CISE need input from domain scientists to advance science and domain projects need CISE to advance computing. So asking for test cases from domain projects is evidence of an excellent collaboration.
\item No, I don't think so. If productive mutually-beneficial collaborations are possible and/or exist, sure. But what CISE-funded people are motivated by is rarely what we (scientists) need: they have very different motivations than we do.
\item Having professional software development and maintenance skills are critical to the success of the projects. Its less important where those skills come from and more important that they are present and integral to the team.
\item I think these are really laudable goals, that are really hard to achieve with the realities of the funding levels. I think that it is worth trying to achieve these goals, though - striking a balance in any CSSI project.
\item There are two separate questions here. 1. CISE collaborators - while this is beneficial, it may put an unnecessary burden on the PIs to find the right match 2. Domain Scientists - this would be useful.
\item I think this combination of CISE and domain science divisions is a good one. In some cases, CISE may want to fund some CI projects which may impact many domain sciences.
\item Yes, although a true collaboration and engagement more likely happens before the proposal is submitted and thus scientists are involved and champions of the project.
\item Yes. I strongly believe in a industry and academia working together to make sure code quality is high as well as it does what it's suppose to with real world data.
\item domain science projects should be encouraged to take advantage of national networks that try to support ``CISE experts'' that assist domain sciences.
\item Yes, but there has to be funding and/or traditional scientific metrics (publications) to ensure that this will happen (\emph{e.g.,} for end user testing)
\item Collaboration between domain and computer scientists is critical. I'm not sure if it can be effectively forced, though.
\item Yes, I think maintenance and the quality of software in general is very important for domain-science projects.
\item Yes. Otherwise somebody will skip the step about building sustainable software or repeatable results.
\item Requiring such things can cause resistance -- it's better if people see how they will gain from it.
\item I am not familiar enough with the program to answer.
\item Yes. But, needs to be a meaningful collaboration
\item Yes, if the effort is funded at the same level.
\item YES. There are so many benefits to this.
\item This is highly dependent on the project.
\item If this requirement can be sustained.
\item I think so to both.
\item definitely yes.
\item Yes possibly
\item Yes and yes.
\item Not sure.
\item sure
\item yes
\end{itemize}

\subsubsection*{A.5.19 Are deep learning techniques disrupting or impacting the future directions of your discipline? If yes, then how?}
\begin{itemize}
\item There are some efforts in deep learning/machine learning (ML) in my discipline (excited states and nonadiabatic dynamics), and they may be useful from the practical perspectives (\emph{e.g.,} to accelerate some calculations). For instance, one can use ML techniques to accelerate nonadiabatic molecular dynamics calculations, one can use ML to search for some materials with desired properties or classify large datasets of materials and mine new ``guiding principles''. Ultimately the ML methods can not replace the approaches based on more rigorous theoretical grounds. So, ML can be useful to gain some technical advantages and develop some sophisticated chemical heuristic theories, but one should not excessively commit to such approach as a way to discover/understand new physics.
\item Yes! I work on developing software tools for performance analysis, comparison, prediction, validation, and now performance reproducibility. Deep learning is the key technique I use to combine various sources of information about the applications, runtime, hardware counters to gain insights into how these different sources are connected, what story do they tell, how can we learn from the past and project to the future, and how can we continue to learn and iterate our models based on new information. Without deep learning the questions I am interested in---comparing performance across applications and architectures, predicting resource utilization on future systems without ever running an application---would not be possible.
\item We define ``disruptive technology'' as technology that affects the overall ecosystem of development and use. Contrast that with ``sustaining technology'' that incrementally improves the ecosystem capabilities. Deep learning is requiring a reconsideration of how software is developed and deployed and used; and so we consider it disruptive.
\item yes. it's a new disruptive technology. There is pressure to learn, and even apply, DL techniques even before the benefit is clear. This is no different from other disruptive technologies, and can end up in a benefit, however. Just the pace at which DL as a disruptive technology emerges seems unprecedented.
\item Yes, but I'm not sure how. Part of it is based on the perception of what deep learning can/could do. For instance, people are using deep learning to explore supercollider events to search for unexpected results/physics. This is interesting, but the value will depend on what results from this exploration.
\item Chemistry maps well onto deep learning. Local chemical structures determine local properties of a molecule, and the global molecular properties arise from these local properties. Deep learning is beginning to see some use in chemistry, and I expect that it will continue to grow.
\item Yes, we recently created a deep learning log parser. While not the most exciting nor ground breaking thing, it will save time from hand coding parsing rules. Just one example but I strongly believe deep learning will change cybersecurity, but I'm biased as I work at NVIDIA.
\item Yes. Let me illustrate with an example. Before DL, no one cared about large message MPI\_Allreduce. Traditional scientific community was focused on 8-64 byte reductions. DL has made the community focus on very large message reductions as well as GPU-awareness
\item They are impacting the future directions of computational chemistry. They are opening the road to thinking of problems in a different way and to finding answers by learning from the years of scientific data and knowledge which we have accumulated.
\item Both disrupting and impacting. Because of the hype, people tend to apply DL to problems that is not appropriate for that technology. In the meantime, some high quality research do stand out from the crowd.
\item Partially, the prevalence of deep learning techniques has created an environment that now is overhyped to do machine learning, which has decreased the incentive to do software development in the sciences.
\item Cynic here. Lots of claim that ML will solve all kinds of problems. it will solve some. At present, ML via DL doesn't produce ``explainable'' results. For me, that means you don't have science.
\item Deep learning techniques are gaining popularity in the neuroscience discipline however it is early to say if the techniques are disrupting the directions of neuroscience research.
\item They have had some impact, but not transformative. They have been and will probably be most effective in routine applications, but not likely in innovative or high-impact areas.
\item I'm trying to understand this across many disciplines and am keenly interested in the answer here - unfortunately, I do not have any insight on my own discipline.
\item Yes. We are working hard to integrate machine learning approaches with mechanistic simulation science. (I'd say we're on the leading edge of that in cancer.)
\item Yes. They are making new applications of machine learning possible, such as detecting infrastructure or environmental conditions automatically.
\item I suppose. But my interests are in why things work, md just not in having the, work. Therefore, deep learning has more limited applications.
\item Not as much in my area due to the combinatorial nature of problems (bioinformatics, graph analytics, \emph{etc.}).
\item Yes. By making it easier to fit models to a bunch of data where it's not clear what the model should be :)
\item So far, deep learning techniques are intriguing, but I do not see supplanting traditional techniques.
\item Yes. It is probably going to impact every discipline, but not dramatically reshape the future.
\item Yes for some problems. \emph{e.g.,} deep learning was applied to predict many materials properties.
\item disrupting, large amount of data, embedding techniques to go beyond simple classification.
\item There is no direct impact of DL in CI software we deal with regarding being disruptive.
\item yes. need new training and perhaps hire people with different skill sets.
\item We're interested from the hardware/software requirements point of view.
\item yes -- I work on using data science and NN for Multiscale coupling
\item Not sure. I think more impactful will be mixed-method approaches.
\item New requirements and opportunities, but not disrupting yet.
\item Some. But there are many aspects where that is not the case.
\item Yes--changing how graph algorithms are used
\item yes, they improve some algorithms.
\item Again, this is project dependent.
\item Possibly, but currently unclear.
\item yes. involved a lot.
\item N.A.
\item No
\end{itemize}

\subsubsection*{A.5.20 What are some of the effective approaches for software dissemination?}
\begin{itemize}
\item Having a central registry helps by providing a place to look. somebody has to manage that directory. It is difficult because you want some kind of sensible indexing, and we don't know how to do that for software very well, let alone for the problem domains to which they apply. Current repositories like GitHub don't have useful indexes in this sense. A major problem with ``software reuse'' is the artifacts are often monolithic because they are overly committed to the particular task or technologies they use. Another approach is find a way to factor software into more reusable parts. This requires research into building solutions from components; the components are more likely to be reused than the entire solution.
\item Publications that discuss software/data to be made Open Access (some outlets cost quite a bit for Open Access, so where would those \$\$ come from?), talk about software in different outlets (social media, conferences, symposiums, domain-based outlets as well), write blogs that are easy to follow than a technical publication, involve the community that the software is targeting. Public engagement is not only a mechanism to provide evidence of the impact but it by itself is a form of impact.
\item For non-domain specific software, having a home in a software foundation is enormously helpful. Implementing a training program, takes it to the next level, when targeted at the right opportunities. For domain specific software, as best as I can tell, using some traditional domain venues for publishing is effective, and supporting a community of users and developers helps build visibility for the domain software.
\item 1. Conducting user group meetings (\emph{e.g.,} http://mug.mvapich.cse.ohio-state.edu/) 2. Conducting tutorials at conferences and workshops 3. Conducting Birds of a Feather session at conferences and workshops 4. Giving talks at various at conferences and workshops
\item Reach out to partners who are interested in co-developing the software, form an open source consortium when there are sufficient partners. Hold workshops at relevant conferences with hands-on training and feedback on software features for future development.
\item Software like technology is most effectively deployed based on trained individuals disseminating the products. This happens rapidly by having testbed and sprint style initiatives that train the community; with participation for all parts of the community.
\item Make software available online. Advertise it in scientific talks. Host new user workshops. Provide clear, concise user documentation. Provide some degree of user support. Make software available on XSEDE or via web portals.
\item The most effective that I have ever found is software-as-a-service, with high-quality web interfaces and REST APIs. That is how people use software outside academia, and how we should deliver much of our software.
\item Identifying target users, engaging with related organizations, identifying barriers, developing success metrics, and allocating resources to implement the plan
\item Opensource, lots of tutorials, webinars \emph{etc.} but bottom line is there has to be an user base who needs it, otherwise dissemination alone will not work.
\item GitHub and alike services, potentially – the educational gateways with such software and suitable training resources, workshops and mini-conferences.
\item Beginning with early adopters; Involving users in the software development from the start; Planning early for adoption and sustainability
\item Publishing scientific articles to describe the code which has been disseminated. Creating an online website for the software.
\item Workshops for (potential) users. Presentations at conferences for domain scientists about the software capabilities.
\item web pages and GitHub. There are many ``software repositories''. We should find ways to cross-reference them.
\item Through open-source sharing on GitHub. This is the only thing that has worked for me so far.
\item Conferences, blogging, and regularly promoting new functionality to gain a user base.
\item I think open source platforms such as Git and Bitbucket are effective and adequate.
\item Free hands-on workshop, ideally in conjunction with domain science conferences.
\item Twitter, GitHub, giving talks at conferences, running training workshops
\item Through git hubs, news groups, presentations, use of social media.
\item online and in person workshops with hands on training sessions
\item Making them available for download, providing a user manual.
\item Building broad user communities that can ``spread the word''.
\item Open source. Web service. One size does not fit all.
\item regularly maintained on open source/data platforms
\item containers connected to high-performance computing
\item Should be made mandatory by funding organization
\item Disseminate within an established community.
\item Make it public in a well known repository.
\item Summer schools with hands-on sessions.
\item Community workshops, meetings, demos.
\item share open source code and documents
\item Conferences, web-portals
\item GitHub works very well.
\item Commercialization
\item twitter
\item GitHub
\end{itemize}

\subsubsection*{A.5.21 How can software and data products be made discoverable, accessible, and usable by the community?}
\begin{itemize}
\item Perhaps, NSF could have built their own gateway/database of such software. This gateway would cover various disciplines and would have a database of registered users, who could add their projects (supported by the NSF or any other). The gateway would have some metrics tools monitoring the use/interest level of each software component. The NSF could then promote the use of the projects they have supported. This can be done at various events or through public outreach channels NSF may have. It would be great to and incentivize/encourage new projects to use/refer to the projects already in the list. One nice example of efforts along these lines is the MolSSI’s database of computational chemistry programs. This is certainly a worthy effort, although such databases should have some proactive components, \emph{e.g.,} one could do better in popularizing the tools registered in the database. Another approach to stimulate the software accessibility and usage by the community is through supporting training events and workshops. This is what is being done already by the CyberTraining program. This program should be continued since it takes a persistent effort to educate the community about what is available and this is what the CyberTraining program aims to support.
\item We need the software institutes that NSF spent resources planning and then never funded in a broad way. They can provide support to users in identifying appropriate products and using them, as well as providing feedback to developers on future innovations that are needed. An online smart library that helps users find research software with the features they need, as well as ratings from other users, would be an easier lift that would be a start.
\item An idea is that there can be one platform with multiple channels, where each channel could be an engagement ground for a community. We can ``register'' our data, software, product \emph{etc.} with some hashtags and key terms to be discoverable. One unified search option that can search across the entire platform or can provide filters for more focused search. Kind of a mixture of google and Slack, but for scientists.
\item I think it is important to have some affiliation with organizations that support and promote the software and data (\emph{e.g.,} NCAR for atmospheric modeling, Unidata for atmospheric science data). This can help raise the visibility of both software and data, especially for domain science, far beyond what a single project or practitioner can achieve.
\item All of the above holds good for this question as well. Adding software/data products mandatorily in a publication will bring it more visibility, dockerization and containerizing of software will make it more accessible and usable, also loading software/data into a repository that's most popular in a field would also help.
\item I'd argue though that what's missing is a ranking of open source software. Once there is a ranking we should see an improvement in the quality of the software and also its use (dissemination). The ranking can be done like in top500 - a software competition every year in certain defined areas.
\item See previous answer. I think there has to be an incentive to place the the products in a centralized repository so that one has sufficient volume to ensure that would be users come look. NSF could arguably run such a repository (even if contracted out commercially).
\item Make software available online. Advertise it in scientific talks. Host new user workshops. Provide clear, concise user documentation. Provide some degree of user support. Make software available on XSEDE or via web portals.
\item I think now a days this is not a big problem given how easily everything is known by everybody via social medial, lots of conferences, webnars, websearch \emph{etc.} Real issue is if there is a need and user base for the software.
\item Having some repository for these tools would be a great option. Not sure what user communities have this. Having information available from professional societies/conferences is also useful.
\item Include cloud-hosted models, find ways to get software into other package management systems. It would be nice if there were truly cross-platform package managers.
\item Software should be open source. Developers should provide examples, tutorials and a mailing list. Information should be available on a dedicated webpage.
\item The above mentioned cross-references could contain agreed metadata to inform about important features for discovery and usability.
\item Discovery through publications and talks. And usable by providing decent documentation, access to well-commented code.
\item With appropriate metadata, in open repositories that are searchable, following the FAIR principles.
\item Conferences, blogging, and regularly promoting new functionality to gain a user base.
\item Twitter, scientific papers, giving talks at conferences, running training workshops
\item symposiums at national conferences just for new software and data products?
\item Deploy them on the community cyberinfrastructure, such as XSEDE resources.
\item Have a research community moderate and curate a software collection.
\item A registry that shows how the products meet community standards.
\item Publish the code in a paper as well as making it open source
\item Create curated searchable repository of research resources.
\item through multi ways, such as publications, websites, ....
\item through open source/data platform, with DOI links \emph{etc.}
\item Keep everything well documented and openly available.
\item Historically, through publication and research talks.
\item As I said we have not cracked the code on this.
\item Every software product should have a DOI
\item Docker hub, singularity hub, Dock Store
\item Be sure they are locatable via Google.
\item software expos, hackathons, workshops
\item social media, GitHub, publications
\item Should be standardized by NSF
\item Website and publications
\item Same answer as above
\item arXiv for software?
\item put on GitHub.
\end{itemize}

\subsubsection*{A.5.22 Hosting software or data on GitHub may not be sufficient for meeting the community engagement needs of the project. What are some of the approaches to engage the community?}
\begin{itemize}
\item Some ideas include adding author survey results such as - how many times has the software/data been used for other academic collaborations, has the software influenced policy by a decision maker, help disseminate information in layman language for easier understanding of the problem and solution (science is not finished until it is communicated), disseminate software/data and results via talks in more than one type of a conference, for example with interdisciplinary science projects it is important to talk about the software in both domain as well computer science outlets/conferences and workshops.
\item ``may not be sufficient'' is definitely spot-on here. Community engagement needs to be multipronged, from providing support through some kind of community user email, through holding Birds of a Feather sessions at appropriate conferences, through tutorials whereever appropriate, and targeting specific sessions at conferences for paper submission and presentation. The use of GitHub alone does not accomplish community engagement, thought it is helpful to make sure that the software is not hidden.
\item Discussions, good requirement elicitation, being involved in truly collaborative efforts where disciplines eventually disappear, communication techniques, recognizing there is a learning process about the vocabulary and incentives across disciplines, institutions and sometimes countries, finding common goals.
\item The problem is that the community needs to be educated in order to be able to contribute. We believe a strong outreach program is a must for CSSI projects. Such programs should not just teach how to use the program, but should also teach the underlying techniques so that others can contribute.
\item GoSciLack -- the platform I just made up would work. On the channels, scientists can engage in discussions as well. Or people can make their own groups (channels) and chat about their future plan.
\item Most software builders want to solve their problem, not contribute to a repository. The research funding has to somehow make this contributory effect obligatory or it basically won't happen.
\item It has to go through domain science researchers. Need to make it as a requirement for the domain science PI/Co-PI in order to request further funding.
\item Hosting events where the software developers and the community meets. Have events like lightening talks, speed-dating of software and domains
\item Wikis, hackathons, training meetings. Online training videos. Cloud-hosted coding environments. cloud-hosted model instantiations.
\item GitHub of course is only a part of the engagement. Publishing papers that demonstrate software is a minimum threshold in science.
\item Social Media, in-person testing and training, collaborations that result in scientific publications that use your software
\item Host new user workshops. Provide some degree of user support. Make software available on XSEDE or via web portals.
\item Train them in the code, make them use the code and rely on the code, so they are motivated to develop it further.
\item Open Source development requires a community and shared community practices, c.f., Apache Software Foundation.
\item Attend workshops in the targeting domain, and publish in journals that is well known to the community.
\item CSSI PI meetings. Other networking project meetings. Holding events at major conferences, such as SC.
\item Conferences, blogging, and regularly promoting new functionality to gain a user base.
\item Why is GitHub not sufficient? I suspect the answers are very project dependent.
\item Workshops, tutorials and hackathons can help increase community engagement.
\item Provide diverse training materials (tutorials, video lectures, workshops).
\item As far as I know, GitHub is the main hosting site. I don't know others.
\item The promise of support, maintenance, stability and interoperability.
\item Develop software that even your grandmother could run...
\item Fair point. We are investigating other approaches.
\item Create forums, Organize workshops and meetings.
\item Webinars, tutorials, videos, workshops, talks.
\item Workshops, collaborative projects, education
\item Advertising at conferences and in meetings.
\item Community-oriented conferences, forums.
\item social media, publications, conferences
\item software expos, hackathons, workshops
\item Don't have a clear comment on this.
\item Hosted by NSF
\end{itemize}

\subsubsection*{A.5.23 How can software be enabled on XSEDE systems?}
\begin{itemize}
\item For application software, negotiating with service providers to install and provide modules (and perhaps containers) is key. Beyond that, making sure that the software is showing up in the XSEDE software search is essential - this is typically pulled from module information. For other kinds of software, \emph{e.g.}, non-domain-specific software with user interfaces, it is much harder, as this is typically not in XSEDE's scope to enable and support this software. Just because it isn't in scope today, though, doesn't mean that it shouldn't be in scope tomorrow.
\item Most high-quality software has lots of dependencies, and XSEDE managers do not do their communities a favor by installing X different compilers and Y different MPI installations. Chances are that if you want to build a package, that one of the dependencies is only available with one compiler and one MPI installation, and that there is possibly another dependency that uses a different compiler. This just makes it unnecessarily hard to install software. In the end, efforts such as xSDK will help provide many of these dependencies.
\item SPs provide software module, user documentation and support. NSF should encourage PI to work with SP to enable the capability and continuing user support. This could be used as a metric for evaluating the impact of the project.
\item One option is to use something like Spack -- a content manager for HPC. It makes it very easy for people to make their software available through Spack to the community.
\item production level APIs, Agave/Tapas needs to be thought of as a production level project with appropriate testing. Pegasus is awesome but complex to install and use.
\item XSEDE systems need to be incentivized to allow research software to be installed on their systems. Adequate funding for enabling software on XSEDE is also needed.
\item We have been prototyping methods to build Jupyter notebook-based GUIs for command-line C++ simulators, and connect them to back-end infrastructure like XSEDE.
\item My first encounter with the term. This looks on the surface like such central repository. Don't know enough about it to judge.
\item By working with organizations contributing resources to XSEDE (\emph{e.g.,} TACC, SDSC) to test and deploy software.
\item Provide a world-accessible directory where research groups can install and maintain their software.
\item I've never done it, but I assume you would first contact XSEDE staff and work from there.
\item We are using XSEDE to validate/tune our software. Not sure what the question is asking.
\item Can NSF-supported software be automatically created as a module by XSEDE?
\item Good compiler/library support. User support from computing center staff.
\item Via Binder and Jupyter notebooks (easy installation and accessibility)
\item Software can be installed on XSEDE resources used by the community
\item I think XSEDE news and webinars are good way to do this.
\item this is important. Need a full time staff to do this.
\item I am not sufficiently familiar with XSEDE to comment.
\item Allowing the software to be a module on XSEDE systems
\item Better staffing support for the entire community
\item XSEDE users should voice their needs to XSEDE.
\item containers and virtual environments
\item Not sure what this question means.
\item I have no experience with XSEDE.
\item Do not have a good answer.
\item no comments
\end{itemize}

\subsubsection*{A.5.24 What are the metrics for evaluating the impact of a project?}
\begin{itemize}
\item The most important metrics for evaluating the impact of domain projects is scientific discovery in that particular domain, with the particular software that is being developed and supported. This is something that takes time to fully appreciate, but certainly literature analysis would be one way to start to understand what may be going on with respect to impact. For non-domain-specific software, \emph{e.g.}, tooling that may be used to develop software, impact is much more difficult to ascertain. I'm very open to brainstorming on this.
\item Growth in users, new projects powered by the software, third-party papers and grants based on the project, third-party contributions back to the project, quality and quantity of software support new users. Interactions with new communities (\emph{e.g.,} at conferences). Social media engagement. Improvements to software usability.
\item These projects should not use traditional metrics like a research project. Number of users or downloads are fine, but not number of citations of the software. The metrics should more focus on the impact on the users, researchers and the community, like the number of publications that used the software/data products.
\item With respect to NSF, traceability from support activities (components, libraries) to publications seems like the key thing to measure. Somehow one needs to measure demand for a component, estimated effort to construct it, and whether it actually got used. Highly trafficked components would then be successful.
\item The most important in most cases would be publications based on the developed software, but this can take years to develop. Downloads and a visible user community (\emph{e.g.,} in a message board or at a user work shop) would provide earlier evidence of impact. Evidence of adoption by industry would also be valuable.
\item The number of collaborations, publications, number of students trained, perhaps letters of support after the project ends to follow up from the partners about how they perceive the impact, technology transfer or interest from the industry or government labs.
\item Usability is one of the primary metrics. Who is using the software and how has that impacted science? Has the software influenced a policy maker? Has the project enabled to do more science? Gather evaluation/feedback from stakeholders,
\item This is where the SI2 program, over time, really created a very restricting requirement. These shouldn't be prescribed metrics. Evidence can come in many forms. Right now, the SI2 program expects bean counting that is unrealistic.
\item 1. Number of downloads/releases 2. Number of organizations using the software stack 3. Number of external contributors 4. Number of publications arising out of the project to measure the innovation component.
\item \# of adopters, ongoing users, user satisfaction, publications, software produced, software downloads, consortia formed, \# of open source contributors, amount of code produced by open source contributors
\item long term: how did it enable science (\# publications referring to the sw, but also highlights describing how the sw enabled a project). Short term: dissemination (download) and use.
\item Usability and usefulness - is this enabling scientists to answer scientific questions? Is this solving a real-world problem and not only improving previous algorithms?
\item Software features, publications, publications by users besides the developers, number of users, outside code contributors, number of people downloading the software.
\item I think number of users (for software/CI projects), publications, usage of other resources (such as HPC, HTC) by the project etc are good measures.
\item Multiple factors: a successful software product, training, new areas where this was used, serving under-represented communities.
\item Deliverables, publication, users, impact on science, solution or advancement of key problems and fulfillment of the grant.
\item Depending on the project, adoption rate, usage, and publications resulting from the project are some metrics
\item platform delivery, code delivery, domain scientists involvement and usage, publications, workshops
\item Publications, open-source/open-data shared online, tutorials, workshops, and presentations
\item High impact publications, size of user base and citations of a software project or database
\item Citations in papers. But how about creating an open review model for software?
\item Active End users, publications, citations, website hits and analytics, \emph{etc.}
\item User adoption, derivative software, citations, downloads, issues filed...
\item 1. Science advancement 2. Adoption 3. Progress towards sustainability
\item reuse, DOIs, citations, future grant funds for further development
\item publications, software usage tracking and analytics, citations
\item Number of publications, citations, downloads on GitHub.
\item For SaaS: uptime, usage, support ticket response time.
\item publications, user download times, and workshops.
\item Very unclear. More guidance from NSF would help.
\item Unique problems addresses, how widely it is used
\item Usability and use. Publications. Citations.
\item Citations. Better metrics would be great.
\item Both publication and software metrics.
\item includes business need and alignment
\item Scientific discoveries and citations
\item citations/downloads/users
\item Usage, citations
\item Adoption
\end{itemize}

\subsubsection*{A.5.25 How can the magnitude of the impact of the CSSI program be amplified?}
\begin{itemize}
\item Hiring a science communicator (or pay for one in a grant) who will have the right skills and abilities to amplify the impact. Publications that discuss software/data to be made Open Access (some outlets cost quite a bit for Open Access, so where would those \$\$ come from?), talk about software in different outlets (social media, conferences, symposiums, domain-based outlets as well), write blogs that are easy to follow than a technical publication, involve the community that the software is targeting. Public engagement is not only a mechanism to provide evidence of the impact but it by itself is a form of impact.
\item With so many projects over the years, it is definitely hard to see the impact arising from the projects, as this is not raising out of the trees very well. Some things to think about: using the NSF news mechanism to publicize successes arising from CSSI investments, this mechanism certainly raises the visibility of many important discoveries (which are are always fascinating!) and taking an approach which tries to make sure we amplify the successes arising from the CSSI investments would help amplify the impact, I believe. This is not the only mechanism, but it is one that people pay attention to.
\item I think there is a need for facilitating cross-cutting team building---that is, enabling people from different backgrounds to discover each other's needs and forging collaborations so that they can accomplish larger and more impactful projects, than tiny ones. The tiny ones are good for early career people though, for establishing themselves. I really think if CSSI to amplify its impact, one way would be to fill this gap---facilitate all hands from different backgrounds so that we can do bigger and more challenging things. This workshop is a step towards that.
\item Look for ways to break monolithic components into smaller, more reusable pieces and make those widely available. NSF needs to invest more in understanding and enable software composition techniques to ensure that these smaller components become explicit artifacts. To avoid overcommitment, components must not be merely code, but packages of design decisions that drive the code choice.
\item Make it clear to PIs and reviewers that building ``high quality, long lasting'' software is the main metric, not the ``scientific'' innovation. From my experience being a frequent reviewer, a lot of proposals get high ratings despite it being obvious that the PIs have no experience in, and don't care about, best practices in software engineering and making their software widely usable.
\item It has to be related to the impact on the research enabled by each project. Unlike research projects, NSF should spend more efforts on reviewing the products from the projects through peer-review process in the community or user surveys. NSF ought to factor this into consideration of the project renewal and continuing support.
\item The CSSI should advertise what software are available, organized in such a way that potential users could easily navigate to their field of interest. Additionally, if there could be enforcement of the quality of public software in terms of usability and accuracy, I expect that usage would increase.
\item Enable more small teams with faster evaluation and funding cycles. Shared software engineering resources (national core services?) Fast funding cycles for very small prototyping to complement slower funding/evaluation cycles for more expensive efforts.
\item As an academic pursuit, it must fundamentally promote the training and education of students and postdocs so they are more productive in their field. This is very natural for some disciplines and seems a distraction for others.
\item CSSI projects should include a specific plan to promote the developed software to its user community. User support and documentation make software more usable, and therefore also ``amplify'' impact.
\item offer funds to integrate software from multiple CSSI projects into interoperable frameworks, enable software and data publications that can be used for academic evaluations
\item Address sustainability seriously. Fund less research and more development of software on which application groups can depend. Communicate CSSI successes and capabilities.
\item through better publicity to people who are meant to use the software--conferences, demos, \emph{etc.} Provide opportunities for potential users to review the software
\item Promote and participate in other meetings/conferences in addition to the XSEDE conference. Keep up current emphasis on student participation (long-term impact).
\item Supporting successful projects. Only few will succeed but they should be supported. Often projects die just when they start producing results.
\item 1. Incorporation into larger frameworks 2. Leveraging established entities such as the supercomputing centers, XSEDE and software institutes.
\item This is a difficult question. One way would be to continue funding projects that have had a lasting impact on the scientific community.
\item Funding projects which themselves act as productivity levers, \emph{e.g.,} extensible frameworks, automation tools, \emph{etc.}
\item open forum for discussing challenges, including social/personal issues beyond the technical
\item build a consortium and have annual meetings for grantees and others and monthly webinars
\item Sustained long-term support for a number of collaborative ``moon shot'' type projects
\item I think how many users using a software should be used as a measuring criteria.
\item More partnership with industry and more promotion of SW to foster adoption.
\item providing tutorials and examples to make it easy for other disciples to use
\item Find and communicate highlights of the use of CSSI products.
\item need to relate content knowledge to real-world applications.
\item By collaborating with other federal agencies DOE, NIH etc
\item Moving science towards applications with societal impact.
\item have funding to be directed towards software development
\item Engaging with broader software development communities.
\item Identify strategic goals and focus on those?
\item I first need to learn more bout the program.
\item more domain science community involvements
\item greater industry engagement
\item no comments
\item Not sure.
\end{itemize}

\subsubsection*{A.5.26 NSF has invested millions of dollars - across multiple directorates - for building web-portals that support computation, data analyses, and data management from the convenience of the web-browser. Are you aware of such investments, and if needed, will you be able to comfortably leverage these investments in your current or future projects?}
\begin{itemize}
\item Yes, one of such portals will be developed in one of my recently funded projects. Thanks to NSF. Other notable efforts are those of MolSSI and potentially some of the older NanoHub developments. What MolSSI has been doing so far is very helpful and it makes impact on us as well. I can potentially leverage some of the NanoHub resources, but so far we did not have immediate need for them. I’m not aware of other web-portals, so it would be great to have NSF’s-wide database of such projects with some metrics of their use/impact publicly available and constantly advertised.
\item NSF has funded the project like SGCI and PIs should be strongly encouraged to work with the centralized service like SGCI for related science gateway development. Yes. CHARMM-GUI is a web-portal that I used in almost all my projects for system setup. While it represents a tiny amount among all NSF investments in web-portals, it is a very successful project that impact thousands of researchers in the field of biophysics and computational chemistry.
\item As an intermediate to advanced bioinformatics lab, we do not use these web portals (I think they are impractical for our research, our HPCC workflows are much faster and easier to use on a day-to-day basis). However, I would like to use them for teaching and training.
\item I am not aware of those investments, but would like to learn more about them. I would be very happy to leverage any such opportunity, if that falls in the scope of what I am doing.
\item My field is not very externally data-driven; I have not felt much impact from web-based portals (except the EMSL Basis Set Exchange and of course Google Docs).
\item Not a lot of them but probably a few, the question prompted me to search more of these. I would be definitely comfortable leveraging these investments
\item Some, but knowing what's available for particular purposes and understanding the advantages and disadvantages of them for my use case is difficult.
\item I'm aware of the investments, and would definitely want to start there on any future projects that require web accessibility.
\item These developments are often of poor quality and reinvent capabilities available elsewhere. 
\item I am not aware of these. Many companies will not be able to use these portals, due to IP and security issues.
\item As a small lab, the only way to survive in infrastructure development is to leverage the work of others.
\item Aware to a limited extent. Haven't been using/exploring these but should look at it more closely.
\item Need more awareness, but would love to break free of proprietary lock-in to leverage these.
\item I am aware of this as a general trend, but I have not seen it in my own field of work yet.
\item I was not aware of this investment but would like to use them in future projects.
\item Yes I am aware of certain projects and may use in the future if still available
\item I don't use them much because I primarily run applications on supercomputers.
\item I am aware. At present, they provide very little in terms of use.
\item Not aware. To be fair, I don't work exclusively in this space.
\item I am not aware as much about these investments/efforts.
\item I am aware of the developments, but have not used them.
\item Some but NSF can do a better job in advertising this.
\item not much aware and need more educated or publicized.
\item Yes. We are currently trying to do just that.
\item I am less aware than I probably should be
\item I mostly think the answer is yes for me
\item I think I am aware of some of them.
\item yes, but maybe not all of them
\item yes--these would be helpful
\item yes mostly involving XSEDE.
\item I'm not aware and maybe
\item no. I don't know them.
\item Somewhat
\item No.
\item yes
\end{itemize}

\subsubsection*{A.5.27 Would you like to provide any other comments?}
\begin{itemize}
\item Originally, the SI2 program had a very simple mission: Fund software as an infrastructure much like hardware (\emph{e.g.,} through the NSF supercomputing centers) is funded as infrastructure. A few years in, the focus seemed to shift more towards ``how do we build happy communities.'' Now, there seem to be all kinds of requirements to up-front state software engineering methods to be followed, \emph{etc.} We are academics. If we wanted to plan to the degree that is now required by NSF as part of the grant proposal for CSSI, we would be in industry. Some of us have a demonstrated history of delivering high quality software. Evaluated if the software we propose to develop is of value to the community and within the scope of CSSI. If it is, fund us and get out of the way!
\item These questions are fundamental and were discussed years ago at the seed meetings I attended. I am surprised to see they are still outstanding after 511 funded proposals and \$314 million has been spent. Why aren't we discussing experiences including success and failures from the funded proposals? There is existing data to analyze and what are the metrics that have been used for those?
\item I think it is a mistake to believe that sustainability of scientific software is fundamentally different that sustainability of software in general. Investment by NSF in ``software sustainability'' in general would have a much broader effect, and would likely bring a much larger community to the problem with corresponding scale benefits.
\item I'm hoping we have a really productive 2 days, a lot of good has come from the investments already made, but I think that we have to think hard about how to amplify that good and help the investments achieve and sustain the broad usage and impact after the initial investment is finished.
\item Need to be able to support teams lead by domain expertise paired with CISE collaborators. Need to help encourage best practices without stifling creativity.
\item This seems like a great opportunity to build a network among people who are likely participants of a future CSSI grant. Thank you for organizing!
\item I often feel like this community is more advanced in collaborative ideas than others I travel in, such as the science community.
\item MCB has done a great job advertising new tools or new develops from their division via MCB blogs. Perhaps CISE can do that too.
\item I m eager to learn more and contribute based on my biological research-centric perspective.
\item These questions are well thought. Thank you for the efforts.
\item Important program that I hope NSF continues to invest.
\item i hope to see the CSSI program a success.
\item Nothing further at this time.
\item No thanks.
\item No.

\end{itemize}

\end{document}